\documentclass[a4paper,preprint,superscriptaddress,nofootinbib,pra,11pt]{revtex4-1}
\usepackage[dvipdfmx]{graphicx}
\usepackage{bm}
\usepackage{amssymb}
\usepackage{amsmath}
\usepackage{color}
\usepackage{cancel}
\usepackage{comment}
\usepackage{here}
\usepackage{multirow}
\usepackage{url}
\usepackage{braket}
\usepackage{tikz}
\usepackage{float}
\usepackage{tikz-feynhand}
\usepackage{mathrsfs}
\usepackage[colorlinks=true,allcolors=blue]{hyperref}

\begin{document} 

\title{Electroweak baryogenesis via top-charm mixing}

\author{Shinya Kanemura}
\email{kanemu@het.phys.sci.osaka-u.ac.jp }
\affiliation{Department of Physics, Osaka University, Toyonaka, Osaka 560-0043, Japan}

\author{Yushi Mura}
\email{y\_mura@het.phys.sci.osaka-u.ac.jp }
\affiliation{Department of Physics, Osaka University, Toyonaka, Osaka 560-0043, Japan}


\preprint{OU-HET-1174}

\begin{abstract}
We investigate a scenario of electroweak baryogenesis in the two Higgs doublet model with quark flavor mixing.
In general, off-diagonal components of quark Yukawa interactions with additional Higgs bosons are strongly constrained by the data for flavor changing neutral currents.
However, top-charm quark mixing is not the case, so that a large off-diagonal element can be taken, which can contribute to generating baryon asymmetry of the universe.
It turns out that CP violating phases of the off-diagonal element in the source term in the Boltzmann equation are eliminated by the rephasing.
This result is somewhat different from previous works on electroweak baryogenesis by flavor off-diagonal Yukawa couplings.
Instead, we find that the absolute value of the top-charm off-diagonal element enhances CP violating phases in the Higgs potential, by which sufficient amount of the baryon number can be generated to explain the observed baryon asymmetry of the universe.
We find that such a scenario is viable under the current experimental data.
The model can be tested by the current and future measurements of various flavor experiments like Kaon rare decays, in addition to high energy collider experiments as well as gravitational wave observations.
Characteristic predictions of our model would be deviations in Kaon rare decays.
Branching ratios of $K^+ \to \pi^+ \nu \overline{\nu}$ and $K_L \to \pi^0 \nu \overline{\nu}$ can deviate by the order of 10\% and 1\%, respectively, which may be tested at future Kaon experiments such as NA62 and KOTO step-2.
\end{abstract}

\maketitle

\section{Introduction\label{sec:intro}}

Baryon Asymmetry of the Universe (BAU) is one of the unsolved big questions of particle physics and cosmology~\cite{ParticleDataGroup:2022pth}.
It is plausible that the BAU was generated in the early universe after the era of cosmic inflation before that of the big-bang nucleosynthesis, which is so called baryogenesis.
In order to realize baryogenesis, Sakharov's conditions~\cite{Sakharov:1967dj} must be satisfied; i.e., (i) baryon number violation, (ii) C and CP violation, (iii) out of thermal equilibrium.
It has turned out that the Standard Model (SM) does not satisfy the conditions (ii) and (iii)~\cite{Huet:1994jb,Kajantie:1996mn,DOnofrio:2015gop}, while (i) is realized by sphaleron at high temperatures~\cite{Klinkhamer:1984di}.
Therefore, a new theory beyond the SM is necessary, and a lot of scenarios of baryogenesis have been proposed up to now such as GUT baryogenesis~\cite{Yoshimura:1978ex, Weinberg:1979bt}, leptogenesis~\cite{Fukugita:1986hr}, Electroweak Baryogenesis (EWBG)~\cite{Kuzmin:1985mm}, and so on.

EWBG is a promising scenario, in which Sakharov's conditions are satisfied as follows.
(i) Baryon number is changed by the sphaleron process, (ii) CP violation in the new particle interactions is used, and (iii) thermal non-equilibrium is realized by strongly first order electroweak phase transition.
Models with extended Higgs sectors can basically satisfy these conditions, so that they are candidates for the scenario of EWBG.
Non-vanishing CP violating phases appear in the Yukawa interaction and the Higgs potential.
On the other hand, the extended Higgs sector can make electroweak phase transition to be strongly first order.
As EWBG relies on the physics of electroweak symmetry breaking, models for EWBG predict many characteristic phenomenological consequences below the TeV scale.
Therefore, such a scenario of EWBG can be tested at various current and future experiments.

Two Higgs Doublet Model (2HDM) is one of the simplest models where the scenario of EWBG can be realized~\cite{Turok:1990zg,Cline:1995dg,Fromme:2006cm,Cline:2011mm,Tulin:2011wi,Liu:2011jh,Ahmadvand:2013sna,Chiang:2016vgf,Guo:2016ixx,Fuyuto:2017ewj,Dorsch:2016nrg,Modak:2018csw,Basler:2021kgq,Enomoto:2021dkl,Enomoto:2022rrl,Zhou:2020irf}.
In general, scenarios of EWBG in the 2HDM with CP violation are strongly constrained by the current data from experiments of the Electric Dipole Moment (EDM)~\cite{ACME:2018yjb,Abel:2020pzs,Roussy:2022cmp}.
However, it has been shown that in the CP violating 2HDM the current data of electron EDM can be satisfied by destructive interference among Barr--Zee type diagrams~\cite{Kanemura:2020ibp}.
Recently, it has been shown that the observed BAU can be explained in this model~\cite{Enomoto:2021dkl,Enomoto:2022rrl}.

In this scenario of EWBG, rich phenomenological predictions are obtained as follows.
Effects of the additional CP violating phases can be tested by future EDM experiments, various flavor experiments and high energy collider experiments like LHC and future lepton colliders such as the International Linear Collider (ILC)~\cite{Fujii:2015jha, Bambade:2019fyw}.
Physics of strongly first order phase transition via the quantum effects of additional scalar bosons can be tested by the future measurements of the triple Higgs boson coupling~\cite{Kanemura:2004ch} at the future high energy colliders such as High Luminosity LHC (HL-LHC)~\cite{Cepeda:2019klc} and energy extensions of the ILC or the Compact Linear Collider (CLIC)~\cite{CLICdp:2018cto}.
Furthermore, the nature of the strongly first order phase transition is expected to be tested at the future space-based Gravitational Wave (GW) interferometers~\cite{Grojean:2006bp,Kakizaki:2015wua,Hashino:2016rvx,Hashino:2018wee,Kanemura:2022txx} like LISA~\cite{LISA:2017pwj}, DECIGO~\cite{Seto:2001qf}, etc.

Generated BAU can be evaluated by solving the transport equations which are deduced from Boltzmann equations.
In many scenarios of EWBG including the above scenario of the 2HDM, due to the largest Yukawa coupling constant with CP violating phases, top quarks play an important role in the transport equations; i.e., the top quark transport scenario~\cite{Joyce:1994fu,Joyce:1994zn, Joyce:1994zt, Fromme:2006wx,Fromme:2006cm, Cline:2011mm,Enomoto:2021dkl,Enomoto:2022rrl}.
However, there are possibilities where other particles than top quarks become important, i.e., the bottom quark transport or the tau lepton transport scenario.
Furthermore, if off-diagonal components of Yukawa interaction matrices are not small, charge transport scenario using the flavor mixing such as $\tau$-$\mu$ and top-charm mixing can also be considered for EWBG~\cite{Cline:2000nw,Prokopec:2003pj,Prokopec:2004ic,Chiang:2016vgf,Fuyuto:2017ewj,Guo:2016ixx}.
In general, off-diagonal components of quark Yukawa interactions with additional Higgs bosons are strongly constrained by the data for flavor changing neutral currents.
However, top-charm quark mixing is not the case, so that a large off-diagonal element can be taken, which can contribute to generating baryon asymmetry of the universe.
In the literature, BAU was evaluated solving transport equations based on closed time path formalism~\cite{Riotto:1998zb} in the charge transport scenario with flavor mixing, i.e., $\tau$-$\mu$ mixing~\cite{Chiang:2016vgf,Guo:2016ixx} and top-charm mixing~\cite{Fuyuto:2017ewj}.

In this paper, we evaluate the contribution from the top-charm mixing to the generated BAU by using the Boltzmann equations which are deduced based on the semi-classical force mechanism (WKB method)~\cite{Joyce:1994fu, Joyce:1994zn, Joyce:1994zt, Cline:2000nw, Fromme:2006wx,Fromme:2006cm,Cline:2020jre}.
It turns out that CP violating phases of the off-diagonal element disappear in the source term in the Boltzmann equation at the leading order.
This result is somewhat different from previous works on electroweak baryogenesis by flavor off-diagonal Yukawa couplings~\cite{Fuyuto:2017ewj,Chiang:2016vgf}.
Instead, we find that the absolute value of the off-diagonal elements in the Yukawa matrices of additional Higgs doublet contributes to BAU by the effect of CP violating Vacuum Expectation Value (VEV) which appears due to CP violation in the Higgs potential.

We find that the top-charm EWBG scenario is viable under the current experimental data.
The model can be tested by the current and future measurements of various flavor experiments like Kaon rare decays, in addition to high energy collider experiments as well as gravitational wave observations.
For example, top-charm mixing and its CP violating phase are important parameters to test the model via the signals at direct search experiments~\cite{ATLAS:2018jqi,CMS:2017bhz,CMS:2017ocm,CMS:2019rvj,ATLAS:2018rvc,CMS:2019pzc,ATLAS:2021upq,ATLAS:2022xpz} and various transition processes of $K$ mesons and $B$ mesons~\cite{UTfit:2007eik,UTfit2018,Chen:2018ytc,Haller:2018nnx,HFLAV:2022pwe,CMS:2022mgd,KOTO:2018dsc,NA62:2021zjw,NA62:2022hqi,Iguro:2023jju}.
In particular, they would be testable via the golden modes $K \to \pi \nu \overline{\nu}$~\cite{Iguro:2019zlc,Hou:2022qvx} at KOTO experiment at J-PARK~\cite{KOTO:2018dsc,Aoki:2021cqa} and NA62 experiment at the CERN SPS~\cite{NA62:2021zjw,NA62:2022hqi}. 
In our model, branching ratios of $K^+ \to \pi^+ \nu \overline{\nu}$ and $K_L \to \pi^0 \nu \overline{\nu}$ can deviate by the order of 10\% and 1\%, respectively, which may be tested at future Kaon experiments such as NA62 and KOTO step-2.

The paper is organized as follows.
In section~\ref{sec:model}, we define the general 2HDM with CP violation.
In section~\ref{sec:const}, various experiments which can constrain parameters of the model are discussed.
In section~\ref{sec:source}, the source terms for the transport equation are deduced by using the WKB method, and a benchmark scenario is proposed where the observed BAU is reproduced.
In section~\ref{sec:pheno}, predictions in our scenario are discussed, and relationships of top-charm mixing couplings and observables in the $K$ meson processes are given.
In section~\ref{sec:discuss}, we give some discussions for the results obtained in sec.~\ref{sec:source} and sec.\ref{sec:pheno}.
Finally, conclusions are given in section~\ref{sec:conclusion}.

\section{The model\label{sec:model}}
We consider a general Two Higgs Doublet Model (2HDM), which has two scalar $SU(2)_L$ doublet $\Phi_k = (\phi_k^+, \phi_k^0)^T~(k=1,2)$ with hypercharge $Y = 1/2$.
We can take the Higgs basis~\cite{Davidson:2005cw} by rotating the doublet fields without changing the form of the Higgs potential.
In this basis, only one of the scalar fields has a VEV, denoted by $\langle \phi_1^0 \rangle = v/\sqrt{2}$ and $\langle \phi_2^0 \rangle =0$~with~$v\simeq 246~\mathrm{GeV}$.
In the scalar potential without the $Z_2$ symmetry, we impose the Higgs alignment condition $\lambda_6 = 0$~\cite{Kanemura:2020ibp}.
Under this condition, the two scalar doublets $\Phi_1$ and $\Phi_2$ can be expressed as 
\begin{align}
    \Phi_1 = 
    \begin{pmatrix}
       G^+ \\
       \frac{1}{\sqrt{2}} ( v + H_1 + i G^0)
    \end{pmatrix},
    ~~
    \Phi_2 = 
    \begin{pmatrix}
       H^+ \\
       \frac{1}{\sqrt{2}} ( H_2 + i H_3)
    \end{pmatrix}.
\end{align}
The components $G^0$, $G^+$, $H^\pm$ and $H_k ~(k=1,2,3)$ are the Nambu--Goldstone bosons, physical charged scalar bosons and neutral scalar bosons which are mass eigenstates, respectively.
In the Higgs alignment scenario, $H_1$ is identified as the SM Higgs boson with the mass of 125 GeV, and $H_1WW$, $H_1ZZ$ and $H_1ff$~couplings coincide with the SM ones at the tree level.
We note that mixing of the neutral scalar bosons caused by non-zero $\lambda_6$ changes these couplings from the SM, and it is constrained by LHC results~\cite{ATLAS:2019nkf, CMS:2018uag}.
In this paper, we consider the scalar potential with the Higgs alignment condition.

The Yukawa interaction is given by
\begin{equation}
    \mathcal{L}_{y} = - \sum_{k=1}^2 \sum_{ij} \left( \overline{Q^\prime_{i,L}} Y^{u\dagger}_{k,ij} \tilde{\Phi}_k u^\prime_{j,R} + \overline{Q^\prime_{i,L}} Y^{d}_{k,ij} \Phi_k d^\prime_{j,R} + \overline{L^\prime_{i,L}} Y^e_{k,ij} \Phi_k e^\prime_{j,R} + \mathrm{h.c.} \right),
    \label{eq:yukawahbs}
\end{equation}
where we define $\tilde{\Phi}_k=i\sigma_2\Phi_k^*$.
Yukawa matrices have flavor indices $i$ and $j$, and the prime of the fermion fields represents gauge eigenstates in the flavor space.
In the following, we call the basis of the gauge eigenstates as the weak basis while the basis based on the mass eigenstates as the mass basis.
Fermion fields in the mass basis $f_i~(f=u,d,e)$, which have diagonal Yukawa interactions $Y_{\mathrm{d}}^f = \mathrm{diag}(y_{f_1}, y_{f_2}, y_{f_3})$ related to $\Phi_1$, are defined by
\begin{align}
    f_{i,L} = V^f_{L,ij} f_{j,L}^\prime, ~
    f_{i,R} = V^f_{R,ij} f_{j,R}^\prime,
\label{eq:weakandmass}
\end{align}
where $V_{L,R}^f$ are unitary matrices and satisfy $Y_{\mathrm{d}}^u = V_L^u Y^{u\dagger}_1 V_R^{u\dagger}$ and $Y_{\mathrm{d}}^{d,e}= V_L^{d,e} Y_1^{d,e} V_R^{d,e \dagger}$.
The Cabbibo--Kobayashi--Masukawa~(CKM) matrix is given by $V^u_L V^{d \dagger}_R \equiv V_{\mathrm{CKM}}$.
We also define additional Yukawa interactions related to $\Phi_2$ as $V^u_L Y_2^{u \dagger} V^{u \dagger}_R = \rho^u$ and $V^{d,e}_L Y^{d,e}_2 V^{d,e \dagger}_R = \rho^{d,e}$ in the mass basis.
In general, these matrices have off-diagonal elements.
Then, Yukawa interaction involving the physical scalar bosons is given by
\begin{align}
    \mathcal{L}_y \supset \sum_{ij} \Bigg\{ &\sum_{k=1}^3 - \overline{f_{iL}} g_{k,ij}^f f_{jR} H_k  \notag \\
    &+ \Big\{ \overline{u_{iR}} (\rho^{u \dagger} V_{\mathrm{CKM}})_{ij} d_{jL} - \overline{u_{iL}} (V_{\mathrm{CKM}} \rho^d)_{ij} d_{jR} -\overline{\nu_{iL}} \rho^e_{ij} e_{jR} \Big\} H^+ \bigg\} + \mathrm{h.c.} ,
    \label{eq:yukawatoH}
\end{align}
where,
\begin{align}
    \begin{array}{ccc}
        g^f_1 = \frac{1}{\sqrt{2}}Y_{\mathrm{d}}^f, & g^f_2 = \frac{1}{\sqrt{2}}\rho^f,\\
        g^u_3 = -\frac{i}{\sqrt{2}}\rho^u, & g^{d,e}_3 = \frac{i}{\sqrt{2}}\rho^{d,e}.
    \end{array}
\end{align}

We parametrize the additional Yukawa couplings as
 \begin{align}
    \rho^u = 
    \begin{pmatrix}
    \rho_{uu} & \rho_{uc} & \rho_{ut} \\
    \rho_{cu} & \rho_{cc} & \rho_{ct} \\
    \rho_{tu} & \rho_{tc} & \rho_{tt}
    \end{pmatrix}, 
    ~~
    \rho^d = 
    \begin{pmatrix}
    \rho_{dd} & \rho_{ds} & \rho_{db} \\
    \rho_{sd} & \rho_{ss} & \rho_{sb} \\
    \rho_{bd} & \rho_{bs} & \rho_{bb}
    \end{pmatrix},
    ~~
    \rho^e = 
    \begin{pmatrix}
    \rho_{ee} & \rho_{e\mu} & \rho_{e\tau} \\
    \rho_{\mu e} & \rho_{\mu \mu} & \rho_{\mu \tau} \\
    \rho_{\tau e} & \rho_{\tau \mu} & \rho_{\tau \tau}
    \end{pmatrix},
\end{align}
and define the phases of these complex couplings $\rho_{ij} \in \mathbb{C}$ as $\theta_{ij} = \mathrm{arg}(\rho_{ij})$.
Off-diagonal components of $\rho^f$ is constrained by flavor experiments, so that we set these to 0 except for the second and third generations of up-type quark.
In addition, for the discussions of EWBG with top-charm mixing, we do not take into account $\rho_{dd}$, $\rho_{ss}$ and $\rho^e$, unless otherwise noted.

\section{Constraints on the model\label{sec:const}}

In this section, we discuss experimental and theoretical constraints on the model.

The additional Yukawa couplings $\rho^u$ and $\rho^d$ affect $B^0$-$\overline{B^0}$ and $K^0$-$\overline{K^0}$ mixing, $B \to X_s \gamma$ and $B_s \to \mu \mu $ processes.
The UTfit results~\cite{UTfit:2007eik,UTfit2018} give constraints on $B_d$-$\overline{B_d}$ and $B_s$-$\overline{B_s}$ mixing amplitudes as $C_{B_d} = [0.83, 1.29]$, $\phi_{B_d} = [-6.0^\circ, 1.5^\circ]$ and $C_{B_s} = [0.942, 1.288]$ at 95\%~C.L., where $C_{B_q} e^{2i\phi_{B_q}} = \bra{B_q^0} H_{\mathrm{eff}}^{\mathrm{full}} \ket{\overline{B_q^0}} / \bra{B_q^0} H_{\mathrm{eff}}^{\mathrm{SM}} \ket{\overline{B_q^0}}~(q=d,s)$.
The upper limit of the 2HDM contribution to the indirect CP violation in $K^0$-$\overline{K^0}$ mixing is $|\epsilon_K^{\mathrm{2HDM}}|<4.0 \times 10^{-4}$~\cite{Chen:2018ytc}.
The formulae of Wilson coefficients of these $\Delta F = 2$ processes in our model are given by ref.~\cite{Crivellin:2013wna}, and we take into account QCD running effects with the B parameters in hadronic matrix elements shown in refs.~\cite{Becirevic:2001jj, UTfit:2007eik, Ciuchini:1998ix}.
The observed branching fraction of $B \to X_s \gamma$ is given by $\mathcal{B} (B \to X_s \gamma)_{\mathrm{EXP}} = (3.32 \pm 0.15) \times 10^{-4}$ with photon energy cut $E_\gamma > 1.6$ GeV~\cite{Haller:2018nnx,HFLAV:2022pwe}.
The SM prediction is evaluated by $\mathcal{B} (B \to X_s \gamma)_{\mathrm{SM}} = (3.36 \pm 0.23) \times 10^{-4}$ at NNLO in QCD~\cite{Czakon:2015exa}.
We define the theoretical prediction in our model as $\mathcal{B}(B \to X_s \gamma)_{\mathrm{th}} = R_{\mathrm{th}} \cdot \mathcal{B} (B \to X_s \gamma)_{\mathrm{SM}}$ where $R_{\mathrm{th}} = \mathcal{B}(B \to X_s \gamma)_{\mathrm{2HDM}} / \mathcal{B}(B \to X_s \gamma)_{\rho_u = \rho_d = 0}$~\cite{Modak:2018csw, Enomoto:2022rrl}, and we calculate this by using LO formulae shown in ref.~\cite{Crivellin:2013wna}.
The observed branching ratio of $B_s \to \mu \mu$ is given by latest CMS result~\cite{CMS:2022mgd} as $\mathcal{B}(B_s \to \mu \mu)_{\mathrm{EXP}}=\big(3.83^{+0.38}_{-0.36}~^{+0.24}_{-0.21} \big)\times 10^{-9}$.
The first and second uncertainties are statistical and systematical ones, respectively.
We have set $\rho_{\mu\mu}$ to 0 for simplicity, so that only the Wilson coefficient $C_{10}$, which is relevant to $Z$ and $\gamma$ penguin diagrams involving charged scalar bosons at 1 loop level, contributes to this process~\cite{Crivellin:2019dun}.
We define the theoretical value as~\cite{Iguro:2019zlc}
\begin{equation}
    \mathcal{B}(B_s \to \mu \mu) = \mathcal{B}(B_s \to \mu \mu)_{\mathrm{SM}} \left| 1 + \frac{C_{10}^{\mathrm{2HDM}}}{C_{10}^{\mathrm{SM}}} \right|^2,
\end{equation}
with $\mathcal{B}(B_s \to \mu \mu)_{\mathrm{SM}} = 3.66 \times 10^{-9}$~\cite{Beneke:2019slt} and use $C_{10}^{\mathrm{SM}}$ evaluated at NNLO in QCD~\cite{Bobeth:2013uxa}.
We require the theoretical predictions of $B \to X_s \gamma$ and $B_s \to \mu\mu$ to be within 2$\sigma$ deviation from the experimental values.

We next discuss constraints from direct searches for the additional scalar bosons.
Here we only focus on the absolute values of $\rho_{tc}$ and $\rho_{tt}$ because the other couplings are small in our benchmark points shown in the following sections.
The off-diagonal element $\rho_{ct}$ receives stronger constraints from the flavor experiment than $\rho_{tc}$, so that $\rho_{ct}$ cannot be large to be tested at the high energy collider experiments.

The off-diagonal element $\rho_{tc}$ contributes to the top quark decay process $t \to c H_1$ in non-alignment case.
When the mixing angle $\gamma$ among the neutral scalar bosons satisfies $\cos\gamma \simeq 0.3$, $\rho_{tc} \gtrsim 0.3$ is excluded~\cite{Altunkaynak:2015twa, Hou:2020tnc} by ATLAS~\cite{ATLAS:2018jqi} and CMS~\cite{CMS:2017bhz} data.
In our alignment scenario, the branching ratio coincides with the SM one at the tree level.

The additional Yukawa couplings $\rho_{tc}$ and $\rho_{tt}$ affect production processes of the heavy neutral scalar bosons via $gg \to H_{2,3}$ and $gc \to H_{2,3} t$ and their decay processes into $tc$ or $tt$.
Especially $\rho_{tc}$ produces same sign top quarks via $gc \to H_{2,3} t \to tt\overline{c}$ process, and this is constrained by the control region for $t\overline{t}W$ background (CRW) in the SM four top production searches~\cite{Kohda:2017fkn, Iguro:2018qzf, Hou:2019gpn, Hou:2018zmg, Hou:2020tnc, ATLAS:2022xpz}.
When $m_{H_3} = 350$~GeV and $H_2$ decouples, ref.~\cite{Hou:2020tnc} gives an upper bound $\rho_{tc} \sim 0.5$ by using the CMS data~\cite{CMS:2017ocm, CMS:2019rvj}.
If the mass and decay width of them are degenerated, this constraint is weakened due to the interference between $cg \to tH_2 \to tt\overline{c}$ and $cg \to tH_3 \to tt\overline{c}$ processes~\cite{Kohda:2017fkn,Hou:2018zmg}.
In our benchmark points with $m_{H_2} = m_{H_3}$, for $\rho_{tc} \gtrsim 0.4$ almost all of them decay into $t \overline{c}$ or $\overline{t} c$, so that a difference of their total widths is $O(10^{-1})$~GeV.
As a result, the constraint on $\rho_{tc}$ from CRW vanishes in our benchmark points.
On the other hand, $gc \to H_{2,3} t \to tt \overline{t}$ process induced by $\rho_{tc}$ and $\rho_{tt}$ was discussed in ref.~\cite{Hou:2019gpn}, and our benchmark points are not excluded by the CMS four top searches~\cite{CMS:2017ocm, CMS:2019rvj}.
When $m_{H_2},m_{H_3} > 2m_t$ with small $\rho_{tc}$, $H_{2,3} \to t\overline{t}$ searches by ATLAS~\cite{ATLAS:2018rvc} and CMS~\cite{CMS:2019pzc} constrain the additional Yukawa coupling $\rho_{tt}$~\cite{Enomoto:2022rrl}.
In sec.~\ref{sec:discuss}, we will mention these prospects at future collider experiments such as HL-LHC.

The parameter $\rho_{tt}$ is also constrained by $H^\pm \to tb$ search~\cite{ATLAS:2021upq} via the charged scalar bosons production $gb \to H^\pm t$.
If we take $\rho_{tc} = 0$, $|\rho_{tt}| \gtrsim 0.6$ is excluded with $m_{H^\pm} =350$~GeV from this constraint~\cite{Enomoto:2022rrl}.
The parameter $\rho_{tc}$ enhances the $H^\pm \to cb$ decay and suppresses $\mathcal{B}(H^\pm \to tb)$, so that the constraint on $\rho_{tt}$ is weakened for large $\rho_{tc}$.

We also consider oblique parameters~$S,T$ and $U$~\cite{Peskin:1990zt,Peskin:1991sw} constrained from the electroweak fitting results~\cite{Baak:2012kk}.
In our model, $\lambda_4 - \mathrm{Re}[\lambda_5]$ and $\mathrm{Im}[\lambda_7]$ terms in the scalar potential violate the custordial symmetry~\cite{Sikivie:1980hm,Haber:1992py,Pomarol:1993mu,Gerard:2007kn,Haber:2010bw,Grzadkowski:2010dj,Aiko:2020atr}.
These terms make a deviation in the $T$ parameter from the SM value.
In order to satisfy the constraint from the $T$ parameter, we simply take $m_{H^\pm}=m_{H_3}$, since $\lambda_4 - \mathrm{Re}[\lambda_5]$ is proportional to $m_{H^\pm}^2 - m_{H_3}^2$.
At the one loop level, the $T$ parameter is not affected by the effects from the $\mathrm{Im}[\lambda_7]$ term~\cite{Pomarol:1993mu,Haber:2010bw}.

The CP violation in our model is constrained by EDM experiments.
The electron EDM (eEDM) measured by ACME~\cite{ACME:2018yjb} gives strong bound.
We now take $\rho^e = 0$, such that the diagrams which contain $\rho^e$ disappear (e.g. two loop Barr--Zee type diagrams).
Even if we consider non-zero $\rho_{ee}$ which is relevant to the eEDM, we can avoid the constraint by using destructive interference among independent CP phases in our model~\cite{Kanemura:2020ibp, Enomoto:2021dkl, Enomoto:2022rrl}.
For example, in addition to our benchmark points shown below, if we set $|\rho_{tt}|=0.15$ and $|\rho_{ee}|=3\times 10^{-8}$~(corresponding to $|\zeta_e|\simeq 10^{-2}$ in ref.~\cite{Enomoto:2021dkl, Enomoto:2022rrl}), the ACME bound is satisfied when $2.7 \lesssim \theta_{ee} \lesssim 3.9$.\footnote{
The Cornell group reported latest eEDM bound which is about one-half of the ACME by using trapped $\mathrm{HfF}^+$ molecular ions~\cite{Roussy:2022cmp}.
This bound reduces the allowed $\theta_{ee}$ region by about half.
}
For simplicity, we neglect $\rho_{uu}$ and $\rho_{dd}$, so that contributions to the neutron EDM (nEDM) from Barr--Zee type diagrams vanish.
Although Weinberg operator produced by $\rho_{tt}$ and $\rho_{bb}$ contributes to the nEDM with large theoretical uncertainty~\cite{Demir:2002gg,Jung:2013hka}, we have confirmed that in our benchmark points the contribution is one order smaller than the current nEDM bound given by NEDM collaboration~\cite{Abel:2020pzs}.

As theoretical constraints, we take into account bounds from perturbative unitarity~\cite{Kanemura:1993hm,Akeroyd:2000wc,Ginzburg:2005dt,Kanemura:2015ska}, vacuum stability~\cite{Klimenko:1984qx,Sher:1988mj,Nie:1998yn,Ferreira:2004yd} and triviality~\cite{Flores:1982pr,Kominis:1993zc,Kanemura:1999xf,Ferreira:2009jb,Dorsch:2016nrg}.
Formulae for the bounds from perturbative unitarity and vacuum stability in 2HDM without $Z_2$ symmetry are given by refs.~\cite{Kanemura:2015ska, Ferreira:2004yd}.
We employ the renormalization group analysis, and in our benchmark points, the Landau pole at which the scalar self couplings diverge is above 3 TeV by considering the threshold effect of heavy scalar bosons~\cite{Dorsch:2016nrg}.

For numerical analyses, we use SM input parameters at the scale of the $Z$ boson mass shown in ref.~\cite{Enomoto:2022rrl}.
We here use \texttt{RunDec v3}~\cite{Herren:2017osy} which is a \texttt{mathematica} package to calculate the values at other energy scale.

\section{CP violating source terms and baryogenesis\label{sec:source}}

\begin{figure}[t]
    \centering
    \includegraphics[width=0.8\linewidth]{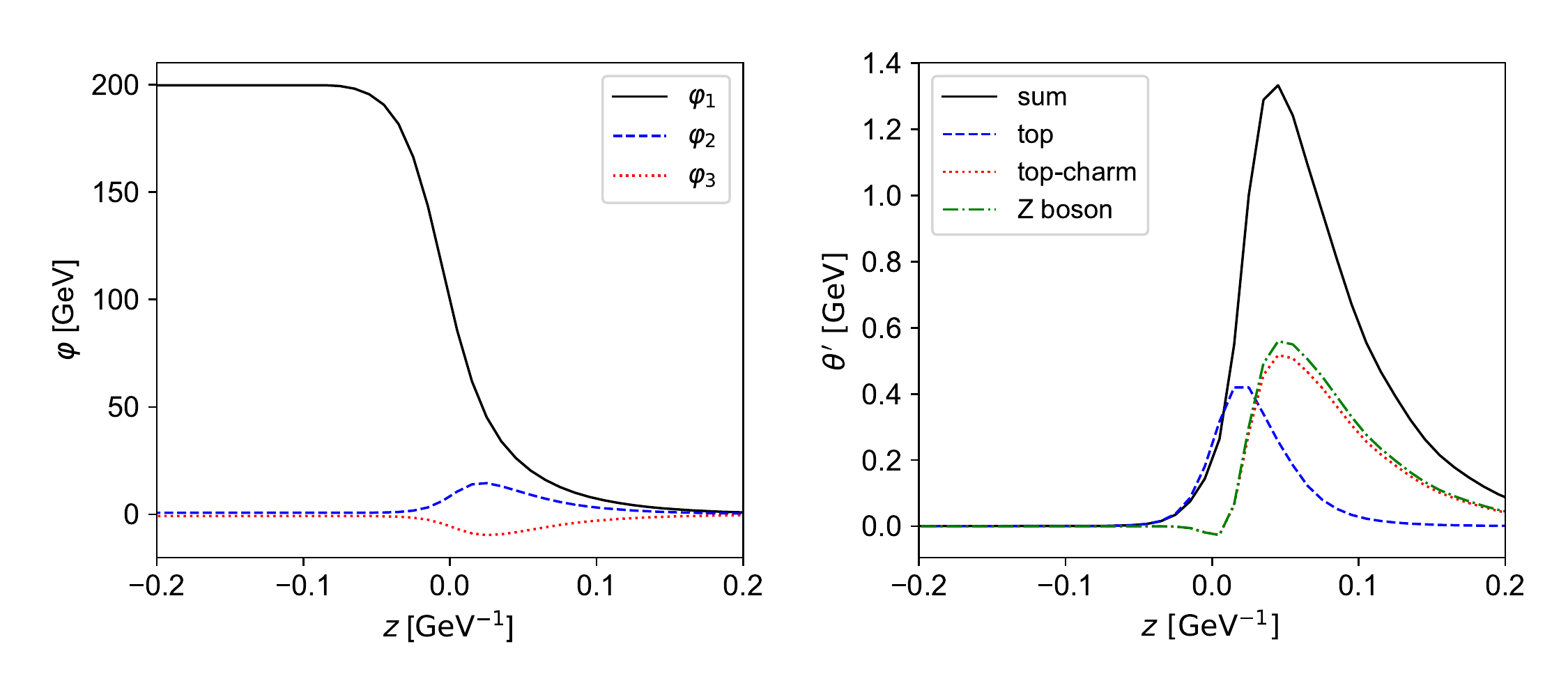}
    \caption{
    Left: bounce solutions in the benchmark point shown in eq.~(\ref{eq:imputpotential}).
    Black solid and blue dashed lines are CP conserving VEVs $\varphi_1$ and $\varphi_2$ respectively, and red dotted line is CP violating VEV $\varphi_3$.
    Right: black solid line shows derivative of the phase $\theta^\prime$ in the benchmark point with $|\rho_{tc}|=1.0$ and $|\rho_{tt}|=0.1$.
    Blue dashed (red dotted) and green dashdot lines are the top (top-charm) transport source and the source from chiral interaction with $Z$ boson and top and charm quarks, respectively.}
    \label{fig:profphis}
\end{figure}

First, we discuss strongly first order phase transition for EWBG.
We consider the CP violating effective potential~\cite{Enomoto:2021dkl, Enomoto:2022rrl} and calculate phase transition in the potential by using \texttt{CosmoTransitions}~\cite{Wainwright:2011kj}, which is a Python module package.
We assume that the bubble wall velocity is constant and set $v_w = 0.1$ in the following discussions.
We neglect the curvature of the wall and define radius coordinate as $z$.
In the Higgs basis, bounce solutions which are classical configurations of neutral scalar fields in the potential can be parametrized by $\langle \phi^0_1 \rangle = \varphi_1(z) / \sqrt{2}$ and $\langle \phi^0_2 \rangle = \big(\varphi_2(z) + i\varphi_3(z)\big) / \sqrt{2}$.
We take input parameters which are relevant to the phase transition as 
\begin{gather}
    m_\Phi \equiv m_{H_2}=m_{H_3} = m_{H^\pm}  = 350~\mathrm{GeV},~ M=20~\mathrm{GeV}, \notag \\
    \lambda_2 = 0.01, ~|\lambda_7| = 1.0,~ \mathrm{arg}(\lambda_7) = -2.4,~ |\rho_{tt}| = 0.1,~\theta_{tt} = -0.2,
    \label{eq:imputpotential}
\end{gather}
and we numerically obtain the bounce solutions at this benchmark point.
The left panel of fig.~\ref{fig:profphis} shows the solutions.
The center of the bubble $z=0$ is defined by the spacial point maximizing $d v(z) / dz$, where $v(z) \equiv \sqrt{\sum_i\varphi_i^2(z)}$.
The black solid and blue dashed lines are CP conserving VEV $\varphi_1$ and $\varphi_2$, respectively, and the red dotted line is CP violating VEV $\varphi_3$.
At this benchmark point, the ratio of the VEV inside the wall $v_n \equiv v(-\infty)$ and the nucleation temperature $T_n$ is $v_n / T_n =2.4$ with $T_n=84.6$~GeV, so that the sphaleron process inside the wall sufficiently decouples.
The wall width $L_w$, which can be obtained by fitting $v(z)$ with the function $\frac{v_n}{2}\big(1-\tanh\frac{z}{L_w}\big)$, satisfies $L_w T_n =3.3$, so that the derivative expansion in the WKB approximation is still valid~\cite{Fromme:2006cm}.
In the following analyses, we use this bubble profiles for the calculations of BAU related to the top-charm mixing couplings.

Second, we discuss CP violating source terms in the Boltzmann equation which are based on the semi-classical force mechanism with the WKB approximation~\cite{Joyce:1994fu, Joyce:1994zn, Joyce:1994zt, Cline:2000nw, Fromme:2006wx, Cline:2020jre}.
According to ref.~\cite{Cline:2000nw}, we can derive the source terms in the top-charm quarks system.
From eq.~(\ref{eq:yukawatoH}) the mass term is written by $\mathcal{L}_{\mathrm{mass}} = -\overline{q_L} M(z) q_R + \mathrm{h.c.}$ with two flavor quark $q = (c,t)^T$.
The space dependent mass matrix is given by
\begin{align}
    M(z) = \frac{1}{\sqrt{2}} 
    \begin{pmatrix}
    y_c \varphi_1 + \rho_{cc} (\varphi_2 -i \varphi_3) & \rho_{ct} (\varphi_2 - i\varphi_3) \\ \rho_{tc} (\varphi_2 - i\varphi_3) & y_t \varphi_1 + \rho_{tt} (\varphi_2 - i\varphi_3)
    \end{pmatrix}.
    \label{eq:massmatrix}
\end{align}
The off-diagonal elements of the matrix arise due to the non-zero VEVs $\varphi_{2,3}$ along to the wall.

We take a basis in the flavor space which diagonalizes $MM^\dagger$ with a local unitary matrix $U$, so that $U M M^\dagger U^\dagger \equiv \mathrm{diag} (m_+^{2}, m_-^{2})$.\footnote{
We here have defined $m_+^2 \ge m_-^2$.
} 
At the first order derivative expansion, the CP violating source terms are given by diagonal elements of $\mathrm{Im}(A_1) \equiv \mathrm{Im}(U M \partial_z M^{-1} U^\dagger )$~\cite{Cline:2000nw}.
If we assume a hierarchy of the mass matrix in eq.~(\ref{eq:massmatrix}) as $|M_{11}|/|M_{22}| \simeq |M_{12}|/|M_{22}| \equiv \delta$, we obtain $(A_1)_{++} = (M_{22}^* M_{22}^\prime + M_{21}^* M_{21}^\prime) \big(1 + O(\delta^2) \big) / m_+^2$, where the prime denotes a derivative of $z$.
We define $\mathrm{Im}(A_1)_{++} \equiv -\partial \theta(z) / \partial z (= - \theta^\prime)$, where $\theta$ corresponds to the phase of the mass of $(1,1)$ component of the field in the locally diagonalized basis.
When $\sqrt{\varphi_2^2 + \varphi_3^2} \ll \varphi_1$, $|\rho_{cc}| \ll y_t$ and $|\rho_{ct}| \ll y_t$, this $O(\delta^2)$ correction is negligibly small: $\delta_{\mathrm{max}}\simeq 0.06$ with the bubble profile in the left panel of fig.~\ref{fig:profphis} and $|\rho_{cc}|, |\rho_{ct}| < 0.1$.
Neglecting this correction of at most $0.4 \%$, we obtain the source term of the heavy component as 
\begin{align}
    \mathrm{Im}(A_1)_{++} = &-\frac{1}{2m_+^2}~ \Bigg\{
    (|\rho_{tc}|^2 + |\rho_{tt}|^2 ) ( \varphi_3 \varphi_2^\prime  - \varphi_2 \varphi_3^\prime ) \notag \\
    &\qquad\qquad\qquad\qquad+ y_t |\rho_{tt}| \Big(  (\varphi_3 \varphi_1^\prime - \varphi_1 \varphi_3^\prime) \cos\theta_{tt} + (\varphi_1 \varphi_2^\prime - \varphi_2 \varphi_1^\prime ) \sin\theta_{tt} \Big) \Bigg\} \notag \\
    &-\frac{1}{\varphi_1^2 + \varphi_2^2+\varphi_3^2}(\varphi_3 \varphi_2^\prime  - \varphi_2 \varphi_3^\prime ), \label{eq:topsource}
\end{align}
where,
\begin{equation}
m_+^2 = \frac{1}{2}\Big( y_t^2 \varphi_1^2 + (|\rho_{tc}|^2 + |\rho_{tt}|^2)(\varphi_2^2+\varphi_3^2) + 2y_t |\rho_{tt}| \varphi_1 \big(\varphi_2 \cos\theta_{tt} + \varphi_3\sin\theta_{tt} \big) \Big).
    \label{eq:topmass}
\end{equation}
The second term of eq.~(\ref{eq:topsource}), which has no dependence of $\rho_{tc}$ and $\rho_{tt}$, is stemmed from the axial vector interaction of charm and top quarks with $Z$ boson~\cite{Cline:2011mm}.
For the transport equations, we replace the source terms of the top quarks in ref.~\cite{Enomoto:2022rrl} as
\begin{equation}
    S_{lt} \to -\gamma v_w x^\prime Q_{lt}^8 + \gamma v_w x (m_+^2)^\prime Q_{lt}^9, ~~(l = 1,2)
    \label{eq:CPVsource}
\end{equation}
where $x \equiv - m_+^2 \mathrm{Im}(A_1)_{++}$.
The source terms of the light fermion in the local flavor basis are proportional to $m_- \simeq 0$, so that we neglect it.

We give a comment on the CP violating source terms shown in eqs.~(\ref{eq:topsource})-(\ref{eq:CPVsource}).
In the limit of vanishing the top-charm mixing coupling $\rho_{tc}\to 0$, eq.~(\ref{eq:CPVsource}) coincides with the source terms in the top transport scenario~\cite{Cline:2011mm,Enomoto:2021dkl,Enomoto:2022rrl}.
The source term with top-charm mixing only depends on the absolute value of $\rho_{tc}$ through the CP violating VEV $\varphi_3$.
Therefore, the phase of $\rho_{tc}$ does not affect the BAU up to $O(\delta^2)$, and the $|\rho_{tc}|$ dependence of the BAU vanishes with the CP conserving VEV limit $\varphi_3 \to 0$.
These results have obtained by using the WKB method in the calculation of the source term of the Boltzmann equation.

Although our analysis of evaluating the BAU is based on the WKB method in this paper, to see the consistency we also have examined the source terms in the VEV insertion approximation~(VIA)~\cite{Riotto:1998zb},\footnote{
    Recently, it has been pointed out that the VIA source terms within leading order in derivative expansion exactly vanish by performing correct resummation of 1PI self energy~\cite{Kainulainen:2021oqs,Postma:2022dbr}.
} which are based on the closed time path formalism.
As shown in appendix~\ref{section:app}, the CP violating source terms do not have any $\rho_{tc}$ dependence for $\varphi_3 \to 0$ at leading order in the VIA.
Even if we consider the case of $\varphi_3 \neq 0$, $|\rho_{tc}|$ contributes to the CP violating source terms while the phase does not.

However, these results for the source terms are different from the previous works in refs.~\cite{Chiang:2016vgf,Fuyuto:2017ewj}, where the source terms are calculated at leading order in the VIA.
For example in ref.~\cite{Fuyuto:2017ewj}, they have assumed that VEVs induced by the CP conserving Higgs potential are real, and the phase of $\rho_{tc}$ generates the BAU.
However, we have found that if one focuses on only one flavor in the weak basis as considered in refs.~\cite{Chiang:2016vgf,Fuyuto:2017ewj}, the VIA source terms depend on the rotation matrices $V_L$ or $V_R$ in the flavor space which can be taken to be arbitrary.
Namely, the phase effect of $\rho_{tc}$ on the CP violating source terms is unphysical.

In other words, if we consider contributions from the other flavor in the weak basis in a consistent way, we could easily see that the effect of the phase of $\rho_{tc}$ is unphysical.
For example, in ref.~\cite{Fuyuto:2017ewj}, it seems that the authors have not included the source terms of left-handed charm quarks (second generation) defined in the weak basis.\footnote{In ref.~\cite{Chiang:2016vgf}, only left-handed $\tau$ leptons are considered for evaluating the source terms for the lepton flavor mixing scenario of EWBG.
Therefore, the similar problem as the top-charm mixing is seen.
If $\mu_L$ in the weak basis is included in the transport equations, the effect of CP violating phases in the $\tau$-$\mu$ element of Yukawa matrix should disappear.}
If one sum up the source terms of both the top and charm quarks in the weak basis, the source terms have to coincide with the ones calculated in the mass basis in which independence of the phase of $\rho_{tc}$ is manifest, as we discuss in appendix~\ref{section:app}.
We have also explicitly confirmed that the source terms calculated by the WKB methods are basis independent in the leading order approximation.
We also discuss this issue in sec.~\ref{sec:discuss} and appendix~\ref{section:app}.

In the right panel of fig.~\ref{fig:profphis}, $\theta^\prime$ is shown as a function of $z$ by the black solid line, when we take $|\rho_{tc}|=1.0,~|\rho_{tt}|=0.1$ and use the bubble profiles shown in the left panel of fig.~\ref{fig:profphis}.
The blue dashed (red dotted) line is a contribution from top (top-charm) transport scenario which can be obtained by taking $\rho_{tc}~(\rho_{tt})\to 0$ in the first term in eq.~(\ref{eq:topsource}).
The green dashdot line is a contribution from the second term in eq.~(\ref{eq:topsource}), which is CP violating source caused by the interaction with $Z$ boson current and top and charm quarks~\cite{Cline:2011mm}.

\begin{figure}[t]
    \centering
    \includegraphics[width=1.0\linewidth]{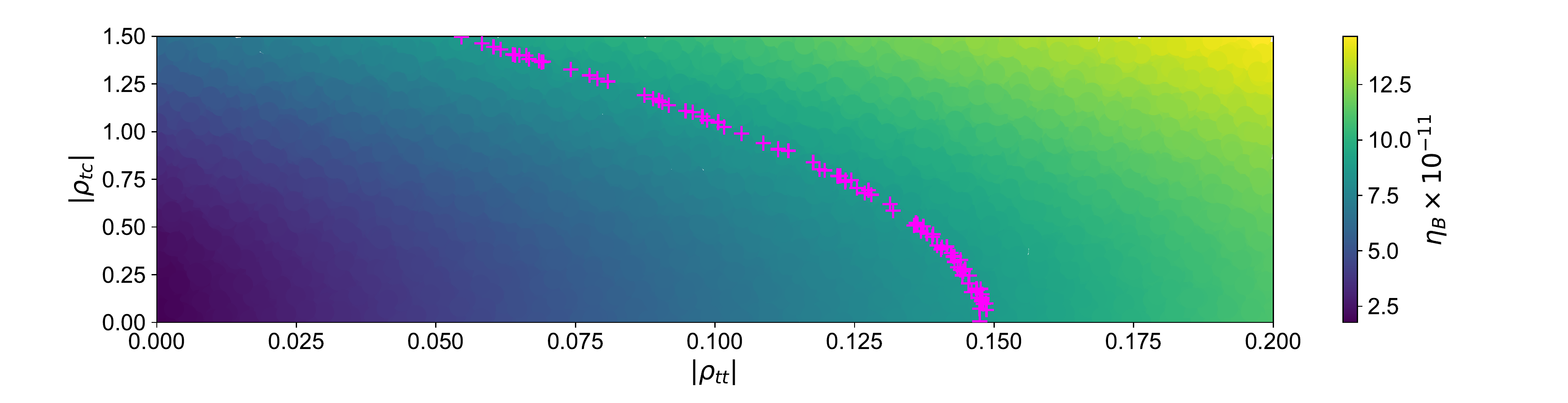}
    \caption{
    $|\rho_{tc}|$ and $|\rho_{tt}|$ dependence of baryon number density $\eta_B$ with $\theta_{tt} = -0.2$.
    Colored points represent magnitude of $\eta_B$, and magenta points satisfy $\eta_B = (8.65 \mathrm{-} 8.74) \times 10^{-11}$.}
    \label{fig:baryonscat1}
\end{figure}
Finally we discuss baryogenesis in this scenario.
The BAU parameter $\eta_B = (n_B - n_{\overline{B}}) / s$, where $n_B ~(n_{\overline{B}})$ and $s$ are the (anti-) baryon density and the entropy density, respectively, is invariant in the adiabatic expanding universe.
The observed value is $\eta_B^{\mathrm{obs}} \simeq 8.7 \times 10^{-11}$~\cite{ParticleDataGroup:2022pth}.
By using the source terms in eq.~(\ref{eq:CPVsource}) based on the relativistic semi-classical force mechanism~\cite{Cline:2020jre}, we have calculated the BAU as in ref.~\cite{Enomoto:2022rrl}.
We have fixed the phase as $\theta_{tt} = -0.2$, and calculated $\eta_B$ at many points in the $|\rho_{tc}|$-$|\rho_{tt}|$ plane.
The result is shown in fig.~\ref{fig:baryonscat1}.
The color in the figure corresponds to the magnitude of the BAU, and the magenta points satisfy the observed value $\eta_B = (8.65 \mathrm{-} 8.74) \times 10^{-11}$.
In fig.~\ref{fig:baryonscat1}, the impact of $|\rho_{tc}|$ to the BAU is shown.
It is seen that regions of large $|\rho_{tt}|$ and $|\rho_{tc}|$ generate large BAU.
Even in a small $|\rho_{tt}|$, the non-zero effect of the top-charm mixing coupling $|\rho_{tc}|$ gives a sufficient BAU by picking up the contributions from the CP violating VEV $\varphi_2^\prime \varphi_3 - \varphi_2 \varphi_3^\prime$.
We will mention the behavior of the BAU with other benchmark points in sec.~\ref{sec:discuss}.

\section{Phenomenological consequences\label{sec:pheno}}

In this section, we discuss phenomenological predictions in our scenario of EWBG with the top-charm mixing.
As the prediction of physics of strongly first order electroweak phase transition, there is a large deviation in the triple Higgs boson coupling, which can be tested by measuring the triple Higgs boson coupling at future high energy colliders~\cite{Kanemura:2004ch}.
The physics of the strongly first order phase transition can also be tested by GWs produced at the first order phase transition, which may be observed at future space-based gravitational wave interferometers~\cite{Grojean:2006bp,Kakizaki:2015wua}.
The effect of the additional CP violating phase can appear in various observables which would be measured in future EDM, flavor or collider experiments.
We will thoroughly discuss these common phenomenological predictions and testabilities of the scenario by using various future experiments in sec.~\ref{sec:discuss}.
In this paper, however, we concentrate on the discussion of the consequences from the top-charm mixing couplings of our model and discuss testabilities for future flavor experiments.

In general, the top-charm mixing couplings gives significant contributions of $K$ meson processes via the loop induced penguin or box diagrams.
In our model, we consider rare decay processes $K_L \to \pi^0 \nu \overline{\nu}$ and $K^+ \to \pi^+ \nu \overline{\nu}$ which are sensitive to additional Yukawa couplings.
These processes are produced by four fermi operators which are induced by penguin diagrams involving the charged scalar bosons~\cite{Iguro:2019zlc,Hou:2022qvx}.

The observed branching fraction of $K^+ \to \pi^+ \nu \overline{\nu}$ is given from the data collected from 2016 to 2019 in NA62 experiment at the CERN SPS as~\cite{NA62:2021zjw}
\begin{align}
    \mathcal{B}  (K^+ \to \pi^+\nu \overline{\nu})_{\mathrm{EXP}} = ( 10.6^{+4.0}_{-3.4} \pm 0.9) \times 10^{-11},
    \label{eq:Kplusexp}
\end{align}
where the first and second uncertainties are statistical and systematical errors, respectively.
$O(10)$\% accuracy is expected by the end of NA62 experiments with the data collected from 2021~\cite{NA62:2022hqi}.

For the process $K_L \to \pi^0 \nu \overline{\nu}$, KOTO experiment at J-PARK gives an upper bound on this branching fraction as~\cite{KOTO:2018dsc}
\begin{align}
    \mathcal{B}  (K_L \to \pi^0 \nu \overline{\nu})_{\mathrm{EXP}} < 3.0 \times 10^{-9},
    \label{eq:KLexp}
\end{align}
at 90\% C.L. 
This upper limit is greater than the Grossman--Nir bound~\cite{Grossman:1997sk}.
It is expected that KOTO step-2, which is extended version of KOTO expected to be launched from 2029 in the earliest scenario, achieves the accuracy predicted by SM and observes the events with $4.2\sigma$ significance~\cite{Aoki:2021cqa}.
KLEVER experiment at the CERN SPS expected to be operated after LHC run3 also aim to observe this process with 20 \% accuracy of SM branching fraction~\cite{Moulson:2019ifj}.

We define quantities relevant to these processes~\cite{Hou:2022qvx}.
\begin{align}
    \Delta \mathcal{R}^+_{\nu} \equiv \mathcal{R}_\nu^+ - 1, ~~~~\Delta \mathcal{R}^0_{\nu} \equiv \mathcal{R}_\nu^0 - 1,
\end{align}
where,
\begin{equation}
    \mathcal{R}_\nu^+ = \frac{\mathcal{B}(K^+ \to \pi^+ \nu \overline{\nu})}{\mathcal{B}(K^+ \to \pi^+ \nu \overline{\nu})_{\mathrm{SM}}},~~~~\mathcal{R}_\nu^0 = \frac{\mathcal{B}(K_L \to \pi^0 \nu \overline{\nu})}{\mathcal{B}(K_L \to \pi^0 \nu \overline{\nu})_{\mathrm{SM}}}.
\end{equation}
SM predictions of these processes in our input parameters are $\mathcal{B}(K^+ \to \pi^+ \nu \overline{\nu})_{\mathrm{SM}}=9.3\times 10^{-11}$ and $\mathcal{B}(K_L \to \pi^0 \nu \overline{\nu})_{\mathrm{SM}} = 3.1 \times 10^{-11}$, and these are consistent with the ones shown in ref.~\cite{Buras:2015qea} within 1$\sigma$.

The additional Yukawa couplings in our model also affect the direct CP violation $\epsilon^\prime / \epsilon$ in $K_L \to 2\pi$ process~\cite{Iguro:2019zlc}.
The observed value is given by $(\epsilon^\prime / \epsilon)_{\mathrm{EXP}} = (16.6 \pm 2.3) \times 10^{-4}$~\cite{NA48:2002tmj, KTeV:2002qqy, KTeV:2010sng}.
Lattice results give the SM prediction as $(\epsilon^\prime / \epsilon)_{\mathrm{SM}} = (21.7 \pm 8.4)\times 10^{-4}$~\cite{Blum:2015ywa, RBC:2020kdj, Hou:2022qvx}, while a result of chiral perturbation gives $(\epsilon^\prime / \epsilon)_{\mathrm{SM}}  = (14 \pm 5)\times 10^{-4}$~\cite{Cirigliano:2019cpi}.

Relevant parameters about the BAU discussed in ref.~\ref{sec:source} are $|\rho_{tt}|$, $|\rho_{tc}|,$ and $\theta_{tt}$.
In addition to eq.~(\ref{eq:imputpotential}), we take a benchmark point about the other Yukawa parameters for the $K$ meson observables as 
\begin{gather}
    |\rho_{cc}| =0.09, ~ |\rho_{ct}|=0.05, ~ \theta_{cc}=0, ~ \theta_{ct}=-2.8, ~ \theta_{tc} = -0.2, \notag \\
    |\rho_{bb}|=1.0 \times 10^{-3}, ~ \theta_{bb}= 1.5.
    \label{eq:inputyukawa}
  \end{gather}
\begin{figure}[t]
    \centering
    \includegraphics[width=1.0\linewidth]{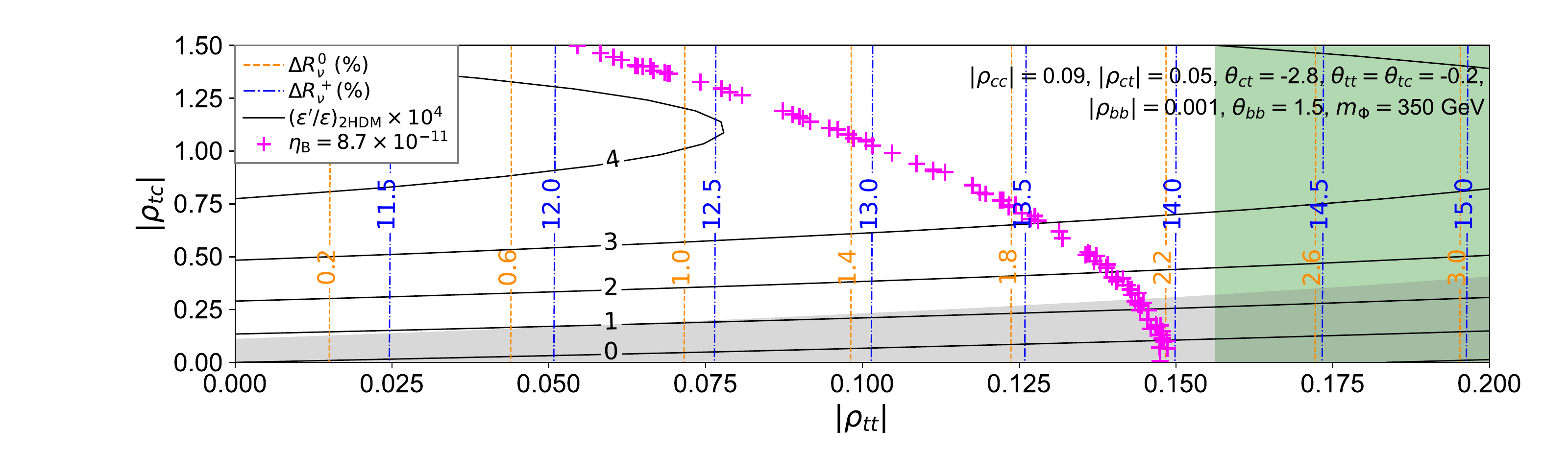}
    \caption{
    Constraints from flavor experiments and predictions on $K$ meson rare decays in eqs.~(\ref{eq:imputpotential}) and (\ref{eq:inputyukawa}).
    Green and gray regions are excluded region by the data of $B_s \to \mu \mu$ and $\epsilon_K$, respectively.
    Blue dotted dash and orange dashed lines correspond to $\Delta \mathcal{R}^+_{\nu} \times 10^2$ and $\Delta \mathcal{R}^0_{\nu} \times 10^2$, respectively, which are defined in the text.
    Black solid lines are the contour of $(\epsilon^\prime / \epsilon)_{\mathrm{2HDM}} \times 10^4$.
    Magenta points are the same as shown in fig.~\ref{fig:baryonscat1} and satisfy $\eta_B = (8.65 \mathrm{-} 8.74) \times 10^{-11}$.}
    \label{fig:baryonscat2}
\end{figure}
In fig.~\ref{fig:baryonscat2}, the green and gray regions are excluded by the data of $B_s \to \mu \mu$ and $|\epsilon_K|$, respectively.
The other flavor constraints are out of this figure, and we have confirmed $|\rho_{tc}| \gtrsim 1.5$ is excluded by the data of $B^0$-$\overline{B^0}$ mixing.
As shown in fig.~\ref{fig:baryonscat2}, $B_s \to \mu \mu$ constrains $\rho_{tt} \gtrsim 0.16$.
Since $B_s \to \mu \mu$ process mainly depends on $\rho_{ct} \rho_{tt}$ in our setup, this process only sets the upper bound on $|\rho_{tt}|$.
On the other hand, $|\epsilon_K|$ constrains the lower region of $|\rho_{tc}|$ in this benchmark point, and $|\rho_{tc}| \lesssim 0.25$ is excluded with $|\rho_{tt}| \simeq 0.1$.
This behavior of $|\epsilon_K|$ constraint in $|\rho_{tc}|$-$|\rho_{tt}|$ plane changes with other values of $\rho_{cc}$ and $\rho_{ct}$.
In fig.~\ref{fig:baryonscat2}, predictions of $\Delta \mathcal{R}^+_{\nu} \times 10^2$ (blue dotted dash) and $\Delta \mathcal{R}^0_{\nu} \times 10^2$ (orange dash) at the benchmark point are shown.
The new physics contribution of the direct CP violation $(\epsilon^\prime / \epsilon)_{\mathrm{2HDM}} \times 10^4$ (black solid) is also shown in fig.~\ref{fig:baryonscat2}.
We have used formulae shown in ref.~\cite{Iguro:2019zlc} to calculate these processes.
The magenta points are the same as shown in fig.~\ref{fig:baryonscat1}, corresponding to the observed baryon asymmetry.
The processes $K_L \to \pi^0 \nu \overline{\nu}$ and $K^+ \to \pi^+ \nu \overline{\nu}$ mainly depend on $\rho_{ct}\rho_{tt}$~\cite{Iguro:2019zlc}, so that only $|\rho_{tt}|$ changes the values in fig.~\ref{fig:baryonscat2}.
At the point $|\rho_{tt}| \simeq 0.15$, predicted deviations from SM branching fractions of $K_L \to \pi^0 \nu \overline{\nu}$ and $K^+ \to \pi^+ \nu \overline{\nu}$ are $\Delta \mathcal{R}^0_{\nu} \simeq 2.2\%$ and $\Delta \mathcal{R}^+_{\nu} \simeq 14\%$, respectively.
Therefore, the branching fractions are greater than the SM predictions at this point.
On the other hand, for $|\rho_{tt}|\simeq 0.05$, they are about $0.8\%$ and $12\%$, respectively. 
The direct CP violation $(\epsilon^\prime / \epsilon)_{\mathrm{2HDM}}$ 
depends not only on $\rho_{ct}\rho_{tt}$ but also on $\rho_{tc}\rho_{cc}$, so that it changes along $|\rho_{tc}|$ axis.
In the region where the observed BAU can be explained under the experimental constraints, allowed maximal value of $|\rho_{tt}|$ predicts $(\epsilon^\prime / \epsilon)_{\mathrm{2HDM}} \simeq 1 \times 10^{-4}$, while it is about $4\times 10^{-4}$ for $|\rho_{tt}| \simeq 0.05$.
We note that if we set $\rho_{ct}$ and $\rho_{cc}$ to be 0 in fig.~\ref{fig:baryonscat2}, only $|\rho_{tc}| \gtrsim 0.8$ is excluded by $B \to X_s \gamma$ constraint, almost without depending on $|\rho_{tt}|$.
In this case, $\Delta \mathcal{R}^0_{\nu}$ and $\Delta \mathcal{R}^+_{\nu}$ are small and less than $1\%$.

\section{Discussions \label{sec:discuss}}

We give some comments and discussions in this section.

In sec.~\ref{sec:source}, we have calculated the source term in the WKB approximation for the top-charm mixing EWBG.
From eq.~(\ref{eq:topsource}) it has been shown that phases of off-diagonal elements of additional Yukawa matrices do not contribute to the source terms up to $O(\delta^2)$.
The contributions to the source terms from the top-charm Yukawa couplings are proportional to the square of the absolute values of $\rho_{tc}$ through the CP violating VEV $\varphi_3$.
On the other hand, in the previous work for flavor mixing EWBG studied in refs.~\cite{Chiang:2016vgf,Fuyuto:2017ewj}, the phases of the off-diagonal complex couplings, e.g. $\rho_{\tau \mu}$ or $\rho_{tc}$, play an important role in the source terms evaluated in the VIA method even considering real VEVs.
However, we note that this discrepancy does not come from the difference between WKB and VIA methods~\cite{Cline:2020jre, Cline:2021dkf}.
As we show in appendix~\ref{section:app}, the CP violating source terms calculated at leading order in the VIA in the mass basis do not depend on $\rho_{tc}$ in the CP conserving limit ($\varphi_3 \to 0$).
Furthermore, in appendix~\ref{section:app}, we also show that the VIA source terms are basis independent.
Namely, the source terms calculated in the weak basis coincide with those in the mass basis.
From these considerations, we would conclude that in refs.~\cite{Chiang:2016vgf,Fuyuto:2017ewj} the transport equations defined in the mass basis are considered, but in which the source terms evaluated in the weak basis are used.
Therefore, there would be a mismatch in the transport equations, and the phase effects of the off-diagonal Yukawa couplings discussed in refs.~\cite{Chiang:2016vgf,Fuyuto:2017ewj} may be unphysical.

In fig.~\ref{fig:baryonscat1}, at the points $(|\rho_{tt}|,|\rho_{tc}|) = (0,0)$ and $(0.2,1.5)$, we have obtained $\eta_{B} \simeq 1.8 \times 10^{-11}$ and $1.5\times 10^{-10}$, respectively.
At the former point, only the CP 
violating VEV $\varphi_3$ which is induced by a complex coupling $\lambda_7$ in the general 2HDM with the Higgs alignment~\cite{Enomoto:2021dkl, Enomoto:2022rrl} produces the BAU.
On the other hand, at the latter one, $\rho_{tc}$ and $\rho_{tt}$ also contribute to the BAU.
The magnitude depends on $\theta_{tt}$, and the maximal BAU $\eta_{B}\simeq 2.8 \times 10^{-10}$ at the point is given by $\theta_{tt} \simeq 1.0$ with the other fixed input parameters.
In this case, magenta points will be shifted to left from fig.~\ref{fig:baryonscat1}, and for example, $(|\rho_{tt}|,|\rho_{tc}|) \simeq (0.06,0)$ and $(0.025,1.5)$ satisfy the observed BAU.
We note that some $\theta_{tt}$ at the point give negative $\eta_B$.

The rare decay processes of $K$ meson, $K^+ \to \pi^+ \nu \overline{\nu}$ and $K_L \to \pi^0 \nu \overline{\nu}$, are sensitive to $\rho_{tt}\rho_{ct}$, and we have considered a non-zero $\rho_{ct}$ coupling in fig.~\ref{fig:baryonscat2}.
In this benchmark point, the branching fraction of $K^+ \to \pi^+ \nu \overline{\nu}$ is up to about $14 \%$ larger than the SM value.
This is consistent with the current experimental value in eq.~(\ref{eq:Kplusexp}) within 1$\sigma$ level.
At the future NA62 experiment, about 10\% precision is expected~\cite{NA62:2022hqi}, so that this benchmark point would be tested.
In our benchmark points, $K_L \to \pi^0 \nu \overline{\nu}$ process, which KOTO step-2~\cite{Aoki:2021cqa} and KLEVER~\cite{Moulson:2019ifj} experiments aim to measure, is about up to $2.2 \%$ larger than the SM value.
If these excesses in the $K^+$ and $K_L$ rare decays are detected in the future experiments, we can confirm non-zero $\rho_{tt}\rho_{ct}$, and then it is expected that the scenario of top transport EWBG with $\rho_{tt}$ is realized in our model.
In addition, if we observe these observables with about $1\%$ experimental uncertainties, we can know from the magenta points in fig.~\ref{fig:baryonscat2} how $|\rho_{tc}|$ coupling affects the BAU.
However, we note that these branching fractions vary in $\rho_{ct}$.
In order to know the effect of $|\rho_{tc}|$ to the BAU with arbitrary $\rho_{ct}$, $\epsilon_K$ and $\epsilon^\prime / \epsilon$ which contain $\rho_{tc}$ dependence become important observables.
These results motivate us to improve the accuracies of theoretical predictions and experimental observations about $K$ meson physics.

As we have mentioned in sec.~\ref{sec:pheno}, the upper bound on $|\rho_{tt}|$ is given by the data of $B_s \to \mu \mu$ in fig.~\ref{fig:baryonscat2}.
The expected total uncertainty of this process at the LHCb~\cite{LHCb:2012myk} is 4.4\% with the integrated luminosity of 300 $\mathrm{fb^{-1}}$~\cite{Cerri:2018ypt}.
Also, from our benchmark point if we take $\rho_{ct} = 0$ without changing other parameters, the upper bound on $|\rho_{tc}|$ is given by the data of $B \to X_s \gamma$.
The relative uncertainty of this measurement expected to be reduced about 3.2\% at the Belle-II~\cite{Belle-II:2018jsg} with 50~$\mathrm{ab^{-1}}$.
Therefore, $B$ meson physics is equivalently important as $K$ meson physics for testing our scenario.

The off-diagonal $\rho_{tc}$ coupling predicts some interesting signals at future high energy collider experiments.
In our benchmark point where the heavy neutral scalar bosons are degenerated, even at the HL-LHC it is difficult to measure the processes of $gc \to tH_{2,3} \to tt\overline{c}$ due to the interference effect.
It is known that if the neutral scalar bosons have a mass difference as $m_{H_2} - m_{H_3} = 30$~GeV, the interference effect enhances such processes~\cite{Kohda:2017fkn, Hou:2020tnc}.
In the case that only $H_3$ has the mass of 350~GeV while $H_2$ is much heavier, $\rho_{tc} \gtrsim 0.2$ will be excluded at 2$\sigma$ level by the same-sign top search at the HL-LHC~\cite{Hou:2020tnc,Hou:2020ciy}.
On the other hand, triple top production processes $gc \to t H_{2,3} \to tt\overline{t}$ are not disturbed by that interference, so that we may be able to detect the signals even in the case of $m_{H_2}=m_{H_3}$~\cite{Hou:2019gpn}.
About the charged scalar production induced by $\rho_{tc}$, the process of $pp \to H^\pm \to bc$ has a sensitivity for the large $\rho_{tc}$ coupling~\cite{Iguro:2022uzz}.
In addition, signatures of $cg \to H^+ b \to bt\overline{b}$ may also be tested at the HL-LHC~\cite{Iguro:2017ysu,Ghosh:2019exx,Hou:2020tnc}.

Our top-charm mixing scenario for EWBG in the aligned 2HDM can be tested by detecting the common features in EWBG scenarios like strongly first order phase transition and CP violation.
In our benchmark points shown in eqs.~(\ref{eq:imputpotential}) and (\ref{eq:inputyukawa}), $v_n / T_n$ is almost same as BP1 in ref.~\cite{Enomoto:2022rrl}, so that our scenario predicts the similar predictions on the electroweak phase transition.
Physics of the strongly first order phase transition predicts a large deviation in the triple Higgs boson coupling from the SM prediction~\cite{Kanemura:2004ch} and the decay branching ratio of $H_1 \to \gamma \gamma$~\cite{Ellis:1975ap, Shifman:1979eb}.
We can estimate the deviation of the triple Higgs coupling from the SM by ref.~\cite{Enomoto:2022rrl}, and obtain $\lambda_{hhh} - \lambda^{\mathrm{SM}}_{hhh} = 55$~\% at one loop level.
This deviation may be tested at the HL-LHC~\cite{Cepeda:2019klc} and ILC with $\sqrt{s}=$ 500 GeV and 1 TeV~\cite{Fujii:2015jha, Bambade:2019fyw}.
For our benchmark point, we predict $\sigma \mathcal{B}(H_1 \to \gamma \gamma) = 104 \pm 5~\mathrm{fb}$, where $\sigma$ is the production cross section of the SM Higgs boson, and the uncertainty comes from theoretical errors of the production cross section.

In our scenario, we have assumed $v_w = 0.1$.
It is known that large wall velocity enhances possibilities to observe GWs produced by first order phase transition~\cite{Grojean:2006bp, Kakizaki:2015wua,Hashino:2016rvx,Hashino:2018wee,Kanemura:2022txx} at LISA~\cite{LISA:2017pwj}, DECIGO~\cite{Seto:2001qf} and BBO~\cite{Corbin:2005ny}.
On the other side, the BAU decreases as the velocity approaches the speed of light~\cite{Cline:2020jre}.
For example, if we set the velocity to 0.45, the BAU decreases, and all the magenta points shown in figs.~\ref{fig:baryonscat1} and \ref{fig:baryonscat2} move to right.
In this case, some points constrained by the data of $B_s \to \mu \mu$ and $|\epsilon_K|$, but the other points, e.g. $(|\rho_{tt}|, |\rho_{tc}|)=(0.15,0.8)$, satisfy $\eta_B = 8.65\mathrm{-}8.74 \times 10^{-11}$ under the experimental constraints.
The shape of GW spectra predicted at these points is similar to the ones in ref.~\cite{Enomoto:2022rrl}, and we have obtained the energy density of GW at the peak point as $h^2 \Omega_{\mathrm{GW}}(f_{\mathrm{peak}})=6\times 10^{-14}$ with $f_{\mathrm{peak}} = 0.14~\mathrm{Hz}$.
This point is above the peak integrated sensitivity curves of DECIGO and BBO shown in ref.~\cite{Cline:2021iff}, so that the points are testable by the future GWs observations.

In this paper, we have neglected $(1,1)$ components of the $3 \times 3$ matrices $\rho^u$ and $\rho^d$, which are relevant parameters to the nEDM.
The CP violating phases of the top-charm mixing couplings can contribute to the nEDM through the Weinberg operator induced by the charm chromo EDM~\cite{Sala:2013osa}.
Although these phases are not directly connected with the BAU, some parameter regions may be constrained by the current nEDM bound~\cite{Abel:2020pzs}. 
In this paper, we have set $\rho^e = 0$.
If we consider $\rho^e$ to be a non-zero $3\times 3$ matrix, especially its $(1,1)$ component of $\rho_{ee}$, the eEDM becomes sensitive to the CP violation in our model.
Even if we consider $\rho_{ee}$, parameter regions which are allowed under the ACME bound~\cite{ACME:2018yjb} exist thanks to the destructive interferences among additional CP phases~\cite{Kanemura:2020ibp,Enomoto:2021dkl,Enomoto:2022rrl}.
Clearly, if accuracies are substantially improved at future EDM experiments~\cite{ACME:2018yjb,TUCAN:2022koi}, wider parameter regions can be explored.

\section{Conclusion \label{sec:conclusion}}

We have investigated a scenario of electroweak baryogenesis in the two Higgs doublet model with quark flavor mixing.
In general, off-diagonal components of quark Yukawa interactions with additional Higgs bosons are strongly constrained by the data for flavor changing neutral currents.
However, top-charm quark mixing is not the case, so that a large off-diagonal element can be taken, which can contribute to generating baryon asymmetry of the universe.
As we have shown in sec.~\ref{sec:source}, it has turned out that CP violating phases of the off-diagonal element in the source term in the Boltzmann equation are eliminated by the rephasing.
This result is somewhat different from previous works on electroweak baryogenesis by flavor off-diagonal Yukawa couplings.
Instead, we have found that the absolute value of the top-charm off-diagonal element enhances CP violating phases in the Higgs potential, by which sufficient amount of the baryon number can be generated to explain the observed baryon asymmetry of the universe.
We have found that such a scenario is viable under the current experimental data.
The model can be tested by the current and future measurements of various flavor experiments like Kaon rare decays, in addition to high energy collider experiments as well as gravitational wave observations.
Characteristic predictions of our model would be deviations in Kaon rare decays.
As we have shown in sec.~\ref{sec:pheno}, branching ratios of $K^+ \to \pi^+ \nu \overline{\nu}$ and $K_L \to \pi^0 \nu \overline{\nu}$  can deviate by the order of 10\% and 1\%, respectively, which may be tested at future Kaon experiments such as NA62 and KOTO step-2.

\section*{Acknowledgments}
The work of S. K. was supported by the JSPS KAKENHI Grant No.~20H00160.
The work of Y. M. was supported by JST SPRING, Grant No.~JPMJSP2138.
The authors would like to thank G. W.-S. Hou and E. Senaha for useful discussions.

\appendix

\section{Basis independency of source terms\label{section:app}}

In order to reinforce the discussion in sec.~\ref{sec:source}, we here show the basis independency of the source terms in the Boltzmann equation.

First we show that the VIA source term is same in the weak and mass basis.
The kinetic equation of fermion $\psi$ is given by~\cite{Riotto:1998zb}
\begin{align}
    \partial_\mu^X j^\mu_{\psi} = - \int d^3 \bm{w} \int_{-\infty}^{T} dw^0 ~\mathrm{Tr}\Big[ &\Sigma_\psi^>(X,w) G_\psi^<(w,X) - \Sigma_\psi^<(X,w) G_\psi^>(w,X) \notag \\
    &- G_\psi^>(X,w) \Sigma_\psi^<(w,X) + G_\psi^<(X,w) \Sigma_\psi^>(w,X) \Big],
    \label{eq:psisource}
\end{align}
where $G_\psi^{<,>}$ and $\Sigma_\psi^{<,>}$ are Wightman functions and self energies, respectively, following notations in ref.~\cite{Prokopec:2003pj}.
The trace is taken in the spinor space.
By definition, sum of divergence of currents of left-handed top and charm quarks is basis independent as 
\begin{equation}
    \partial_\mu j^\mu_{t_L^\prime} + \partial_\mu j^\mu_{c_L^\prime} = \partial_\mu j^\mu_{t_L} + \partial_\mu j^\mu_{c_L}.
\end{equation}
As shown in following, the right hand side of eq.~(\ref{eq:psisource}) is also basis independent at the leading order in VIA.

From eq.~(\ref{eq:yukawahbs}), the relevant lagrangian of the top and charm quarks is given by 
\begin{equation}
    \mathcal{L}_y = -\overline{u_{i,L}^\prime} \big( Y^u_{1,ij}  \phi^{0*}_1  + Y^u_{2,ij} \phi^{0*}_2 \big) u^\prime_{j,R}+ \mathrm{h.c.}.
    \label{eq:yukawaforsource}
\end{equation}
The relations between Yukawa matrices in the weak basis $Y_1^u, Y_2^u$ and the mass basis $Y^u_{\mathrm{d}}, \rho^u$ are given by eq.~(\ref{eq:weakandmass}).
Here we define matrices as
\begin{align}
    A(u) &\equiv Y_1^u v_1^*(u) +  Y_2^u v_2^*(u) \notag \\
    &= V_L^{u\dagger} \big( Y^u_{\mathrm{d}} v_1^*(u) + \rho^u v_2^*(u)  \big) V_R^{u} \notag \\
    &\equiv V_L^{u \dagger} B(u) V_R^u,
    \label{eq:transA}
\end{align}
where the space-time dependent VEVs $v_k(u) = \sqrt{2} \langle \phi^0_k \rangle~ (k=1,2)$ are defined in the Higgs basis.

\begin{figure}[t]
    \centering
    \setlength{\feynhandlinesize}{0.7pt}
    \setlength{\feynhandarrowsize}{5pt}
    \begin{tikzpicture}
    \begin{feynhand}
    \vertex (a) at (0,0); \vertex[particle] (b) at (-1,-1){$t^\prime_L$};
    \vertex[particle] (c) at (-1,1){$v_a(v)$};
    \vertex (d) at (1.5,0);
    \vertex[particle] (e) at (2.5,1){$v_b(u)$};
    \vertex[particle] (f) at (2.5,-1){$t_L^\prime$};
    \propag[fer] (a) to (b);\propag[sca] (a) to (c);
    \propag[fer] (d) to [edge label=$t^\prime_R\mathrm{,}c^\prime_R$] (a);
    \propag[sca] (d) to (e);
    \propag[fer] (f) to (d);
    \end{feynhand}
    \end{tikzpicture}
    ~~
    \begin{tikzpicture}
        \begin{feynhand}
        \vertex (a) at (0,0); \vertex[particle] (b) at (-1,-1){$c^\prime_L$};
        \vertex[particle] (c) at (-1,1){$v_a(v)$};
        \vertex (d) at (1.5,0);
        \vertex[particle] (e) at (2.5,1){$v_b(u)$};
        \vertex[particle] (f) at (2.5,-1){$c_L^\prime$};
        \propag[fer] (a) to (b);\propag[sca] (a) to (c);
        \propag[fer] (d) to [edge label=$t^\prime_R\mathrm{,}c^\prime_R$] (a);
        \propag[sca] (d) to (e);
        \propag[fer] (f) to (d);
        \end{feynhand}
        \end{tikzpicture}
    \caption{Self energy of left-handed top and charm quarks in the weak basis~($a,b=1,2$).}
    \label{fig:topselfweak}
\end{figure}
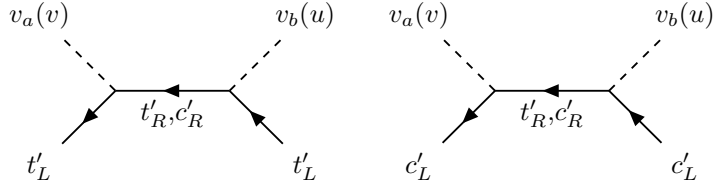
We calculate the first term in right hand side of eq.~(\ref{eq:psisource}).
At the 2nd order in VIA as shown in the left panel of fig.~\ref{fig:topselfweak}, the self energy of the left-handed top quarks in the weak basis is written by
\begin{align}
    \Sigma^>_{t_L^\prime}(u,v) &= -\frac{1}{2} \Big( A_{tt}(u) G^>_{t^\prime_R t^\prime_R}(u,v) A^*_{tt}(v) + A_{tc}(u) G^>_{c^\prime_R c^\prime_R}(u,v) A^*_{tc}(v) \Big) \notag \\
    &= -\frac{1}{2} \sum_{q=c,t}  \Big( A_{tq}(u) G^>_{q_R^\prime q_R^\prime}(u,v) A^*_{tq}(v) \Big).
\end{align}
We define a $2\times 2$ matrix which has flavor indices as 
\begin{equation}
    G^>_{R}(u,v) = G^>_{q_{iR} q_{jR} }(u,v) = -i \langle q_{iR}(u) \overline{q_{jR}}(v) \rangle,
\end{equation}
and the basis transformation is given by 
\begin{equation}
    G^>_{R^\prime}(u,v) = G^>_{q_{iR}^\prime q_{jR}^\prime }(u,v) = V_{R,ik}^* G^>_{q_{kR} q_{mR}}(u,v) V_{R,mj} = V_R^\dagger G^>_{R}(u,v) V_R.
    \label{eq:transGR}
\end{equation}
Here we have omitted the up-type subscript in the rotation matrix $V_R$.
By using eqs.~(\ref{eq:transA}) and (\ref{eq:transGR}), the self energy can be written by 
\begin{equation}
    \Sigma^>_{t_L^\prime}(u,v) = -\sum_{p,q=c,t} \frac{1}{2} \Big( A_{tp}(u) G^>_{p_R^\prime q_R^\prime}(u,v) A^*_{tq}(v) \Big) = -\frac{1}{2}\big(V_L^\dagger B(u) G^>_R (u,v) B^\dagger(v) V_L\big)_{tt},
\end{equation}
at the level at which we are considering in VIA.
We have used the fact that $G_{q_{iR}q_{jR}}^>~(i\neq j)$ first appear at 2nd order in VIA.
Therefore, we find that
\begin{align}
    \Sigma_{t_L^\prime}^>(X,w) G_{t_L^\prime t_L^\prime}^<(w,X) &= -\frac{1}{2} \big(V_L^\dagger B(X) G^>_R (X,w) B^\dagger(w) V_L\big)_{tt} G^<_{t_L^\prime t_L^\prime}(w, X) \notag \\
    &= -\frac{1}{2}\big(V_L^\dagger B(X) G^>_R (X,w) B^\dagger(w) V_L G^<_{L^\prime}(w,X) \big)_{tt} \notag \\
    &= -\frac{1}{2}\big(V_L^\dagger B(X) G^>_R (X,w) B^\dagger(w) G^<_{L}(w,X) V_L \big)_{tt},
\end{align}
where we have also defined the matrix $G^<_{L}$ for the left-handed fields, and we have used the transformation low $G^<_{L^\prime}(u,v) = V_L^\dagger G^<_{L}(u,v) V_L$.

In the same way, we obtain the contribution from the self energy of the left-handed charm quarks in the weak basis shown in the right panel of fig.~\ref{fig:topselfweak} as
\begin{align}
    \Sigma_{c_L^\prime}^>(X,w) G_{c_L^\prime c_L^\prime}^<(w,X) &= -\frac{1}{2}\big(V_L^\dagger B(X) G^>_R (X,w) B^\dagger(w) G^<_{L}(w,X) V_L \big)_{cc},
\end{align}
so that the sum of the contributions of the top and charm quarks is given by 
\begin{align}
    \Sigma_{t_L^\prime}^> G_{t_L^\prime t_L^\prime}^< + \Sigma_{c_L^\prime}^> G_{c_L^\prime c_L^\prime}^< &= -\frac{1}{2}\mathrm{Tr}\Big[ V_L^\dagger B G^>_R  B^\dagger G^<_{L} V_L \Big] \notag \\
    &= -\frac{1}{2} \mathrm{Tr}\Big[ B G^>_R  B^\dagger G^<_{L} \Big].
\end{align}
The trace is taken in the flavor space.
Similarly we can show that $V_L$ and $V_R$ dependencies in the other terms in eq.~(\ref{eq:psisource}) disappear.
As a result, the right hand side of eq.~(\ref{eq:psisource}) calculated in the weak basis coincides with the one in the mass basis.
This result is a consequence of summing up the contribution of the charm quark defined in the weak basis.

When we consider CP conserving Higgs potential and $v_1, v_2 \in \mathbb{R}$, the source term relevant to $\rho_{tc}$ is given by 
\begin{equation}
    S_{t_L}^{\mathrm{VIA}} \supset \frac{1}{2} |\rho_{tc}|^2 v_2^2(X) \int d^3 \bm{w} \int_{-\infty}^T dw^0  \mathrm{Re}\Big[ \mathrm{Tr}\big[ G^>_{c_R} G^<_{t_R} - G^<_{c_R} G^>_{t_R} \big] \Big],
\end{equation}
so that CP violating source term does not appear at the leading order in VIA,\footnote{If we consider 4th order in VIA, the phases of the off-diagonal couplings may appear in the self energy. However, such contributions are suppressed about $O(\delta^2)$.} and only CP conserving source term arises from the kinetic equation.
This source just generates scattering terms for a process $t_L \leftrightarrow c_R$ induced by the collision with the bubble wall.
However, these effects are much smaller than the other chirality flip processes due to the smallness of $v_2 (\ll v_1)$.
If we assume CP violating VEV $v_2 \in \mathbb{C}$, $\mathrm{arg}(v_2)$ causes CP violating source terms, which are proportional to the absolute value of $\rho_{tc}$.

Second we show the basis independency of the source term derived by the semi-classical force mechanism in the WKB approximation.
We denote mass matrices which can be obtained in the weak and mass basis as $M^\prime$ and $M$, respectively.
From eq.~(\ref{eq:weakandmass}), the relation of these matrices can be written by
\begin{equation}
    V_L M^\prime  V_R^\dagger = M.
    \label{eq:massmatrixweakandmass}
\end{equation}
Dirac equations of these basis are given by
\begin{align}
    &(i\cancel{\partial} - M^{\prime\dagger} P_L - M^\prime P_R ) \psi^\prime = 0, &~&(\mathrm{weak~basis}) \notag \\
    &(i\cancel{\partial} - M^\dagger P_L - M P_R ) \psi = 0, &~&(\mathrm{mass~basis}) 
\end{align}
where
\begin{align}
    &\psi_i^\prime = (u_{iL}^\prime, u_{iR}^\prime)^T = \big((V_L^\dagger u_L)_i, (V_R^\dagger u_R)_i \big)^T, \notag \\
    &\psi_i = (u_{iL}, u_{iR})^T.
\end{align}
With a same discussion in ref.~\cite{Cline:2000nw}, we obtain the equation in the weak basis as
\begin{align}
    \Big\{ \omega^2 + \partial_z^2 - M^\prime M^{\prime \dagger} + is (M^\prime \partial_z M^{\prime -1})(\omega - is \partial_z) \Big\} L^\prime_s = 0,
\end{align}
where $L_s^{\prime} = (c^{\prime}_L, t^{\prime}_L)$.
The only difference between the two basis is primed or not, so that the equation in the mass basis can be obtained by omitting the primes.
From the weak and mass basis to locally diagonalized basis by unitary matrix $U$, the respective equations are
\begin{align}
    &\Big\{ \omega^2 + \partial_z^2 - m_D^{\prime2} +2U^\prime_1 \partial_z + U_2 +isA^\prime_1 (\omega - is\partial_z ) + A^\prime_2 \Big\} L_s^{D \prime} = 0, ~~~ &(\mathrm{weak~basis})\notag \\
    &\Big\{ \omega^2 + \partial_z^2 - m_D^2 +2U_1 \partial_z + U_2 +isA_1 (\omega - is\partial_z ) + A_2 \Big\} L_s^D = 0, ~~~ &(\mathrm{mass~basis})
    \label{eq:WKBweakandmasslocal}
\end{align}
where
\begin{align}
    &U^{(\prime)}_1 \equiv U^{(\prime)} \partial_z U^{(\prime)^\dagger},& &A_1^{(\prime)} \equiv U^{(\prime)} (M^{(\prime)} \partial_z M^{(\prime)-1}) U^{(\prime)\dagger}, \notag \\
    &U_2^{(\prime)} \equiv U^{(\prime)} \partial_z^2 U^{(\prime)\dagger},& &A_2^{(\prime)} \equiv A_1^{(\prime)} U_1^{(\prime)}, \notag \\
    &L_s^{(\prime)D} \equiv U^{(\prime)} L_s^{(\prime)},& &m_D^{(\prime)2} \equiv U^{(\prime)}M^{(\prime)}M^{(\prime)\dagger} U^{(\prime)\dagger} = \mathrm{diag} (m_+^{(\prime)2}, m_-^{(\prime)2}).
    \label{eq:defWKB}
\end{align}
The diagonalized mass matrix from the weak basis is written by 
\begin{equation}
    m_D^{\prime 2} = U^\prime M^\prime M^{\prime \dagger} U^{\prime\dagger} = U^\prime V_L^\dagger MM^\dagger V_L U^{\prime\dagger},
\end{equation}
so that $m_D^{\prime 2} = m_D^2$ is concluded by uniqueness of eigenvalue, but $U^\prime V_L^\dagger = U$ is not.
The relation between $L_s^D$ and $L_s^{D\prime}$ is given by 
\begin{equation}
    L_s^D = U L_s = U V_L L_s^\prime = UV_L U^{\prime \dagger} L_s^{D\prime} \equiv QL_s^{D\prime}.
    \label{eq:defQ}
\end{equation}

By using $L_s^{D\prime}$, the equation of the mass basis in eq.~(\ref{eq:WKBweakandmasslocal}) can be written by 
\begin{align}
    Q \Big\{ \omega^2 &+ \partial_z^2 - m_D^2 + Q^\dagger \partial_z^2 Q +2Q^\dagger (\partial_z Q) \partial_z +2 Q^\dagger U_1 Q \partial_z + 2 Q^\dagger U_1 (\partial_z Q) \notag \\
    &+ Q^\dagger U_2 Q  +isQ^\dagger A_1 Q (\omega - is\partial_z ) +s Q^\dagger A_1 (\partial_z Q) + Q^\dagger A_2 Q \Big\} L_s^{D \prime} = 0.
\end{align}
At the leading order in derivative expansion, CP violating source terms arise from the second term in the second line, so that from eqs.~(\ref{eq:massmatrixweakandmass}), (\ref{eq:defWKB}) and (\ref{eq:defQ}), we obtain
\begin{equation}
    Q^\dagger A_1 Q = A_1^\prime.
\end{equation}

\bibliography{references}

\begin{thebibliography}{139}%
\makeatletter
\providecommand \@ifxundefined [1]{%
 \@ifx{#1\undefined}
}%
\providecommand \@ifnum [1]{%
 \ifnum #1\expandafter \@firstoftwo
 \else \expandafter \@secondoftwo
 \fi
}%
\providecommand \@ifx [1]{%
 \ifx #1\expandafter \@firstoftwo
 \else \expandafter \@secondoftwo
 \fi
}%
\providecommand \natexlab [1]{#1}%
\providecommand \enquote  [1]{``#1''}%
\providecommand \bibnamefont  [1]{#1}%
\providecommand \bibfnamefont [1]{#1}%
\providecommand \citenamefont [1]{#1}%
\providecommand \href@noop [0]{\@secondoftwo}%
\providecommand \href [0]{\begingroup \@sanitize@url \@href}%
\providecommand \@href[1]{\@@startlink{#1}\@@href}%
\providecommand \@@href[1]{\endgroup#1\@@endlink}%
\providecommand \@sanitize@url [0]{\catcode `\\12\catcode `\$12\catcode
  `\&12\catcode `\#12\catcode `\^12\catcode `\_12\catcode `\%12\relax}%
\providecommand \@@startlink[1]{}%
\providecommand \@@endlink[0]{}%
\providecommand \url  [0]{\begingroup\@sanitize@url \@url }%
\providecommand \@url [1]{\endgroup\@href {#1}{\urlprefix }}%
\providecommand \urlprefix  [0]{URL }%
\providecommand \Eprint [0]{\href }%
\providecommand \doibase [0]{http://dx.doi.org/}%
\providecommand \selectlanguage [0]{\@gobble}%
\providecommand \bibinfo  [0]{\@secondoftwo}%
\providecommand \bibfield  [0]{\@secondoftwo}%
\providecommand \translation [1]{[#1]}%
\providecommand \BibitemOpen [0]{}%
\providecommand \bibitemStop [0]{}%
\providecommand \bibitemNoStop [0]{.\EOS\space}%
\providecommand \EOS [0]{\spacefactor3000\relax}%
\providecommand \BibitemShut  [1]{\csname bibitem#1\endcsname}%
\let\auto@bib@innerbib\@empty
\bibitem [{\citenamefont {Workman}\ \emph {et~al.}(2022)\citenamefont {Workman}
  \emph {et~al.}}]{ParticleDataGroup:2022pth}%
  \BibitemOpen
  \bibfield  {author} {\bibinfo {author} {\bibfnamefont {R.~L.}\ \bibnamefont
  {Workman}} \emph {et~al.} (\bibinfo {collaboration} {Particle Data Group}),\
  }\href {\doibase 10.1093/ptep/ptac097} {\bibfield  {journal} {\bibinfo
  {journal} {PTEP}\ }\textbf {\bibinfo {volume} {2022}},\ \bibinfo {pages}
  {083C01} (\bibinfo {year} {2022})}\BibitemShut {NoStop}%
\bibitem [{\citenamefont {Sakharov}(1967)}]{Sakharov:1967dj}%
  \BibitemOpen
  \bibfield  {author} {\bibinfo {author} {\bibfnamefont {A.~D.}\ \bibnamefont
  {Sakharov}},\ }\href {\doibase 10.1070/PU1991v034n05ABEH002497} {\bibfield
  {journal} {\bibinfo  {journal} {Pisma Zh. Eksp. Teor. Fiz.}\ }\textbf
  {\bibinfo {volume} {5}},\ \bibinfo {pages} {32} (\bibinfo {year}
  {1967})}\BibitemShut {NoStop}%
\bibitem [{\citenamefont {Huet}\ and\ \citenamefont
  {Sather}(1995)}]{Huet:1994jb}%
  \BibitemOpen
  \bibfield  {author} {\bibinfo {author} {\bibfnamefont {P.}~\bibnamefont
  {Huet}}\ and\ \bibinfo {author} {\bibfnamefont {E.}~\bibnamefont {Sather}},\
  }\href {\doibase 10.1103/PhysRevD.51.379} {\bibfield  {journal} {\bibinfo
  {journal} {Phys. Rev. D}\ }\textbf {\bibinfo {volume} {51}},\ \bibinfo
  {pages} {379} (\bibinfo {year} {1995})},\ \Eprint
  {http://arxiv.org/abs/hep-ph/9404302} {arXiv:hep-ph/9404302} \BibitemShut
  {NoStop}%
\bibitem [{\citenamefont {Kajantie}\ \emph {et~al.}(1996)\citenamefont
  {Kajantie}, \citenamefont {Laine}, \citenamefont {Rummukainen},\ and\
  \citenamefont {Shaposhnikov}}]{Kajantie:1996mn}%
  \BibitemOpen
  \bibfield  {author} {\bibinfo {author} {\bibfnamefont {K.}~\bibnamefont
  {Kajantie}}, \bibinfo {author} {\bibfnamefont {M.}~\bibnamefont {Laine}},
  \bibinfo {author} {\bibfnamefont {K.}~\bibnamefont {Rummukainen}}, \ and\
  \bibinfo {author} {\bibfnamefont {M.~E.}\ \bibnamefont {Shaposhnikov}},\
  }\href {\doibase 10.1103/PhysRevLett.77.2887} {\bibfield  {journal} {\bibinfo
   {journal} {Phys. Rev. Lett.}\ }\textbf {\bibinfo {volume} {77}},\ \bibinfo
  {pages} {2887} (\bibinfo {year} {1996})},\ \Eprint
  {http://arxiv.org/abs/hep-ph/9605288} {arXiv:hep-ph/9605288} \BibitemShut
  {NoStop}%
\bibitem [{\citenamefont {D'Onofrio}\ and\ \citenamefont
  {Rummukainen}(2016)}]{DOnofrio:2015gop}%
  \BibitemOpen
  \bibfield  {author} {\bibinfo {author} {\bibfnamefont {M.}~\bibnamefont
  {D'Onofrio}}\ and\ \bibinfo {author} {\bibfnamefont {K.}~\bibnamefont
  {Rummukainen}},\ }\href {\doibase 10.1103/PhysRevD.93.025003} {\bibfield
  {journal} {\bibinfo  {journal} {Phys. Rev. D}\ }\textbf {\bibinfo {volume}
  {93}},\ \bibinfo {pages} {025003} (\bibinfo {year} {2016})},\ \Eprint
  {http://arxiv.org/abs/1508.07161} {arXiv:1508.07161 [hep-ph]} \BibitemShut
  {NoStop}%
\bibitem [{\citenamefont {Klinkhamer}\ and\ \citenamefont
  {Manton}(1984)}]{Klinkhamer:1984di}%
  \BibitemOpen
  \bibfield  {author} {\bibinfo {author} {\bibfnamefont {F.~R.}\ \bibnamefont
  {Klinkhamer}}\ and\ \bibinfo {author} {\bibfnamefont {N.~S.}\ \bibnamefont
  {Manton}},\ }\href {\doibase 10.1103/PhysRevD.30.2212} {\bibfield  {journal}
  {\bibinfo  {journal} {Phys. Rev. D}\ }\textbf {\bibinfo {volume} {30}},\
  \bibinfo {pages} {2212} (\bibinfo {year} {1984})}\BibitemShut {NoStop}%
\bibitem [{\citenamefont {Yoshimura}(1978)}]{Yoshimura:1978ex}%
  \BibitemOpen
  \bibfield  {author} {\bibinfo {author} {\bibfnamefont {M.}~\bibnamefont
  {Yoshimura}},\ }\href {\doibase 10.1103/PhysRevLett.41.281} {\bibfield
  {journal} {\bibinfo  {journal} {Phys. Rev. Lett.}\ }\textbf {\bibinfo
  {volume} {41}},\ \bibinfo {pages} {281} (\bibinfo {year} {1978})},\ \bibinfo
  {note} {[Erratum: Phys.Rev.Lett. 42, 746 (1979)]}\BibitemShut {NoStop}%
\bibitem [{\citenamefont {Weinberg}(1979)}]{Weinberg:1979bt}%
  \BibitemOpen
  \bibfield  {author} {\bibinfo {author} {\bibfnamefont {S.}~\bibnamefont
  {Weinberg}},\ }\href {\doibase 10.1103/PhysRevLett.42.850} {\bibfield
  {journal} {\bibinfo  {journal} {Phys. Rev. Lett.}\ }\textbf {\bibinfo
  {volume} {42}},\ \bibinfo {pages} {850} (\bibinfo {year} {1979})}\BibitemShut
  {NoStop}%
\bibitem [{\citenamefont {Fukugita}\ and\ \citenamefont
  {Yanagida}(1986)}]{Fukugita:1986hr}%
  \BibitemOpen
  \bibfield  {author} {\bibinfo {author} {\bibfnamefont {M.}~\bibnamefont
  {Fukugita}}\ and\ \bibinfo {author} {\bibfnamefont {T.}~\bibnamefont
  {Yanagida}},\ }\href {\doibase 10.1016/0370-2693(86)91126-3} {\bibfield
  {journal} {\bibinfo  {journal} {Phys. Lett. B}\ }\textbf {\bibinfo {volume}
  {174}},\ \bibinfo {pages} {45} (\bibinfo {year} {1986})}\BibitemShut
  {NoStop}%
\bibitem [{\citenamefont {Kuzmin}\ \emph {et~al.}(1985)\citenamefont {Kuzmin},
  \citenamefont {Rubakov},\ and\ \citenamefont {Shaposhnikov}}]{Kuzmin:1985mm}%
  \BibitemOpen
  \bibfield  {author} {\bibinfo {author} {\bibfnamefont {V.~A.}\ \bibnamefont
  {Kuzmin}}, \bibinfo {author} {\bibfnamefont {V.~A.}\ \bibnamefont {Rubakov}},
  \ and\ \bibinfo {author} {\bibfnamefont {M.~E.}\ \bibnamefont
  {Shaposhnikov}},\ }\href {\doibase 10.1016/0370-2693(85)91028-7} {\bibfield
  {journal} {\bibinfo  {journal} {Phys. Lett. B}\ }\textbf {\bibinfo {volume}
  {155}},\ \bibinfo {pages} {36} (\bibinfo {year} {1985})}\BibitemShut
  {NoStop}%
\bibitem [{\citenamefont {Turok}\ and\ \citenamefont
  {Zadrozny}(1991)}]{Turok:1990zg}%
  \BibitemOpen
  \bibfield  {author} {\bibinfo {author} {\bibfnamefont {N.}~\bibnamefont
  {Turok}}\ and\ \bibinfo {author} {\bibfnamefont {J.}~\bibnamefont
  {Zadrozny}},\ }\href {\doibase 10.1016/0550-3213(91)90356-3} {\bibfield
  {journal} {\bibinfo  {journal} {Nucl. Phys. B}\ }\textbf {\bibinfo {volume}
  {358}},\ \bibinfo {pages} {471} (\bibinfo {year} {1991})}\BibitemShut
  {NoStop}%
\bibitem [{\citenamefont {Cline}\ \emph {et~al.}(1996)\citenamefont {Cline},
  \citenamefont {Kainulainen},\ and\ \citenamefont {Vischer}}]{Cline:1995dg}%
  \BibitemOpen
  \bibfield  {author} {\bibinfo {author} {\bibfnamefont {J.~M.}\ \bibnamefont
  {Cline}}, \bibinfo {author} {\bibfnamefont {K.}~\bibnamefont {Kainulainen}},
  \ and\ \bibinfo {author} {\bibfnamefont {A.~P.}\ \bibnamefont {Vischer}},\
  }\href {\doibase 10.1103/PhysRevD.54.2451} {\bibfield  {journal} {\bibinfo
  {journal} {Phys. Rev. D}\ }\textbf {\bibinfo {volume} {54}},\ \bibinfo
  {pages} {2451} (\bibinfo {year} {1996})},\ \Eprint
  {http://arxiv.org/abs/hep-ph/9506284} {arXiv:hep-ph/9506284} \BibitemShut
  {NoStop}%
\bibitem [{\citenamefont {Fromme}\ \emph {et~al.}(2006)\citenamefont {Fromme},
  \citenamefont {Huber},\ and\ \citenamefont {Seniuch}}]{Fromme:2006cm}%
  \BibitemOpen
  \bibfield  {author} {\bibinfo {author} {\bibfnamefont {L.}~\bibnamefont
  {Fromme}}, \bibinfo {author} {\bibfnamefont {S.~J.}\ \bibnamefont {Huber}}, \
  and\ \bibinfo {author} {\bibfnamefont {M.}~\bibnamefont {Seniuch}},\ }\href
  {\doibase 10.1088/1126-6708/2006/11/038} {\bibfield  {journal} {\bibinfo
  {journal} {JHEP}\ }\textbf {\bibinfo {volume} {11}},\ \bibinfo {pages} {038}
  (\bibinfo {year} {2006})},\ \Eprint {http://arxiv.org/abs/hep-ph/0605242}
  {arXiv:hep-ph/0605242} \BibitemShut {NoStop}%
\bibitem [{\citenamefont {Cline}\ \emph {et~al.}(2011)\citenamefont {Cline},
  \citenamefont {Kainulainen},\ and\ \citenamefont {Trott}}]{Cline:2011mm}%
  \BibitemOpen
  \bibfield  {author} {\bibinfo {author} {\bibfnamefont {J.~M.}\ \bibnamefont
  {Cline}}, \bibinfo {author} {\bibfnamefont {K.}~\bibnamefont {Kainulainen}},
  \ and\ \bibinfo {author} {\bibfnamefont {M.}~\bibnamefont {Trott}},\ }\href
  {\doibase 10.1007/JHEP11(2011)089} {\bibfield  {journal} {\bibinfo  {journal}
  {JHEP}\ }\textbf {\bibinfo {volume} {11}},\ \bibinfo {pages} {089} (\bibinfo
  {year} {2011})},\ \Eprint {http://arxiv.org/abs/1107.3559} {arXiv:1107.3559
  [hep-ph]} \BibitemShut {NoStop}%
\bibitem [{\citenamefont {Tulin}\ and\ \citenamefont
  {Winslow}(2011)}]{Tulin:2011wi}%
  \BibitemOpen
  \bibfield  {author} {\bibinfo {author} {\bibfnamefont {S.}~\bibnamefont
  {Tulin}}\ and\ \bibinfo {author} {\bibfnamefont {P.}~\bibnamefont
  {Winslow}},\ }\href {\doibase 10.1103/PhysRevD.84.034013} {\bibfield
  {journal} {\bibinfo  {journal} {Phys. Rev. D}\ }\textbf {\bibinfo {volume}
  {84}},\ \bibinfo {pages} {034013} (\bibinfo {year} {2011})},\ \Eprint
  {http://arxiv.org/abs/1105.2848} {arXiv:1105.2848 [hep-ph]} \BibitemShut
  {NoStop}%
\bibitem [{\citenamefont {Liu}\ \emph {et~al.}(2012)\citenamefont {Liu},
  \citenamefont {Ramsey-Musolf},\ and\ \citenamefont {Shu}}]{Liu:2011jh}%
  \BibitemOpen
  \bibfield  {author} {\bibinfo {author} {\bibfnamefont {T.}~\bibnamefont
  {Liu}}, \bibinfo {author} {\bibfnamefont {M.~J.}\ \bibnamefont
  {Ramsey-Musolf}}, \ and\ \bibinfo {author} {\bibfnamefont {J.}~\bibnamefont
  {Shu}},\ }\href {\doibase 10.1103/PhysRevLett.108.221301} {\bibfield
  {journal} {\bibinfo  {journal} {Phys. Rev. Lett.}\ }\textbf {\bibinfo
  {volume} {108}},\ \bibinfo {pages} {221301} (\bibinfo {year} {2012})},\
  \Eprint {http://arxiv.org/abs/1109.4145} {arXiv:1109.4145 [hep-ph]}
  \BibitemShut {NoStop}%
\bibitem [{\citenamefont {Ahmadvand}(2014)}]{Ahmadvand:2013sna}%
  \BibitemOpen
  \bibfield  {author} {\bibinfo {author} {\bibfnamefont {M.}~\bibnamefont
  {Ahmadvand}},\ }\href {\doibase 10.1142/S0217751X14500900} {\bibfield
  {journal} {\bibinfo  {journal} {Int. J. Mod. Phys. A}\ }\textbf {\bibinfo
  {volume} {29}},\ \bibinfo {pages} {1450090} (\bibinfo {year} {2014})},\
  \Eprint {http://arxiv.org/abs/1308.3767} {arXiv:1308.3767 [hep-ph]}
  \BibitemShut {NoStop}%
\bibitem [{\citenamefont {Chiang}\ \emph {et~al.}(2016)\citenamefont {Chiang},
  \citenamefont {Fuyuto},\ and\ \citenamefont {Senaha}}]{Chiang:2016vgf}%
  \BibitemOpen
  \bibfield  {author} {\bibinfo {author} {\bibfnamefont {C.-W.}\ \bibnamefont
  {Chiang}}, \bibinfo {author} {\bibfnamefont {K.}~\bibnamefont {Fuyuto}}, \
  and\ \bibinfo {author} {\bibfnamefont {E.}~\bibnamefont {Senaha}},\ }\href
  {\doibase 10.1016/j.physletb.2016.09.052} {\bibfield  {journal} {\bibinfo
  {journal} {Phys. Lett. B}\ }\textbf {\bibinfo {volume} {762}},\ \bibinfo
  {pages} {315} (\bibinfo {year} {2016})},\ \Eprint
  {http://arxiv.org/abs/1607.07316} {arXiv:1607.07316 [hep-ph]} \BibitemShut
  {NoStop}%
\bibitem [{\citenamefont {Guo}\ \emph {et~al.}(2017)\citenamefont {Guo},
  \citenamefont {Li}, \citenamefont {Liu}, \citenamefont {Ramsey-Musolf},\ and\
  \citenamefont {Shu}}]{Guo:2016ixx}%
  \BibitemOpen
  \bibfield  {author} {\bibinfo {author} {\bibfnamefont {H.-K.}\ \bibnamefont
  {Guo}}, \bibinfo {author} {\bibfnamefont {Y.-Y.}\ \bibnamefont {Li}},
  \bibinfo {author} {\bibfnamefont {T.}~\bibnamefont {Liu}}, \bibinfo {author}
  {\bibfnamefont {M.}~\bibnamefont {Ramsey-Musolf}}, \ and\ \bibinfo {author}
  {\bibfnamefont {J.}~\bibnamefont {Shu}},\ }\href {\doibase
  10.1103/PhysRevD.96.115034} {\bibfield  {journal} {\bibinfo  {journal} {Phys.
  Rev. D}\ }\textbf {\bibinfo {volume} {96}},\ \bibinfo {pages} {115034}
  (\bibinfo {year} {2017})},\ \Eprint {http://arxiv.org/abs/1609.09849}
  {arXiv:1609.09849 [hep-ph]} \BibitemShut {NoStop}%
\bibitem [{\citenamefont {Fuyuto}\ \emph {et~al.}(2018)\citenamefont {Fuyuto},
  \citenamefont {Hou},\ and\ \citenamefont {Senaha}}]{Fuyuto:2017ewj}%
  \BibitemOpen
  \bibfield  {author} {\bibinfo {author} {\bibfnamefont {K.}~\bibnamefont
  {Fuyuto}}, \bibinfo {author} {\bibfnamefont {W.-S.}\ \bibnamefont {Hou}}, \
  and\ \bibinfo {author} {\bibfnamefont {E.}~\bibnamefont {Senaha}},\ }\href
  {\doibase 10.1016/j.physletb.2017.11.073} {\bibfield  {journal} {\bibinfo
  {journal} {Phys. Lett. B}\ }\textbf {\bibinfo {volume} {776}},\ \bibinfo
  {pages} {402} (\bibinfo {year} {2018})},\ \Eprint
  {http://arxiv.org/abs/1705.05034} {arXiv:1705.05034 [hep-ph]} \BibitemShut
  {NoStop}%
\bibitem [{\citenamefont {Dorsch}\ \emph {et~al.}(2017)\citenamefont {Dorsch},
  \citenamefont {Huber}, \citenamefont {Konstandin},\ and\ \citenamefont
  {No}}]{Dorsch:2016nrg}%
  \BibitemOpen
  \bibfield  {author} {\bibinfo {author} {\bibfnamefont {G.~C.}\ \bibnamefont
  {Dorsch}}, \bibinfo {author} {\bibfnamefont {S.~J.}\ \bibnamefont {Huber}},
  \bibinfo {author} {\bibfnamefont {T.}~\bibnamefont {Konstandin}}, \ and\
  \bibinfo {author} {\bibfnamefont {J.~M.}\ \bibnamefont {No}},\ }\href
  {\doibase 10.1088/1475-7516/2017/05/052} {\bibfield  {journal} {\bibinfo
  {journal} {JCAP}\ }\textbf {\bibinfo {volume} {05}},\ \bibinfo {pages} {052}
  (\bibinfo {year} {2017})},\ \Eprint {http://arxiv.org/abs/1611.05874}
  {arXiv:1611.05874 [hep-ph]} \BibitemShut {NoStop}%
\bibitem [{\citenamefont {Modak}\ and\ \citenamefont
  {Senaha}(2019)}]{Modak:2018csw}%
  \BibitemOpen
  \bibfield  {author} {\bibinfo {author} {\bibfnamefont {T.}~\bibnamefont
  {Modak}}\ and\ \bibinfo {author} {\bibfnamefont {E.}~\bibnamefont {Senaha}},\
  }\href {\doibase 10.1103/PhysRevD.99.115022} {\bibfield  {journal} {\bibinfo
  {journal} {Phys. Rev. D}\ }\textbf {\bibinfo {volume} {99}},\ \bibinfo
  {pages} {115022} (\bibinfo {year} {2019})},\ \Eprint
  {http://arxiv.org/abs/1811.08088} {arXiv:1811.08088 [hep-ph]} \BibitemShut
  {NoStop}%
\bibitem [{\citenamefont {Basler}\ \emph {et~al.}(2023)\citenamefont {Basler},
  \citenamefont {Biermann}, \citenamefont {M\"uhlleitner},\ and\ \citenamefont
  {M\"uller}}]{Basler:2021kgq}%
  \BibitemOpen
  \bibfield  {author} {\bibinfo {author} {\bibfnamefont {P.}~\bibnamefont
  {Basler}}, \bibinfo {author} {\bibfnamefont {L.}~\bibnamefont {Biermann}},
  \bibinfo {author} {\bibfnamefont {M.}~\bibnamefont {M\"uhlleitner}}, \ and\
  \bibinfo {author} {\bibfnamefont {J.}~\bibnamefont {M\"uller}},\ }\href
  {\doibase 10.1140/epjc/s10052-023-11192-9} {\bibfield  {journal} {\bibinfo
  {journal} {Eur. Phys. J. C}\ }\textbf {\bibinfo {volume} {83}},\ \bibinfo
  {pages} {57} (\bibinfo {year} {2023})},\ \Eprint
  {http://arxiv.org/abs/2108.03580} {arXiv:2108.03580 [hep-ph]} \BibitemShut
  {NoStop}%
\bibitem [{\citenamefont {Enomoto}\ \emph
  {et~al.}(2022{\natexlab{a}})\citenamefont {Enomoto}, \citenamefont
  {Kanemura},\ and\ \citenamefont {Mura}}]{Enomoto:2021dkl}%
  \BibitemOpen
  \bibfield  {author} {\bibinfo {author} {\bibfnamefont {K.}~\bibnamefont
  {Enomoto}}, \bibinfo {author} {\bibfnamefont {S.}~\bibnamefont {Kanemura}}, \
  and\ \bibinfo {author} {\bibfnamefont {Y.}~\bibnamefont {Mura}},\ }\href
  {\doibase 10.1007/JHEP01(2022)104} {\bibfield  {journal} {\bibinfo  {journal}
  {JHEP}\ }\textbf {\bibinfo {volume} {01}},\ \bibinfo {pages} {104} (\bibinfo
  {year} {2022}{\natexlab{a}})},\ \Eprint {http://arxiv.org/abs/2111.13079}
  {arXiv:2111.13079 [hep-ph]} \BibitemShut {NoStop}%
\bibitem [{\citenamefont {Enomoto}\ \emph
  {et~al.}(2022{\natexlab{b}})\citenamefont {Enomoto}, \citenamefont
  {Kanemura},\ and\ \citenamefont {Mura}}]{Enomoto:2022rrl}%
  \BibitemOpen
  \bibfield  {author} {\bibinfo {author} {\bibfnamefont {K.}~\bibnamefont
  {Enomoto}}, \bibinfo {author} {\bibfnamefont {S.}~\bibnamefont {Kanemura}}, \
  and\ \bibinfo {author} {\bibfnamefont {Y.}~\bibnamefont {Mura}},\ }\href
  {\doibase 10.1007/JHEP09(2022)121} {\bibfield  {journal} {\bibinfo  {journal}
  {JHEP}\ }\textbf {\bibinfo {volume} {09}},\ \bibinfo {pages} {121} (\bibinfo
  {year} {2022}{\natexlab{b}})},\ \Eprint {http://arxiv.org/abs/2207.00060}
  {arXiv:2207.00060 [hep-ph]} \BibitemShut {NoStop}%
\bibitem [{\citenamefont {Zhou}\ and\ \citenamefont
  {Bian}(2022)}]{Zhou:2020irf}%
  \BibitemOpen
  \bibfield  {author} {\bibinfo {author} {\bibfnamefont {R.}~\bibnamefont
  {Zhou}}\ and\ \bibinfo {author} {\bibfnamefont {L.}~\bibnamefont {Bian}},\
  }\href {\doibase 10.1016/j.physletb.2022.137105} {\bibfield  {journal}
  {\bibinfo  {journal} {Phys. Lett. B}\ }\textbf {\bibinfo {volume} {829}},\
  \bibinfo {pages} {137105} (\bibinfo {year} {2022})},\ \Eprint
  {http://arxiv.org/abs/2001.01237} {arXiv:2001.01237 [hep-ph]} \BibitemShut
  {NoStop}%
\bibitem [{\citenamefont {Andreev}\ \emph {et~al.}(2018)\citenamefont {Andreev}
  \emph {et~al.}}]{ACME:2018yjb}%
  \BibitemOpen
  \bibfield  {author} {\bibinfo {author} {\bibfnamefont {V.}~\bibnamefont
  {Andreev}} \emph {et~al.} (\bibinfo {collaboration} {ACME}),\ }\href
  {\doibase 10.1038/s41586-018-0599-8} {\bibfield  {journal} {\bibinfo
  {journal} {Nature}\ }\textbf {\bibinfo {volume} {562}},\ \bibinfo {pages}
  {355} (\bibinfo {year} {2018})}\BibitemShut {NoStop}%
\bibitem [{\citenamefont {Abel}\ \emph {et~al.}(2020)\citenamefont {Abel} \emph
  {et~al.}}]{Abel:2020pzs}%
  \BibitemOpen
  \bibfield  {author} {\bibinfo {author} {\bibfnamefont {C.}~\bibnamefont
  {Abel}} \emph {et~al.},\ }\href {\doibase 10.1103/PhysRevLett.124.081803}
  {\bibfield  {journal} {\bibinfo  {journal} {Phys. Rev. Lett.}\ }\textbf
  {\bibinfo {volume} {124}},\ \bibinfo {pages} {081803} (\bibinfo {year}
  {2020})},\ \Eprint {http://arxiv.org/abs/2001.11966} {arXiv:2001.11966
  [hep-ex]} \BibitemShut {NoStop}%
\bibitem [{\citenamefont {Roussy}\ \emph {et~al.}(2022)\citenamefont {Roussy}
  \emph {et~al.}}]{Roussy:2022cmp}%
  \BibitemOpen
  \bibfield  {author} {\bibinfo {author} {\bibfnamefont {T.~S.}\ \bibnamefont
  {Roussy}} \emph {et~al.},\ }\href@noop {} {\  (\bibinfo {year} {2022})},\
  \Eprint {http://arxiv.org/abs/2212.11841} {arXiv:2212.11841
  [physics.atom-ph]} \BibitemShut {NoStop}%
\bibitem [{\citenamefont {Kanemura}\ \emph {et~al.}(2020)\citenamefont
  {Kanemura}, \citenamefont {Kubota},\ and\ \citenamefont
  {Yagyu}}]{Kanemura:2020ibp}%
  \BibitemOpen
  \bibfield  {author} {\bibinfo {author} {\bibfnamefont {S.}~\bibnamefont
  {Kanemura}}, \bibinfo {author} {\bibfnamefont {M.}~\bibnamefont {Kubota}}, \
  and\ \bibinfo {author} {\bibfnamefont {K.}~\bibnamefont {Yagyu}},\ }\href
  {\doibase 10.1007/JHEP08(2020)026} {\bibfield  {journal} {\bibinfo  {journal}
  {JHEP}\ }\textbf {\bibinfo {volume} {08}},\ \bibinfo {pages} {026} (\bibinfo
  {year} {2020})},\ \Eprint {http://arxiv.org/abs/2004.03943} {arXiv:2004.03943
  [hep-ph]} \BibitemShut {NoStop}%
\bibitem [{\citenamefont {Fujii}\ \emph {et~al.}(2015)\citenamefont {Fujii}
  \emph {et~al.}}]{Fujii:2015jha}%
  \BibitemOpen
  \bibfield  {author} {\bibinfo {author} {\bibfnamefont {K.}~\bibnamefont
  {Fujii}} \emph {et~al.},\ }\href@noop {} {\  (\bibinfo {year} {2015})},\
  \Eprint {http://arxiv.org/abs/1506.05992} {arXiv:1506.05992 [hep-ex]}
  \BibitemShut {NoStop}%
\bibitem [{\citenamefont {Bambade}\ \emph {et~al.}(2019)\citenamefont {Bambade}
  \emph {et~al.}}]{Bambade:2019fyw}%
  \BibitemOpen
  \bibfield  {author} {\bibinfo {author} {\bibfnamefont {P.}~\bibnamefont
  {Bambade}} \emph {et~al.},\ }\href@noop {} {\  (\bibinfo {year} {2019})},\
  \Eprint {http://arxiv.org/abs/1903.01629} {arXiv:1903.01629 [hep-ex]}
  \BibitemShut {NoStop}%
\bibitem [{\citenamefont {Kanemura}\ \emph {et~al.}(2005)\citenamefont
  {Kanemura}, \citenamefont {Okada},\ and\ \citenamefont
  {Senaha}}]{Kanemura:2004ch}%
  \BibitemOpen
  \bibfield  {author} {\bibinfo {author} {\bibfnamefont {S.}~\bibnamefont
  {Kanemura}}, \bibinfo {author} {\bibfnamefont {Y.}~\bibnamefont {Okada}}, \
  and\ \bibinfo {author} {\bibfnamefont {E.}~\bibnamefont {Senaha}},\ }\href
  {\doibase 10.1016/j.physletb.2004.12.004} {\bibfield  {journal} {\bibinfo
  {journal} {Phys. Lett. B}\ }\textbf {\bibinfo {volume} {606}},\ \bibinfo
  {pages} {361} (\bibinfo {year} {2005})},\ \Eprint
  {http://arxiv.org/abs/hep-ph/0411354} {arXiv:hep-ph/0411354} \BibitemShut
  {NoStop}%
\bibitem [{\citenamefont {Cepeda}\ \emph {et~al.}(2019)\citenamefont {Cepeda}
  \emph {et~al.}}]{Cepeda:2019klc}%
  \BibitemOpen
  \bibfield  {author} {\bibinfo {author} {\bibfnamefont {M.}~\bibnamefont
  {Cepeda}} \emph {et~al.},\ }\href {\doibase 10.23731/CYRM-2019-007.221}
  {\bibfield  {journal} {\bibinfo  {journal} {CERN Yellow Rep. Monogr.}\
  }\textbf {\bibinfo {volume} {7}},\ \bibinfo {pages} {221} (\bibinfo {year}
  {2019})},\ \Eprint {http://arxiv.org/abs/1902.00134} {arXiv:1902.00134
  [hep-ph]} \BibitemShut {NoStop}%
\bibitem [{\citenamefont {Charles}\ \emph {et~al.}(2018)\citenamefont {Charles}
  \emph {et~al.}}]{CLICdp:2018cto}%
  \BibitemOpen
  \bibfield  {author} {\bibinfo {author} {\bibfnamefont {T.~K.}\ \bibnamefont
  {Charles}} \emph {et~al.} (\bibinfo {collaboration} {CLICdp, CLIC}),\ }\href
  {\doibase 10.23731/CYRM-2018-002} {\ \textbf {\bibinfo {volume} {2/2018}}
  (\bibinfo {year} {2018}),\ 10.23731/CYRM-2018-002},\ \Eprint
  {http://arxiv.org/abs/1812.06018} {arXiv:1812.06018 [physics.acc-ph]}
  \BibitemShut {NoStop}%
\bibitem [{\citenamefont {Grojean}\ and\ \citenamefont
  {Servant}(2007)}]{Grojean:2006bp}%
  \BibitemOpen
  \bibfield  {author} {\bibinfo {author} {\bibfnamefont {C.}~\bibnamefont
  {Grojean}}\ and\ \bibinfo {author} {\bibfnamefont {G.}~\bibnamefont
  {Servant}},\ }\href {\doibase 10.1103/PhysRevD.75.043507} {\bibfield
  {journal} {\bibinfo  {journal} {Phys. Rev. D}\ }\textbf {\bibinfo {volume}
  {75}},\ \bibinfo {pages} {043507} (\bibinfo {year} {2007})},\ \Eprint
  {http://arxiv.org/abs/hep-ph/0607107} {arXiv:hep-ph/0607107} \BibitemShut
  {NoStop}%
\bibitem [{\citenamefont {Kakizaki}\ \emph {et~al.}(2015)\citenamefont
  {Kakizaki}, \citenamefont {Kanemura},\ and\ \citenamefont
  {Matsui}}]{Kakizaki:2015wua}%
  \BibitemOpen
  \bibfield  {author} {\bibinfo {author} {\bibfnamefont {M.}~\bibnamefont
  {Kakizaki}}, \bibinfo {author} {\bibfnamefont {S.}~\bibnamefont {Kanemura}},
  \ and\ \bibinfo {author} {\bibfnamefont {T.}~\bibnamefont {Matsui}},\ }\href
  {\doibase 10.1103/PhysRevD.92.115007} {\bibfield  {journal} {\bibinfo
  {journal} {Phys. Rev. D}\ }\textbf {\bibinfo {volume} {92}},\ \bibinfo
  {pages} {115007} (\bibinfo {year} {2015})},\ \Eprint
  {http://arxiv.org/abs/1509.08394} {arXiv:1509.08394 [hep-ph]} \BibitemShut
  {NoStop}%
\bibitem [{\citenamefont {Hashino}\ \emph {et~al.}(2016)\citenamefont
  {Hashino}, \citenamefont {Kakizaki}, \citenamefont {Kanemura},\ and\
  \citenamefont {Matsui}}]{Hashino:2016rvx}%
  \BibitemOpen
  \bibfield  {author} {\bibinfo {author} {\bibfnamefont {K.}~\bibnamefont
  {Hashino}}, \bibinfo {author} {\bibfnamefont {M.}~\bibnamefont {Kakizaki}},
  \bibinfo {author} {\bibfnamefont {S.}~\bibnamefont {Kanemura}}, \ and\
  \bibinfo {author} {\bibfnamefont {T.}~\bibnamefont {Matsui}},\ }\href
  {\doibase 10.1103/PhysRevD.94.015005} {\bibfield  {journal} {\bibinfo
  {journal} {Phys. Rev. D}\ }\textbf {\bibinfo {volume} {94}},\ \bibinfo
  {pages} {015005} (\bibinfo {year} {2016})},\ \Eprint
  {http://arxiv.org/abs/1604.02069} {arXiv:1604.02069 [hep-ph]} \BibitemShut
  {NoStop}%
\bibitem [{\citenamefont {Hashino}\ \emph {et~al.}(2019)\citenamefont
  {Hashino}, \citenamefont {Jinno}, \citenamefont {Kakizaki}, \citenamefont
  {Kanemura}, \citenamefont {Takahashi},\ and\ \citenamefont
  {Takimoto}}]{Hashino:2018wee}%
  \BibitemOpen
  \bibfield  {author} {\bibinfo {author} {\bibfnamefont {K.}~\bibnamefont
  {Hashino}}, \bibinfo {author} {\bibfnamefont {R.}~\bibnamefont {Jinno}},
  \bibinfo {author} {\bibfnamefont {M.}~\bibnamefont {Kakizaki}}, \bibinfo
  {author} {\bibfnamefont {S.}~\bibnamefont {Kanemura}}, \bibinfo {author}
  {\bibfnamefont {T.}~\bibnamefont {Takahashi}}, \ and\ \bibinfo {author}
  {\bibfnamefont {M.}~\bibnamefont {Takimoto}},\ }\href {\doibase
  10.1103/PhysRevD.99.075011} {\bibfield  {journal} {\bibinfo  {journal} {Phys.
  Rev. D}\ }\textbf {\bibinfo {volume} {99}},\ \bibinfo {pages} {075011}
  (\bibinfo {year} {2019})},\ \Eprint {http://arxiv.org/abs/1809.04994}
  {arXiv:1809.04994 [hep-ph]} \BibitemShut {NoStop}%
\bibitem [{\citenamefont {Kanemura}\ \emph {et~al.}(2022)\citenamefont
  {Kanemura}, \citenamefont {Nagai},\ and\ \citenamefont
  {Tanaka}}]{Kanemura:2022txx}%
  \BibitemOpen
  \bibfield  {author} {\bibinfo {author} {\bibfnamefont {S.}~\bibnamefont
  {Kanemura}}, \bibinfo {author} {\bibfnamefont {R.}~\bibnamefont {Nagai}}, \
  and\ \bibinfo {author} {\bibfnamefont {M.}~\bibnamefont {Tanaka}},\ }\href
  {\doibase 10.1007/JHEP06(2022)027} {\bibfield  {journal} {\bibinfo  {journal}
  {JHEP}\ }\textbf {\bibinfo {volume} {06}},\ \bibinfo {pages} {027} (\bibinfo
  {year} {2022})},\ \Eprint {http://arxiv.org/abs/2202.12774} {arXiv:2202.12774
  [hep-ph]} \BibitemShut {NoStop}%
\bibitem [{\citenamefont {Amaro-Seoane}\ \emph {et~al.}(2017)\citenamefont
  {Amaro-Seoane} \emph {et~al.}}]{LISA:2017pwj}%
  \BibitemOpen
  \bibfield  {author} {\bibinfo {author} {\bibfnamefont {P.}~\bibnamefont
  {Amaro-Seoane}} \emph {et~al.} (\bibinfo {collaboration} {LISA}),\
  }\href@noop {} {\  (\bibinfo {year} {2017})},\ \Eprint
  {http://arxiv.org/abs/1702.00786} {arXiv:1702.00786 [astro-ph.IM]}
  \BibitemShut {NoStop}%
\bibitem [{\citenamefont {Seto}\ \emph {et~al.}(2001)\citenamefont {Seto},
  \citenamefont {Kawamura},\ and\ \citenamefont {Nakamura}}]{Seto:2001qf}%
  \BibitemOpen
  \bibfield  {author} {\bibinfo {author} {\bibfnamefont {N.}~\bibnamefont
  {Seto}}, \bibinfo {author} {\bibfnamefont {S.}~\bibnamefont {Kawamura}}, \
  and\ \bibinfo {author} {\bibfnamefont {T.}~\bibnamefont {Nakamura}},\ }\href
  {\doibase 10.1103/PhysRevLett.87.221103} {\bibfield  {journal} {\bibinfo
  {journal} {Phys. Rev. Lett.}\ }\textbf {\bibinfo {volume} {87}},\ \bibinfo
  {pages} {221103} (\bibinfo {year} {2001})},\ \Eprint
  {http://arxiv.org/abs/astro-ph/0108011} {arXiv:astro-ph/0108011} \BibitemShut
  {NoStop}%
\bibitem [{\citenamefont {Joyce}\ \emph {et~al.}(1995)\citenamefont {Joyce},
  \citenamefont {Prokopec},\ and\ \citenamefont {Turok}}]{Joyce:1994fu}%
  \BibitemOpen
  \bibfield  {author} {\bibinfo {author} {\bibfnamefont {M.}~\bibnamefont
  {Joyce}}, \bibinfo {author} {\bibfnamefont {T.}~\bibnamefont {Prokopec}}, \
  and\ \bibinfo {author} {\bibfnamefont {N.}~\bibnamefont {Turok}},\ }\href
  {\doibase 10.1103/PhysRevLett.75.1695} {\bibfield  {journal} {\bibinfo
  {journal} {Phys. Rev. Lett.}\ }\textbf {\bibinfo {volume} {75}},\ \bibinfo
  {pages} {1695} (\bibinfo {year} {1995})},\ \bibinfo {note} {[Erratum:
  Phys.Rev.Lett. 75, 3375 (1995)]},\ \Eprint
  {http://arxiv.org/abs/hep-ph/9408339} {arXiv:hep-ph/9408339} \BibitemShut
  {NoStop}%
\bibitem [{\citenamefont {Joyce}\ \emph
  {et~al.}(1996{\natexlab{a}})\citenamefont {Joyce}, \citenamefont {Prokopec},\
  and\ \citenamefont {Turok}}]{Joyce:1994zn}%
  \BibitemOpen
  \bibfield  {author} {\bibinfo {author} {\bibfnamefont {M.}~\bibnamefont
  {Joyce}}, \bibinfo {author} {\bibfnamefont {T.}~\bibnamefont {Prokopec}}, \
  and\ \bibinfo {author} {\bibfnamefont {N.}~\bibnamefont {Turok}},\ }\href
  {\doibase 10.1103/PhysRevD.53.2930} {\bibfield  {journal} {\bibinfo
  {journal} {Phys. Rev. D}\ }\textbf {\bibinfo {volume} {53}},\ \bibinfo
  {pages} {2930} (\bibinfo {year} {1996}{\natexlab{a}})},\ \Eprint
  {http://arxiv.org/abs/hep-ph/9410281} {arXiv:hep-ph/9410281} \BibitemShut
  {NoStop}%
\bibitem [{\citenamefont {Joyce}\ \emph
  {et~al.}(1996{\natexlab{b}})\citenamefont {Joyce}, \citenamefont {Prokopec},\
  and\ \citenamefont {Turok}}]{Joyce:1994zt}%
  \BibitemOpen
  \bibfield  {author} {\bibinfo {author} {\bibfnamefont {M.}~\bibnamefont
  {Joyce}}, \bibinfo {author} {\bibfnamefont {T.}~\bibnamefont {Prokopec}}, \
  and\ \bibinfo {author} {\bibfnamefont {N.}~\bibnamefont {Turok}},\ }\href
  {\doibase 10.1103/PhysRevD.53.2958} {\bibfield  {journal} {\bibinfo
  {journal} {Phys. Rev. D}\ }\textbf {\bibinfo {volume} {53}},\ \bibinfo
  {pages} {2958} (\bibinfo {year} {1996}{\natexlab{b}})},\ \Eprint
  {http://arxiv.org/abs/hep-ph/9410282} {arXiv:hep-ph/9410282} \BibitemShut
  {NoStop}%
\bibitem [{\citenamefont {Fromme}\ and\ \citenamefont
  {Huber}(2007)}]{Fromme:2006wx}%
  \BibitemOpen
  \bibfield  {author} {\bibinfo {author} {\bibfnamefont {L.}~\bibnamefont
  {Fromme}}\ and\ \bibinfo {author} {\bibfnamefont {S.~J.}\ \bibnamefont
  {Huber}},\ }\href {\doibase 10.1088/1126-6708/2007/03/049} {\bibfield
  {journal} {\bibinfo  {journal} {JHEP}\ }\textbf {\bibinfo {volume} {03}},\
  \bibinfo {pages} {049} (\bibinfo {year} {2007})},\ \Eprint
  {http://arxiv.org/abs/hep-ph/0604159} {arXiv:hep-ph/0604159} \BibitemShut
  {NoStop}%
\bibitem [{\citenamefont {Cline}\ \emph {et~al.}(2000)\citenamefont {Cline},
  \citenamefont {Joyce},\ and\ \citenamefont {Kainulainen}}]{Cline:2000nw}%
  \BibitemOpen
  \bibfield  {author} {\bibinfo {author} {\bibfnamefont {J.~M.}\ \bibnamefont
  {Cline}}, \bibinfo {author} {\bibfnamefont {M.}~\bibnamefont {Joyce}}, \ and\
  \bibinfo {author} {\bibfnamefont {K.}~\bibnamefont {Kainulainen}},\ }\href
  {\doibase 10.1088/1126-6708/2000/07/018} {\bibfield  {journal} {\bibinfo
  {journal} {JHEP}\ }\textbf {\bibinfo {volume} {07}},\ \bibinfo {pages} {018}
  (\bibinfo {year} {2000})},\ \Eprint {http://arxiv.org/abs/hep-ph/0006119}
  {arXiv:hep-ph/0006119} \BibitemShut {NoStop}%
\bibitem [{\citenamefont {Prokopec}\ \emph
  {et~al.}(2004{\natexlab{a}})\citenamefont {Prokopec}, \citenamefont
  {Schmidt},\ and\ \citenamefont {Weinstock}}]{Prokopec:2003pj}%
  \BibitemOpen
  \bibfield  {author} {\bibinfo {author} {\bibfnamefont {T.}~\bibnamefont
  {Prokopec}}, \bibinfo {author} {\bibfnamefont {M.~G.}\ \bibnamefont
  {Schmidt}}, \ and\ \bibinfo {author} {\bibfnamefont {S.}~\bibnamefont
  {Weinstock}},\ }\href {\doibase 10.1016/j.aop.2004.06.002} {\bibfield
  {journal} {\bibinfo  {journal} {Annals Phys.}\ }\textbf {\bibinfo {volume}
  {314}},\ \bibinfo {pages} {208} (\bibinfo {year} {2004}{\natexlab{a}})},\
  \Eprint {http://arxiv.org/abs/hep-ph/0312110} {arXiv:hep-ph/0312110}
  \BibitemShut {NoStop}%
\bibitem [{\citenamefont {Prokopec}\ \emph
  {et~al.}(2004{\natexlab{b}})\citenamefont {Prokopec}, \citenamefont
  {Schmidt},\ and\ \citenamefont {Weinstock}}]{Prokopec:2004ic}%
  \BibitemOpen
  \bibfield  {author} {\bibinfo {author} {\bibfnamefont {T.}~\bibnamefont
  {Prokopec}}, \bibinfo {author} {\bibfnamefont {M.~G.}\ \bibnamefont
  {Schmidt}}, \ and\ \bibinfo {author} {\bibfnamefont {S.}~\bibnamefont
  {Weinstock}},\ }\href {\doibase 10.1016/j.aop.2004.06.001} {\bibfield
  {journal} {\bibinfo  {journal} {Annals Phys.}\ }\textbf {\bibinfo {volume}
  {314}},\ \bibinfo {pages} {267} (\bibinfo {year} {2004}{\natexlab{b}})},\
  \Eprint {http://arxiv.org/abs/hep-ph/0406140} {arXiv:hep-ph/0406140}
  \BibitemShut {NoStop}%
\bibitem [{\citenamefont {Riotto}(1998)}]{Riotto:1998zb}%
  \BibitemOpen
  \bibfield  {author} {\bibinfo {author} {\bibfnamefont {A.}~\bibnamefont
  {Riotto}},\ }\href {\doibase 10.1103/PhysRevD.58.095009} {\bibfield
  {journal} {\bibinfo  {journal} {Phys. Rev. D}\ }\textbf {\bibinfo {volume}
  {58}},\ \bibinfo {pages} {095009} (\bibinfo {year} {1998})},\ \Eprint
  {http://arxiv.org/abs/hep-ph/9803357} {arXiv:hep-ph/9803357} \BibitemShut
  {NoStop}%
\bibitem [{\citenamefont {Cline}\ and\ \citenamefont
  {Kainulainen}(2020)}]{Cline:2020jre}%
  \BibitemOpen
  \bibfield  {author} {\bibinfo {author} {\bibfnamefont {J.~M.}\ \bibnamefont
  {Cline}}\ and\ \bibinfo {author} {\bibfnamefont {K.}~\bibnamefont
  {Kainulainen}},\ }\href {\doibase 10.1103/PhysRevD.101.063525} {\bibfield
  {journal} {\bibinfo  {journal} {Phys. Rev. D}\ }\textbf {\bibinfo {volume}
  {101}},\ \bibinfo {pages} {063525} (\bibinfo {year} {2020})},\ \Eprint
  {http://arxiv.org/abs/2001.00568} {arXiv:2001.00568 [hep-ph]} \BibitemShut
  {NoStop}%
\bibitem [{\citenamefont {Aaboud}\ \emph {et~al.}(2019)\citenamefont {Aaboud}
  \emph {et~al.}}]{ATLAS:2018jqi}%
  \BibitemOpen
  \bibfield  {author} {\bibinfo {author} {\bibfnamefont {M.}~\bibnamefont
  {Aaboud}} \emph {et~al.} (\bibinfo {collaboration} {ATLAS}),\ }\href
  {\doibase 10.1007/JHEP05(2019)123} {\bibfield  {journal} {\bibinfo  {journal}
  {JHEP}\ }\textbf {\bibinfo {volume} {05}},\ \bibinfo {pages} {123} (\bibinfo
  {year} {2019})},\ \Eprint {http://arxiv.org/abs/1812.11568} {arXiv:1812.11568
  [hep-ex]} \BibitemShut {NoStop}%
\bibitem [{\citenamefont {Sirunyan}\ \emph
  {et~al.}(2018{\natexlab{a}})\citenamefont {Sirunyan} \emph
  {et~al.}}]{CMS:2017bhz}%
  \BibitemOpen
  \bibfield  {author} {\bibinfo {author} {\bibfnamefont {A.~M.}\ \bibnamefont
  {Sirunyan}} \emph {et~al.} (\bibinfo {collaboration} {CMS}),\ }\href
  {\doibase 10.1007/JHEP06(2018)102} {\bibfield  {journal} {\bibinfo  {journal}
  {JHEP}\ }\textbf {\bibinfo {volume} {06}},\ \bibinfo {pages} {102} (\bibinfo
  {year} {2018}{\natexlab{a}})},\ \Eprint {http://arxiv.org/abs/1712.02399}
  {arXiv:1712.02399 [hep-ex]} \BibitemShut {NoStop}%
\bibitem [{\citenamefont {Sirunyan}\ \emph
  {et~al.}(2018{\natexlab{b}})\citenamefont {Sirunyan} \emph
  {et~al.}}]{CMS:2017ocm}%
  \BibitemOpen
  \bibfield  {author} {\bibinfo {author} {\bibfnamefont {A.~M.}\ \bibnamefont
  {Sirunyan}} \emph {et~al.} (\bibinfo {collaboration} {CMS}),\ }\href
  {\doibase 10.1140/epjc/s10052-018-5607-5} {\bibfield  {journal} {\bibinfo
  {journal} {Eur. Phys. J. C}\ }\textbf {\bibinfo {volume} {78}},\ \bibinfo
  {pages} {140} (\bibinfo {year} {2018}{\natexlab{b}})},\ \Eprint
  {http://arxiv.org/abs/1710.10614} {arXiv:1710.10614 [hep-ex]} \BibitemShut
  {NoStop}%
\bibitem [{\citenamefont {Sirunyan}\ \emph
  {et~al.}(2020{\natexlab{a}})\citenamefont {Sirunyan} \emph
  {et~al.}}]{CMS:2019rvj}%
  \BibitemOpen
  \bibfield  {author} {\bibinfo {author} {\bibfnamefont {A.~M.}\ \bibnamefont
  {Sirunyan}} \emph {et~al.} (\bibinfo {collaboration} {CMS}),\ }\href
  {\doibase 10.1140/epjc/s10052-019-7593-7} {\bibfield  {journal} {\bibinfo
  {journal} {Eur. Phys. J. C}\ }\textbf {\bibinfo {volume} {80}},\ \bibinfo
  {pages} {75} (\bibinfo {year} {2020}{\natexlab{a}})},\ \Eprint
  {http://arxiv.org/abs/1908.06463} {arXiv:1908.06463 [hep-ex]} \BibitemShut
  {NoStop}%
\bibitem [{\citenamefont {Aaboud}\ \emph {et~al.}(2018)\citenamefont {Aaboud}
  \emph {et~al.}}]{ATLAS:2018rvc}%
  \BibitemOpen
  \bibfield  {author} {\bibinfo {author} {\bibfnamefont {M.}~\bibnamefont
  {Aaboud}} \emph {et~al.} (\bibinfo {collaboration} {ATLAS}),\ }\href
  {\doibase 10.1140/epjc/s10052-018-5995-6} {\bibfield  {journal} {\bibinfo
  {journal} {Eur. Phys. J. C}\ }\textbf {\bibinfo {volume} {78}},\ \bibinfo
  {pages} {565} (\bibinfo {year} {2018})},\ \Eprint
  {http://arxiv.org/abs/1804.10823} {arXiv:1804.10823 [hep-ex]} \BibitemShut
  {NoStop}%
\bibitem [{\citenamefont {Sirunyan}\ \emph
  {et~al.}(2020{\natexlab{b}})\citenamefont {Sirunyan} \emph
  {et~al.}}]{CMS:2019pzc}%
  \BibitemOpen
  \bibfield  {author} {\bibinfo {author} {\bibfnamefont {A.~M.}\ \bibnamefont
  {Sirunyan}} \emph {et~al.} (\bibinfo {collaboration} {CMS}),\ }\href
  {\doibase 10.1007/JHEP04(2020)171} {\bibfield  {journal} {\bibinfo  {journal}
  {JHEP}\ }\textbf {\bibinfo {volume} {04}},\ \bibinfo {pages} {171} (\bibinfo
  {year} {2020}{\natexlab{b}})},\ \bibinfo {note} {[Erratum: JHEP 03, 187
  (2022)]},\ \Eprint {http://arxiv.org/abs/1908.01115} {arXiv:1908.01115
  [hep-ex]} \BibitemShut {NoStop}%
\bibitem [{\citenamefont {Aad}\ \emph {et~al.}(2021)\citenamefont {Aad} \emph
  {et~al.}}]{ATLAS:2021upq}%
  \BibitemOpen
  \bibfield  {author} {\bibinfo {author} {\bibfnamefont {G.}~\bibnamefont
  {Aad}} \emph {et~al.} (\bibinfo {collaboration} {ATLAS}),\ }\href {\doibase
  10.1007/JHEP06(2021)145} {\bibfield  {journal} {\bibinfo  {journal} {JHEP}\
  }\textbf {\bibinfo {volume} {06}},\ \bibinfo {pages} {145} (\bibinfo {year}
  {2021})},\ \Eprint {http://arxiv.org/abs/2102.10076} {arXiv:2102.10076
  [hep-ex]} \BibitemShut {NoStop}%
\bibitem [{ATL(2022)}]{ATLAS:2022xpz}%
  \BibitemOpen
  \href@noop {} {\  (\bibinfo {year} {2022})}\BibitemShut {NoStop}%
\bibitem [{\citenamefont {Bona}\ \emph {et~al.}(2008)\citenamefont {Bona} \emph
  {et~al.}}]{UTfit:2007eik}%
  \BibitemOpen
  \bibfield  {author} {\bibinfo {author} {\bibfnamefont {M.}~\bibnamefont
  {Bona}} \emph {et~al.} (\bibinfo {collaboration} {UTfit}),\ }\href {\doibase
  10.1088/1126-6708/2008/03/049} {\bibfield  {journal} {\bibinfo  {journal}
  {JHEP}\ }\textbf {\bibinfo {volume} {03}},\ \bibinfo {pages} {049} (\bibinfo
  {year} {2008})},\ \Eprint {http://arxiv.org/abs/0707.0636} {arXiv:0707.0636
  [hep-ph]} \BibitemShut {NoStop}%
\bibitem [{UTf()}]{UTfit2018}%
  \BibitemOpen
  \href@noop {} {\enquote {\bibinfo {title} {{UTfit Collaboration}},}\
  }\bibinfo {howpublished}
  {\url{http://www.utfit.org/UTfit/ResultsSummer2018NP}}\BibitemShut {NoStop}%
\bibitem [{\citenamefont {Chen}\ and\ \citenamefont
  {Nomura}(2018)}]{Chen:2018ytc}%
  \BibitemOpen
  \bibfield  {author} {\bibinfo {author} {\bibfnamefont {C.-H.}\ \bibnamefont
  {Chen}}\ and\ \bibinfo {author} {\bibfnamefont {T.}~\bibnamefont {Nomura}},\
  }\href {\doibase 10.1007/JHEP08(2018)145} {\bibfield  {journal} {\bibinfo
  {journal} {JHEP}\ }\textbf {\bibinfo {volume} {08}},\ \bibinfo {pages} {145}
  (\bibinfo {year} {2018})},\ \Eprint {http://arxiv.org/abs/1804.06017}
  {arXiv:1804.06017 [hep-ph]} \BibitemShut {NoStop}%
\bibitem [{\citenamefont {Haller}\ \emph {et~al.}(2018)\citenamefont {Haller},
  \citenamefont {Hoecker}, \citenamefont {Kogler}, \citenamefont {M\"onig},
  \citenamefont {Peiffer},\ and\ \citenamefont {Stelzer}}]{Haller:2018nnx}%
  \BibitemOpen
  \bibfield  {author} {\bibinfo {author} {\bibfnamefont {J.}~\bibnamefont
  {Haller}}, \bibinfo {author} {\bibfnamefont {A.}~\bibnamefont {Hoecker}},
  \bibinfo {author} {\bibfnamefont {R.}~\bibnamefont {Kogler}}, \bibinfo
  {author} {\bibfnamefont {K.}~\bibnamefont {M\"onig}}, \bibinfo {author}
  {\bibfnamefont {T.}~\bibnamefont {Peiffer}}, \ and\ \bibinfo {author}
  {\bibfnamefont {J.}~\bibnamefont {Stelzer}},\ }\href {\doibase
  10.1140/epjc/s10052-018-6131-3} {\bibfield  {journal} {\bibinfo  {journal}
  {Eur. Phys. J. C}\ }\textbf {\bibinfo {volume} {78}},\ \bibinfo {pages} {675}
  (\bibinfo {year} {2018})},\ \Eprint {http://arxiv.org/abs/1803.01853}
  {arXiv:1803.01853 [hep-ph]} \BibitemShut {NoStop}%
\bibitem [{\citenamefont {Amhis}\ \emph {et~al.}(2022)\citenamefont {Amhis}
  \emph {et~al.}}]{HFLAV:2022pwe}%
  \BibitemOpen
  \bibfield  {author} {\bibinfo {author} {\bibfnamefont {Y.}~\bibnamefont
  {Amhis}} \emph {et~al.} (\bibinfo {collaboration} {HFLAV}),\ }\href@noop {}
  {\  (\bibinfo {year} {2022})},\ \Eprint {http://arxiv.org/abs/2206.07501}
  {arXiv:2206.07501 [hep-ex]} \BibitemShut {NoStop}%
\bibitem [{CMS(2022)}]{CMS:2022mgd}%
  \BibitemOpen
  \href@noop {} {\enquote {\bibinfo {title} {{Measurement of the
  B$^0_\mathrm{S}$$\to$$\mu^+\mu^-$ decay properties and search for the
  B$^0$$\to$$\mu^+\mu^-$ decay in proton-proton collisions at $\sqrt{s}$ = 13
  TeV}},}\ } (\bibinfo {year} {2022}),\ \Eprint
  {http://arxiv.org/abs/2212.10311} {arXiv:2212.10311 [hep-ex]} \BibitemShut
  {NoStop}%
\bibitem [{\citenamefont {Ahn}\ \emph {et~al.}(2019)\citenamefont {Ahn} \emph
  {et~al.}}]{KOTO:2018dsc}%
  \BibitemOpen
  \bibfield  {author} {\bibinfo {author} {\bibfnamefont {J.~K.}\ \bibnamefont
  {Ahn}} \emph {et~al.} (\bibinfo {collaboration} {KOTO}),\ }\href {\doibase
  10.1103/PhysRevLett.122.021802} {\bibfield  {journal} {\bibinfo  {journal}
  {Phys. Rev. Lett.}\ }\textbf {\bibinfo {volume} {122}},\ \bibinfo {pages}
  {021802} (\bibinfo {year} {2019})},\ \Eprint
  {http://arxiv.org/abs/1810.09655} {arXiv:1810.09655 [hep-ex]} \BibitemShut
  {NoStop}%
\bibitem [{\citenamefont {Cortina~Gil}\ \emph {et~al.}(2021)\citenamefont
  {Cortina~Gil} \emph {et~al.}}]{NA62:2021zjw}%
  \BibitemOpen
  \bibfield  {author} {\bibinfo {author} {\bibfnamefont {E.}~\bibnamefont
  {Cortina~Gil}} \emph {et~al.} (\bibinfo {collaboration} {NA62}),\ }\href
  {\doibase 10.1007/JHEP06(2021)093} {\bibfield  {journal} {\bibinfo  {journal}
  {JHEP}\ }\textbf {\bibinfo {volume} {06}},\ \bibinfo {pages} {093} (\bibinfo
  {year} {2021})},\ \Eprint {http://arxiv.org/abs/2103.15389} {arXiv:2103.15389
  [hep-ex]} \BibitemShut {NoStop}%
\bibitem [{\citenamefont {Zamkovsk\'y}\ \emph {et~al.}(2022)\citenamefont
  {Zamkovsk\'y} \emph {et~al.}}]{NA62:2022hqi}%
  \BibitemOpen
  \bibfield  {author} {\bibinfo {author} {\bibfnamefont {M.}~\bibnamefont
  {Zamkovsk\'y}} \emph {et~al.} (\bibinfo {collaboration} {NA62}),\ }\href
  {\doibase 10.22323/1.405.0070} {\bibfield  {journal} {\bibinfo  {journal}
  {PoS}\ }\textbf {\bibinfo {volume} {DISCRETE2020-2021}},\ \bibinfo {pages}
  {070} (\bibinfo {year} {2022})}\BibitemShut {NoStop}%
\bibitem [{\citenamefont {Iguro}(2023)}]{Iguro:2023jju}%
  \BibitemOpen
  \bibfield  {author} {\bibinfo {author} {\bibfnamefont {S.}~\bibnamefont
  {Iguro}},\ }\href@noop {} {\  (\bibinfo {year} {2023})},\ \Eprint
  {http://arxiv.org/abs/2302.08935} {arXiv:2302.08935 [hep-ph]} \BibitemShut
  {NoStop}%
\bibitem [{\citenamefont {Iguro}\ and\ \citenamefont
  {Omura}(2019)}]{Iguro:2019zlc}%
  \BibitemOpen
  \bibfield  {author} {\bibinfo {author} {\bibfnamefont {S.}~\bibnamefont
  {Iguro}}\ and\ \bibinfo {author} {\bibfnamefont {Y.}~\bibnamefont {Omura}},\
  }\href {\doibase 10.1007/JHEP08(2019)098} {\bibfield  {journal} {\bibinfo
  {journal} {JHEP}\ }\textbf {\bibinfo {volume} {08}},\ \bibinfo {pages} {098}
  (\bibinfo {year} {2019})},\ \Eprint {http://arxiv.org/abs/1905.11778}
  {arXiv:1905.11778 [hep-ph]} \BibitemShut {NoStop}%
\bibitem [{\citenamefont {Hou}\ and\ \citenamefont
  {Kumar}(2022)}]{Hou:2022qvx}%
  \BibitemOpen
  \bibfield  {author} {\bibinfo {author} {\bibfnamefont {W.-S.}\ \bibnamefont
  {Hou}}\ and\ \bibinfo {author} {\bibfnamefont {G.}~\bibnamefont {Kumar}},\
  }\href {\doibase 10.1007/JHEP10(2022)129} {\bibfield  {journal} {\bibinfo
  {journal} {JHEP}\ }\textbf {\bibinfo {volume} {10}},\ \bibinfo {pages} {129}
  (\bibinfo {year} {2022})},\ \Eprint {http://arxiv.org/abs/2207.07030}
  {arXiv:2207.07030 [hep-ph]} \BibitemShut {NoStop}%
\bibitem [{\citenamefont {Aoki}\ \emph {et~al.}(2021)\citenamefont {Aoki} \emph
  {et~al.}}]{Aoki:2021cqa}%
  \BibitemOpen
  \bibfield  {author} {\bibinfo {author} {\bibfnamefont {K.}~\bibnamefont
  {Aoki}} \emph {et~al.},\ }\href@noop {} {\  (\bibinfo {year} {2021})},\
  \Eprint {http://arxiv.org/abs/2110.04462} {arXiv:2110.04462 [nucl-ex]}
  \BibitemShut {NoStop}%
\bibitem [{\citenamefont {Davidson}\ and\ \citenamefont
  {Haber}(2005)}]{Davidson:2005cw}%
  \BibitemOpen
  \bibfield  {author} {\bibinfo {author} {\bibfnamefont {S.}~\bibnamefont
  {Davidson}}\ and\ \bibinfo {author} {\bibfnamefont {H.~E.}\ \bibnamefont
  {Haber}},\ }\href {\doibase 10.1103/PhysRevD.72.099902} {\bibfield  {journal}
  {\bibinfo  {journal} {Phys. Rev. D}\ }\textbf {\bibinfo {volume} {72}},\
  \bibinfo {pages} {035004} (\bibinfo {year} {2005})},\ \bibinfo {note}
  {[Erratum: Phys.Rev.D 72, 099902 (2005)]},\ \Eprint
  {http://arxiv.org/abs/hep-ph/0504050} {arXiv:hep-ph/0504050} \BibitemShut
  {NoStop}%
\bibitem [{\citenamefont {Aad}\ \emph {et~al.}(2020)\citenamefont {Aad} \emph
  {et~al.}}]{ATLAS:2019nkf}%
  \BibitemOpen
  \bibfield  {author} {\bibinfo {author} {\bibfnamefont {G.}~\bibnamefont
  {Aad}} \emph {et~al.} (\bibinfo {collaboration} {ATLAS}),\ }\href {\doibase
  10.1103/PhysRevD.101.012002} {\bibfield  {journal} {\bibinfo  {journal}
  {Phys. Rev. D}\ }\textbf {\bibinfo {volume} {101}},\ \bibinfo {pages}
  {012002} (\bibinfo {year} {2020})},\ \Eprint
  {http://arxiv.org/abs/1909.02845} {arXiv:1909.02845 [hep-ex]} \BibitemShut
  {NoStop}%
\bibitem [{\citenamefont {Sirunyan}\ \emph {et~al.}(2019)\citenamefont
  {Sirunyan} \emph {et~al.}}]{CMS:2018uag}%
  \BibitemOpen
  \bibfield  {author} {\bibinfo {author} {\bibfnamefont {A.~M.}\ \bibnamefont
  {Sirunyan}} \emph {et~al.} (\bibinfo {collaboration} {CMS}),\ }\href
  {\doibase 10.1140/epjc/s10052-019-6909-y} {\bibfield  {journal} {\bibinfo
  {journal} {Eur. Phys. J. C}\ }\textbf {\bibinfo {volume} {79}},\ \bibinfo
  {pages} {421} (\bibinfo {year} {2019})},\ \Eprint
  {http://arxiv.org/abs/1809.10733} {arXiv:1809.10733 [hep-ex]} \BibitemShut
  {NoStop}%
\bibitem [{\citenamefont {Crivellin}\ \emph {et~al.}(2013)\citenamefont
  {Crivellin}, \citenamefont {Kokulu},\ and\ \citenamefont
  {Greub}}]{Crivellin:2013wna}%
  \BibitemOpen
  \bibfield  {author} {\bibinfo {author} {\bibfnamefont {A.}~\bibnamefont
  {Crivellin}}, \bibinfo {author} {\bibfnamefont {A.}~\bibnamefont {Kokulu}}, \
  and\ \bibinfo {author} {\bibfnamefont {C.}~\bibnamefont {Greub}},\ }\href
  {\doibase 10.1103/PhysRevD.87.094031} {\bibfield  {journal} {\bibinfo
  {journal} {Phys. Rev. D}\ }\textbf {\bibinfo {volume} {87}},\ \bibinfo
  {pages} {094031} (\bibinfo {year} {2013})},\ \Eprint
  {http://arxiv.org/abs/1303.5877} {arXiv:1303.5877 [hep-ph]} \BibitemShut
  {NoStop}%
\bibitem [{\citenamefont {Becirevic}\ \emph {et~al.}(2002)\citenamefont
  {Becirevic}, \citenamefont {Ciuchini}, \citenamefont {Franco}, \citenamefont
  {Gimenez}, \citenamefont {Martinelli}, \citenamefont {Masiero}, \citenamefont
  {Papinutto}, \citenamefont {Reyes},\ and\ \citenamefont
  {Silvestrini}}]{Becirevic:2001jj}%
  \BibitemOpen
  \bibfield  {author} {\bibinfo {author} {\bibfnamefont {D.}~\bibnamefont
  {Becirevic}}, \bibinfo {author} {\bibfnamefont {M.}~\bibnamefont {Ciuchini}},
  \bibinfo {author} {\bibfnamefont {E.}~\bibnamefont {Franco}}, \bibinfo
  {author} {\bibfnamefont {V.}~\bibnamefont {Gimenez}}, \bibinfo {author}
  {\bibfnamefont {G.}~\bibnamefont {Martinelli}}, \bibinfo {author}
  {\bibfnamefont {A.}~\bibnamefont {Masiero}}, \bibinfo {author} {\bibfnamefont
  {M.}~\bibnamefont {Papinutto}}, \bibinfo {author} {\bibfnamefont
  {J.}~\bibnamefont {Reyes}}, \ and\ \bibinfo {author} {\bibfnamefont
  {L.}~\bibnamefont {Silvestrini}},\ }\href {\doibase
  10.1016/S0550-3213(02)00291-2} {\bibfield  {journal} {\bibinfo  {journal}
  {Nucl. Phys. B}\ }\textbf {\bibinfo {volume} {634}},\ \bibinfo {pages} {105}
  (\bibinfo {year} {2002})},\ \Eprint {http://arxiv.org/abs/hep-ph/0112303}
  {arXiv:hep-ph/0112303} \BibitemShut {NoStop}%
\bibitem [{\citenamefont {Ciuchini}\ \emph {et~al.}(1998)\citenamefont
  {Ciuchini} \emph {et~al.}}]{Ciuchini:1998ix}%
  \BibitemOpen
  \bibfield  {author} {\bibinfo {author} {\bibfnamefont {M.}~\bibnamefont
  {Ciuchini}} \emph {et~al.},\ }\href {\doibase 10.1088/1126-6708/1998/10/008}
  {\bibfield  {journal} {\bibinfo  {journal} {JHEP}\ }\textbf {\bibinfo
  {volume} {10}},\ \bibinfo {pages} {008} (\bibinfo {year} {1998})},\ \Eprint
  {http://arxiv.org/abs/hep-ph/9808328} {arXiv:hep-ph/9808328} \BibitemShut
  {NoStop}%
\bibitem [{\citenamefont {Czakon}\ \emph {et~al.}(2015)\citenamefont {Czakon},
  \citenamefont {Fiedler}, \citenamefont {Huber}, \citenamefont {Misiak},
  \citenamefont {Schutzmeier},\ and\ \citenamefont
  {Steinhauser}}]{Czakon:2015exa}%
  \BibitemOpen
  \bibfield  {author} {\bibinfo {author} {\bibfnamefont {M.}~\bibnamefont
  {Czakon}}, \bibinfo {author} {\bibfnamefont {P.}~\bibnamefont {Fiedler}},
  \bibinfo {author} {\bibfnamefont {T.}~\bibnamefont {Huber}}, \bibinfo
  {author} {\bibfnamefont {M.}~\bibnamefont {Misiak}}, \bibinfo {author}
  {\bibfnamefont {T.}~\bibnamefont {Schutzmeier}}, \ and\ \bibinfo {author}
  {\bibfnamefont {M.}~\bibnamefont {Steinhauser}},\ }\href {\doibase
  10.1007/JHEP04(2015)168} {\bibfield  {journal} {\bibinfo  {journal} {JHEP}\
  }\textbf {\bibinfo {volume} {04}},\ \bibinfo {pages} {168} (\bibinfo {year}
  {2015})},\ \Eprint {http://arxiv.org/abs/1503.01791} {arXiv:1503.01791
  [hep-ph]} \BibitemShut {NoStop}%
\bibitem [{\citenamefont {Crivellin}\ \emph {et~al.}(2019)\citenamefont
  {Crivellin}, \citenamefont {M\"uller},\ and\ \citenamefont
  {Wiegand}}]{Crivellin:2019dun}%
  \BibitemOpen
  \bibfield  {author} {\bibinfo {author} {\bibfnamefont {A.}~\bibnamefont
  {Crivellin}}, \bibinfo {author} {\bibfnamefont {D.}~\bibnamefont {M\"uller}},
  \ and\ \bibinfo {author} {\bibfnamefont {C.}~\bibnamefont {Wiegand}},\ }\href
  {\doibase 10.1007/JHEP06(2019)119} {\bibfield  {journal} {\bibinfo  {journal}
  {JHEP}\ }\textbf {\bibinfo {volume} {06}},\ \bibinfo {pages} {119} (\bibinfo
  {year} {2019})},\ \Eprint {http://arxiv.org/abs/1903.10440} {arXiv:1903.10440
  [hep-ph]} \BibitemShut {NoStop}%
\bibitem [{\citenamefont {Beneke}\ \emph {et~al.}(2019)\citenamefont {Beneke},
  \citenamefont {Bobeth},\ and\ \citenamefont {Szafron}}]{Beneke:2019slt}%
  \BibitemOpen
  \bibfield  {author} {\bibinfo {author} {\bibfnamefont {M.}~\bibnamefont
  {Beneke}}, \bibinfo {author} {\bibfnamefont {C.}~\bibnamefont {Bobeth}}, \
  and\ \bibinfo {author} {\bibfnamefont {R.}~\bibnamefont {Szafron}},\ }\href
  {\doibase 10.1007/JHEP10(2019)232} {\bibfield  {journal} {\bibinfo  {journal}
  {JHEP}\ }\textbf {\bibinfo {volume} {10}},\ \bibinfo {pages} {232} (\bibinfo
  {year} {2019})},\ \bibinfo {note} {[Erratum: JHEP 11, 099 (2022)]},\ \Eprint
  {http://arxiv.org/abs/1908.07011} {arXiv:1908.07011 [hep-ph]} \BibitemShut
  {NoStop}%
\bibitem [{\citenamefont {Bobeth}\ \emph {et~al.}(2014)\citenamefont {Bobeth},
  \citenamefont {Gorbahn}, \citenamefont {Hermann}, \citenamefont {Misiak},
  \citenamefont {Stamou},\ and\ \citenamefont {Steinhauser}}]{Bobeth:2013uxa}%
  \BibitemOpen
  \bibfield  {author} {\bibinfo {author} {\bibfnamefont {C.}~\bibnamefont
  {Bobeth}}, \bibinfo {author} {\bibfnamefont {M.}~\bibnamefont {Gorbahn}},
  \bibinfo {author} {\bibfnamefont {T.}~\bibnamefont {Hermann}}, \bibinfo
  {author} {\bibfnamefont {M.}~\bibnamefont {Misiak}}, \bibinfo {author}
  {\bibfnamefont {E.}~\bibnamefont {Stamou}}, \ and\ \bibinfo {author}
  {\bibfnamefont {M.}~\bibnamefont {Steinhauser}},\ }\href {\doibase
  10.1103/PhysRevLett.112.101801} {\bibfield  {journal} {\bibinfo  {journal}
  {Phys. Rev. Lett.}\ }\textbf {\bibinfo {volume} {112}},\ \bibinfo {pages}
  {101801} (\bibinfo {year} {2014})},\ \Eprint {http://arxiv.org/abs/1311.0903}
  {arXiv:1311.0903 [hep-ph]} \BibitemShut {NoStop}%
\bibitem [{\citenamefont {Altunkaynak}\ \emph {et~al.}(2015)\citenamefont
  {Altunkaynak}, \citenamefont {Hou}, \citenamefont {Kao}, \citenamefont
  {Kohda},\ and\ \citenamefont {McCoy}}]{Altunkaynak:2015twa}%
  \BibitemOpen
  \bibfield  {author} {\bibinfo {author} {\bibfnamefont {B.}~\bibnamefont
  {Altunkaynak}}, \bibinfo {author} {\bibfnamefont {W.-S.}\ \bibnamefont
  {Hou}}, \bibinfo {author} {\bibfnamefont {C.}~\bibnamefont {Kao}}, \bibinfo
  {author} {\bibfnamefont {M.}~\bibnamefont {Kohda}}, \ and\ \bibinfo {author}
  {\bibfnamefont {B.}~\bibnamefont {McCoy}},\ }\href {\doibase
  10.1016/j.physletb.2015.10.024} {\bibfield  {journal} {\bibinfo  {journal}
  {Phys. Lett. B}\ }\textbf {\bibinfo {volume} {751}},\ \bibinfo {pages} {135}
  (\bibinfo {year} {2015})},\ \Eprint {http://arxiv.org/abs/1506.00651}
  {arXiv:1506.00651 [hep-ph]} \BibitemShut {NoStop}%
\bibitem [{\citenamefont {Hou}\ \emph {et~al.}(2021)\citenamefont {Hou},
  \citenamefont {Modak},\ and\ \citenamefont {Plehn}}]{Hou:2020tnc}%
  \BibitemOpen
  \bibfield  {author} {\bibinfo {author} {\bibfnamefont {W.-S.}\ \bibnamefont
  {Hou}}, \bibinfo {author} {\bibfnamefont {T.}~\bibnamefont {Modak}}, \ and\
  \bibinfo {author} {\bibfnamefont {T.}~\bibnamefont {Plehn}},\ }\href
  {\doibase 10.21468/SciPostPhys.10.6.150} {\bibfield  {journal} {\bibinfo
  {journal} {SciPost Phys.}\ }\textbf {\bibinfo {volume} {10}},\ \bibinfo
  {pages} {150} (\bibinfo {year} {2021})},\ \Eprint
  {http://arxiv.org/abs/2012.03572} {arXiv:2012.03572 [hep-ph]} \BibitemShut
  {NoStop}%
\bibitem [{\citenamefont {Kohda}\ \emph {et~al.}(2018)\citenamefont {Kohda},
  \citenamefont {Modak},\ and\ \citenamefont {Hou}}]{Kohda:2017fkn}%
  \BibitemOpen
  \bibfield  {author} {\bibinfo {author} {\bibfnamefont {M.}~\bibnamefont
  {Kohda}}, \bibinfo {author} {\bibfnamefont {T.}~\bibnamefont {Modak}}, \ and\
  \bibinfo {author} {\bibfnamefont {W.-S.}\ \bibnamefont {Hou}},\ }\href
  {\doibase 10.1016/j.physletb.2017.11.056} {\bibfield  {journal} {\bibinfo
  {journal} {Phys. Lett. B}\ }\textbf {\bibinfo {volume} {776}},\ \bibinfo
  {pages} {379} (\bibinfo {year} {2018})},\ \Eprint
  {http://arxiv.org/abs/1710.07260} {arXiv:1710.07260 [hep-ph]} \BibitemShut
  {NoStop}%
\bibitem [{\citenamefont {Iguro}\ and\ \citenamefont
  {Omura}(2018)}]{Iguro:2018qzf}%
  \BibitemOpen
  \bibfield  {author} {\bibinfo {author} {\bibfnamefont {S.}~\bibnamefont
  {Iguro}}\ and\ \bibinfo {author} {\bibfnamefont {Y.}~\bibnamefont {Omura}},\
  }\href {\doibase 10.1007/JHEP05(2018)173} {\bibfield  {journal} {\bibinfo
  {journal} {JHEP}\ }\textbf {\bibinfo {volume} {05}},\ \bibinfo {pages} {173}
  (\bibinfo {year} {2018})},\ \Eprint {http://arxiv.org/abs/1802.01732}
  {arXiv:1802.01732 [hep-ph]} \BibitemShut {NoStop}%
\bibitem [{\citenamefont {Hou}\ \emph {et~al.}(2019)\citenamefont {Hou},
  \citenamefont {Kohda},\ and\ \citenamefont {Modak}}]{Hou:2019gpn}%
  \BibitemOpen
  \bibfield  {author} {\bibinfo {author} {\bibfnamefont {W.-S.}\ \bibnamefont
  {Hou}}, \bibinfo {author} {\bibfnamefont {M.}~\bibnamefont {Kohda}}, \ and\
  \bibinfo {author} {\bibfnamefont {T.}~\bibnamefont {Modak}},\ }\href
  {\doibase 10.1016/j.physletb.2019.134953} {\bibfield  {journal} {\bibinfo
  {journal} {Phys. Lett. B}\ }\textbf {\bibinfo {volume} {798}},\ \bibinfo
  {pages} {134953} (\bibinfo {year} {2019})},\ \Eprint
  {http://arxiv.org/abs/1906.09703} {arXiv:1906.09703 [hep-ph]} \BibitemShut
  {NoStop}%
\bibitem [{\citenamefont {Hou}\ \emph {et~al.}(2018)\citenamefont {Hou},
  \citenamefont {Kohda},\ and\ \citenamefont {Modak}}]{Hou:2018zmg}%
  \BibitemOpen
  \bibfield  {author} {\bibinfo {author} {\bibfnamefont {W.-S.}\ \bibnamefont
  {Hou}}, \bibinfo {author} {\bibfnamefont {M.}~\bibnamefont {Kohda}}, \ and\
  \bibinfo {author} {\bibfnamefont {T.}~\bibnamefont {Modak}},\ }\href
  {\doibase 10.1016/j.physletb.2018.09.046} {\bibfield  {journal} {\bibinfo
  {journal} {Phys. Lett. B}\ }\textbf {\bibinfo {volume} {786}},\ \bibinfo
  {pages} {212} (\bibinfo {year} {2018})},\ \Eprint
  {http://arxiv.org/abs/1808.00333} {arXiv:1808.00333 [hep-ph]} \BibitemShut
  {NoStop}%
\bibitem [{\citenamefont {Peskin}\ and\ \citenamefont
  {Takeuchi}(1990)}]{Peskin:1990zt}%
  \BibitemOpen
  \bibfield  {author} {\bibinfo {author} {\bibfnamefont {M.~E.}\ \bibnamefont
  {Peskin}}\ and\ \bibinfo {author} {\bibfnamefont {T.}~\bibnamefont
  {Takeuchi}},\ }\href {\doibase 10.1103/PhysRevLett.65.964} {\bibfield
  {journal} {\bibinfo  {journal} {Phys. Rev. Lett.}\ }\textbf {\bibinfo
  {volume} {65}},\ \bibinfo {pages} {964} (\bibinfo {year} {1990})}\BibitemShut
  {NoStop}%
\bibitem [{\citenamefont {Peskin}\ and\ \citenamefont
  {Takeuchi}(1992)}]{Peskin:1991sw}%
  \BibitemOpen
  \bibfield  {author} {\bibinfo {author} {\bibfnamefont {M.~E.}\ \bibnamefont
  {Peskin}}\ and\ \bibinfo {author} {\bibfnamefont {T.}~\bibnamefont
  {Takeuchi}},\ }\href {\doibase 10.1103/PhysRevD.46.381} {\bibfield  {journal}
  {\bibinfo  {journal} {Phys. Rev. D}\ }\textbf {\bibinfo {volume} {46}},\
  \bibinfo {pages} {381} (\bibinfo {year} {1992})}\BibitemShut {NoStop}%
\bibitem [{\citenamefont {Baak}\ \emph {et~al.}(2012)\citenamefont {Baak},
  \citenamefont {Goebel}, \citenamefont {Haller}, \citenamefont {Hoecker},
  \citenamefont {Kennedy}, \citenamefont {Kogler}, \citenamefont {Moenig},
  \citenamefont {Schott},\ and\ \citenamefont {Stelzer}}]{Baak:2012kk}%
  \BibitemOpen
  \bibfield  {author} {\bibinfo {author} {\bibfnamefont {M.}~\bibnamefont
  {Baak}}, \bibinfo {author} {\bibfnamefont {M.}~\bibnamefont {Goebel}},
  \bibinfo {author} {\bibfnamefont {J.}~\bibnamefont {Haller}}, \bibinfo
  {author} {\bibfnamefont {A.}~\bibnamefont {Hoecker}}, \bibinfo {author}
  {\bibfnamefont {D.}~\bibnamefont {Kennedy}}, \bibinfo {author} {\bibfnamefont
  {R.}~\bibnamefont {Kogler}}, \bibinfo {author} {\bibfnamefont
  {K.}~\bibnamefont {Moenig}}, \bibinfo {author} {\bibfnamefont
  {M.}~\bibnamefont {Schott}}, \ and\ \bibinfo {author} {\bibfnamefont
  {J.}~\bibnamefont {Stelzer}},\ }\href {\doibase
  10.1140/epjc/s10052-012-2205-9} {\bibfield  {journal} {\bibinfo  {journal}
  {Eur. Phys. J. C}\ }\textbf {\bibinfo {volume} {72}},\ \bibinfo {pages}
  {2205} (\bibinfo {year} {2012})},\ \Eprint {http://arxiv.org/abs/1209.2716}
  {arXiv:1209.2716 [hep-ph]} \BibitemShut {NoStop}%
\bibitem [{\citenamefont {Sikivie}\ \emph {et~al.}(1980)\citenamefont
  {Sikivie}, \citenamefont {Susskind}, \citenamefont {Voloshin},\ and\
  \citenamefont {Zakharov}}]{Sikivie:1980hm}%
  \BibitemOpen
  \bibfield  {author} {\bibinfo {author} {\bibfnamefont {P.}~\bibnamefont
  {Sikivie}}, \bibinfo {author} {\bibfnamefont {L.}~\bibnamefont {Susskind}},
  \bibinfo {author} {\bibfnamefont {M.~B.}\ \bibnamefont {Voloshin}}, \ and\
  \bibinfo {author} {\bibfnamefont {V.~I.}\ \bibnamefont {Zakharov}},\ }\href
  {\doibase 10.1016/0550-3213(80)90214-X} {\bibfield  {journal} {\bibinfo
  {journal} {Nucl. Phys. B}\ }\textbf {\bibinfo {volume} {173}},\ \bibinfo
  {pages} {189} (\bibinfo {year} {1980})}\BibitemShut {NoStop}%
\bibitem [{\citenamefont {Haber}\ and\ \citenamefont
  {Pomarol}(1993)}]{Haber:1992py}%
  \BibitemOpen
  \bibfield  {author} {\bibinfo {author} {\bibfnamefont {H.~E.}\ \bibnamefont
  {Haber}}\ and\ \bibinfo {author} {\bibfnamefont {A.}~\bibnamefont
  {Pomarol}},\ }\href {\doibase 10.1016/0370-2693(93)90423-F} {\bibfield
  {journal} {\bibinfo  {journal} {Phys. Lett. B}\ }\textbf {\bibinfo {volume}
  {302}},\ \bibinfo {pages} {435} (\bibinfo {year} {1993})},\ \Eprint
  {http://arxiv.org/abs/hep-ph/9207267} {arXiv:hep-ph/9207267} \BibitemShut
  {NoStop}%
\bibitem [{\citenamefont {Pomarol}\ and\ \citenamefont
  {Vega}(1994)}]{Pomarol:1993mu}%
  \BibitemOpen
  \bibfield  {author} {\bibinfo {author} {\bibfnamefont {A.}~\bibnamefont
  {Pomarol}}\ and\ \bibinfo {author} {\bibfnamefont {R.}~\bibnamefont {Vega}},\
  }\href {\doibase 10.1016/0550-3213(94)90611-4} {\bibfield  {journal}
  {\bibinfo  {journal} {Nucl. Phys. B}\ }\textbf {\bibinfo {volume} {413}},\
  \bibinfo {pages} {3} (\bibinfo {year} {1994})},\ \Eprint
  {http://arxiv.org/abs/hep-ph/9305272} {arXiv:hep-ph/9305272} \BibitemShut
  {NoStop}%
\bibitem [{\citenamefont {Gerard}\ and\ \citenamefont
  {Herquet}(2007)}]{Gerard:2007kn}%
  \BibitemOpen
  \bibfield  {author} {\bibinfo {author} {\bibfnamefont {J.~M.}\ \bibnamefont
  {Gerard}}\ and\ \bibinfo {author} {\bibfnamefont {M.}~\bibnamefont
  {Herquet}},\ }\href {\doibase 10.1103/PhysRevLett.98.251802} {\bibfield
  {journal} {\bibinfo  {journal} {Phys. Rev. Lett.}\ }\textbf {\bibinfo
  {volume} {98}},\ \bibinfo {pages} {251802} (\bibinfo {year} {2007})},\
  \Eprint {http://arxiv.org/abs/hep-ph/0703051} {arXiv:hep-ph/0703051}
  \BibitemShut {NoStop}%
\bibitem [{\citenamefont {Haber}\ and\ \citenamefont
  {O'Neil}(2011)}]{Haber:2010bw}%
  \BibitemOpen
  \bibfield  {author} {\bibinfo {author} {\bibfnamefont {H.~E.}\ \bibnamefont
  {Haber}}\ and\ \bibinfo {author} {\bibfnamefont {D.}~\bibnamefont {O'Neil}},\
  }\href {\doibase 10.1103/PhysRevD.83.055017} {\bibfield  {journal} {\bibinfo
  {journal} {Phys. Rev. D}\ }\textbf {\bibinfo {volume} {83}},\ \bibinfo
  {pages} {055017} (\bibinfo {year} {2011})},\ \Eprint
  {http://arxiv.org/abs/1011.6188} {arXiv:1011.6188 [hep-ph]} \BibitemShut
  {NoStop}%
\bibitem [{\citenamefont {Grzadkowski}\ \emph {et~al.}(2011)\citenamefont
  {Grzadkowski}, \citenamefont {Maniatis},\ and\ \citenamefont
  {Wudka}}]{Grzadkowski:2010dj}%
  \BibitemOpen
  \bibfield  {author} {\bibinfo {author} {\bibfnamefont {B.}~\bibnamefont
  {Grzadkowski}}, \bibinfo {author} {\bibfnamefont {M.}~\bibnamefont
  {Maniatis}}, \ and\ \bibinfo {author} {\bibfnamefont {J.}~\bibnamefont
  {Wudka}},\ }\href {\doibase 10.1007/JHEP11(2011)030} {\bibfield  {journal}
  {\bibinfo  {journal} {JHEP}\ }\textbf {\bibinfo {volume} {11}},\ \bibinfo
  {pages} {030} (\bibinfo {year} {2011})},\ \Eprint
  {http://arxiv.org/abs/1011.5228} {arXiv:1011.5228 [hep-ph]} \BibitemShut
  {NoStop}%
\bibitem [{\citenamefont {Aiko}\ and\ \citenamefont
  {Kanemura}(2021)}]{Aiko:2020atr}%
  \BibitemOpen
  \bibfield  {author} {\bibinfo {author} {\bibfnamefont {M.}~\bibnamefont
  {Aiko}}\ and\ \bibinfo {author} {\bibfnamefont {S.}~\bibnamefont
  {Kanemura}},\ }\href {\doibase 10.1007/JHEP02(2021)046} {\bibfield  {journal}
  {\bibinfo  {journal} {JHEP}\ }\textbf {\bibinfo {volume} {02}},\ \bibinfo
  {pages} {046} (\bibinfo {year} {2021})},\ \Eprint
  {http://arxiv.org/abs/2009.04330} {arXiv:2009.04330 [hep-ph]} \BibitemShut
  {NoStop}%
\bibitem [{\citenamefont {Demir}\ \emph {et~al.}(2003)\citenamefont {Demir},
  \citenamefont {Pospelov},\ and\ \citenamefont {Ritz}}]{Demir:2002gg}%
  \BibitemOpen
  \bibfield  {author} {\bibinfo {author} {\bibfnamefont {D.~A.}\ \bibnamefont
  {Demir}}, \bibinfo {author} {\bibfnamefont {M.}~\bibnamefont {Pospelov}}, \
  and\ \bibinfo {author} {\bibfnamefont {A.}~\bibnamefont {Ritz}},\ }\href
  {\doibase 10.1103/PhysRevD.67.015007} {\bibfield  {journal} {\bibinfo
  {journal} {Phys. Rev. D}\ }\textbf {\bibinfo {volume} {67}},\ \bibinfo
  {pages} {015007} (\bibinfo {year} {2003})},\ \Eprint
  {http://arxiv.org/abs/hep-ph/0208257} {arXiv:hep-ph/0208257} \BibitemShut
  {NoStop}%
\bibitem [{\citenamefont {Jung}\ and\ \citenamefont
  {Pich}(2014)}]{Jung:2013hka}%
  \BibitemOpen
  \bibfield  {author} {\bibinfo {author} {\bibfnamefont {M.}~\bibnamefont
  {Jung}}\ and\ \bibinfo {author} {\bibfnamefont {A.}~\bibnamefont {Pich}},\
  }\href {\doibase 10.1007/JHEP04(2014)076} {\bibfield  {journal} {\bibinfo
  {journal} {JHEP}\ }\textbf {\bibinfo {volume} {04}},\ \bibinfo {pages} {076}
  (\bibinfo {year} {2014})},\ \Eprint {http://arxiv.org/abs/1308.6283}
  {arXiv:1308.6283 [hep-ph]} \BibitemShut {NoStop}%
\bibitem [{\citenamefont {Kanemura}\ \emph {et~al.}(1993)\citenamefont
  {Kanemura}, \citenamefont {Kubota},\ and\ \citenamefont
  {Takasugi}}]{Kanemura:1993hm}%
  \BibitemOpen
  \bibfield  {author} {\bibinfo {author} {\bibfnamefont {S.}~\bibnamefont
  {Kanemura}}, \bibinfo {author} {\bibfnamefont {T.}~\bibnamefont {Kubota}}, \
  and\ \bibinfo {author} {\bibfnamefont {E.}~\bibnamefont {Takasugi}},\ }\href
  {\doibase 10.1016/0370-2693(93)91205-2} {\bibfield  {journal} {\bibinfo
  {journal} {Phys. Lett. B}\ }\textbf {\bibinfo {volume} {313}},\ \bibinfo
  {pages} {155} (\bibinfo {year} {1993})},\ \Eprint
  {http://arxiv.org/abs/hep-ph/9303263} {arXiv:hep-ph/9303263} \BibitemShut
  {NoStop}%
\bibitem [{\citenamefont {Akeroyd}\ \emph {et~al.}(2000)\citenamefont
  {Akeroyd}, \citenamefont {Arhrib},\ and\ \citenamefont
  {Naimi}}]{Akeroyd:2000wc}%
  \BibitemOpen
  \bibfield  {author} {\bibinfo {author} {\bibfnamefont {A.~G.}\ \bibnamefont
  {Akeroyd}}, \bibinfo {author} {\bibfnamefont {A.}~\bibnamefont {Arhrib}}, \
  and\ \bibinfo {author} {\bibfnamefont {E.-M.}\ \bibnamefont {Naimi}},\ }\href
  {\doibase 10.1016/S0370-2693(00)00962-X} {\bibfield  {journal} {\bibinfo
  {journal} {Phys. Lett. B}\ }\textbf {\bibinfo {volume} {490}},\ \bibinfo
  {pages} {119} (\bibinfo {year} {2000})},\ \Eprint
  {http://arxiv.org/abs/hep-ph/0006035} {arXiv:hep-ph/0006035} \BibitemShut
  {NoStop}%
\bibitem [{\citenamefont {Ginzburg}\ and\ \citenamefont
  {Ivanov}(2005)}]{Ginzburg:2005dt}%
  \BibitemOpen
  \bibfield  {author} {\bibinfo {author} {\bibfnamefont {I.~F.}\ \bibnamefont
  {Ginzburg}}\ and\ \bibinfo {author} {\bibfnamefont {I.~P.}\ \bibnamefont
  {Ivanov}},\ }\href {\doibase 10.1103/PhysRevD.72.115010} {\bibfield
  {journal} {\bibinfo  {journal} {Phys. Rev. D}\ }\textbf {\bibinfo {volume}
  {72}},\ \bibinfo {pages} {115010} (\bibinfo {year} {2005})},\ \Eprint
  {http://arxiv.org/abs/hep-ph/0508020} {arXiv:hep-ph/0508020} \BibitemShut
  {NoStop}%
\bibitem [{\citenamefont {Kanemura}\ and\ \citenamefont
  {Yagyu}(2015)}]{Kanemura:2015ska}%
  \BibitemOpen
  \bibfield  {author} {\bibinfo {author} {\bibfnamefont {S.}~\bibnamefont
  {Kanemura}}\ and\ \bibinfo {author} {\bibfnamefont {K.}~\bibnamefont
  {Yagyu}},\ }\href {\doibase 10.1016/j.physletb.2015.10.047} {\bibfield
  {journal} {\bibinfo  {journal} {Phys. Lett. B}\ }\textbf {\bibinfo {volume}
  {751}},\ \bibinfo {pages} {289} (\bibinfo {year} {2015})},\ \Eprint
  {http://arxiv.org/abs/1509.06060} {arXiv:1509.06060 [hep-ph]} \BibitemShut
  {NoStop}%
\bibitem [{\citenamefont {Klimenko}(1985)}]{Klimenko:1984qx}%
  \BibitemOpen
  \bibfield  {author} {\bibinfo {author} {\bibfnamefont {K.~G.}\ \bibnamefont
  {Klimenko}},\ }\href {\doibase 10.1007/BF01034825} {\bibfield  {journal}
  {\bibinfo  {journal} {Theor. Math. Phys.}\ }\textbf {\bibinfo {volume}
  {62}},\ \bibinfo {pages} {58} (\bibinfo {year} {1985})}\BibitemShut {NoStop}%
\bibitem [{\citenamefont {Sher}(1989)}]{Sher:1988mj}%
  \BibitemOpen
  \bibfield  {author} {\bibinfo {author} {\bibfnamefont {M.}~\bibnamefont
  {Sher}},\ }\href {\doibase 10.1016/0370-1573(89)90061-6} {\bibfield
  {journal} {\bibinfo  {journal} {Phys. Rept.}\ }\textbf {\bibinfo {volume}
  {179}},\ \bibinfo {pages} {273} (\bibinfo {year} {1989})}\BibitemShut
  {NoStop}%
\bibitem [{\citenamefont {Nie}\ and\ \citenamefont {Sher}(1999)}]{Nie:1998yn}%
  \BibitemOpen
  \bibfield  {author} {\bibinfo {author} {\bibfnamefont {S.}~\bibnamefont
  {Nie}}\ and\ \bibinfo {author} {\bibfnamefont {M.}~\bibnamefont {Sher}},\
  }\href {\doibase 10.1016/S0370-2693(99)00019-2} {\bibfield  {journal}
  {\bibinfo  {journal} {Phys. Lett. B}\ }\textbf {\bibinfo {volume} {449}},\
  \bibinfo {pages} {89} (\bibinfo {year} {1999})},\ \Eprint
  {http://arxiv.org/abs/hep-ph/9811234} {arXiv:hep-ph/9811234} \BibitemShut
  {NoStop}%
\bibitem [{\citenamefont {Ferreira}\ \emph {et~al.}(2004)\citenamefont
  {Ferreira}, \citenamefont {Santos},\ and\ \citenamefont
  {Barroso}}]{Ferreira:2004yd}%
  \BibitemOpen
  \bibfield  {author} {\bibinfo {author} {\bibfnamefont {P.~M.}\ \bibnamefont
  {Ferreira}}, \bibinfo {author} {\bibfnamefont {R.}~\bibnamefont {Santos}}, \
  and\ \bibinfo {author} {\bibfnamefont {A.}~\bibnamefont {Barroso}},\ }\href
  {\doibase 10.1016/j.physletb.2004.10.022} {\bibfield  {journal} {\bibinfo
  {journal} {Phys. Lett. B}\ }\textbf {\bibinfo {volume} {603}},\ \bibinfo
  {pages} {219} (\bibinfo {year} {2004})},\ \bibinfo {note} {[Erratum:
  Phys.Lett.B 629, 114--114 (2005)]},\ \Eprint
  {http://arxiv.org/abs/hep-ph/0406231} {arXiv:hep-ph/0406231} \BibitemShut
  {NoStop}%
\bibitem [{\citenamefont {Flores}\ and\ \citenamefont
  {Sher}(1983)}]{Flores:1982pr}%
  \BibitemOpen
  \bibfield  {author} {\bibinfo {author} {\bibfnamefont {R.~A.}\ \bibnamefont
  {Flores}}\ and\ \bibinfo {author} {\bibfnamefont {M.}~\bibnamefont {Sher}},\
  }\href {\doibase 10.1016/0003-4916(83)90331-7} {\bibfield  {journal}
  {\bibinfo  {journal} {Annals Phys.}\ }\textbf {\bibinfo {volume} {148}},\
  \bibinfo {pages} {95} (\bibinfo {year} {1983})}\BibitemShut {NoStop}%
\bibitem [{\citenamefont {Kominis}\ and\ \citenamefont
  {Chivukula}(1993)}]{Kominis:1993zc}%
  \BibitemOpen
  \bibfield  {author} {\bibinfo {author} {\bibfnamefont {D.}~\bibnamefont
  {Kominis}}\ and\ \bibinfo {author} {\bibfnamefont {R.~S.}\ \bibnamefont
  {Chivukula}},\ }\href {\doibase 10.1016/0370-2693(93)91415-J} {\bibfield
  {journal} {\bibinfo  {journal} {Phys. Lett. B}\ }\textbf {\bibinfo {volume}
  {304}},\ \bibinfo {pages} {152} (\bibinfo {year} {1993})},\ \Eprint
  {http://arxiv.org/abs/hep-ph/9301222} {arXiv:hep-ph/9301222} \BibitemShut
  {NoStop}%
\bibitem [{\citenamefont {Kanemura}\ \emph {et~al.}(1999)\citenamefont
  {Kanemura}, \citenamefont {Kasai},\ and\ \citenamefont
  {Okada}}]{Kanemura:1999xf}%
  \BibitemOpen
  \bibfield  {author} {\bibinfo {author} {\bibfnamefont {S.}~\bibnamefont
  {Kanemura}}, \bibinfo {author} {\bibfnamefont {T.}~\bibnamefont {Kasai}}, \
  and\ \bibinfo {author} {\bibfnamefont {Y.}~\bibnamefont {Okada}},\ }\href
  {\doibase 10.1016/S0370-2693(99)01351-9} {\bibfield  {journal} {\bibinfo
  {journal} {Phys. Lett. B}\ }\textbf {\bibinfo {volume} {471}},\ \bibinfo
  {pages} {182} (\bibinfo {year} {1999})},\ \Eprint
  {http://arxiv.org/abs/hep-ph/9903289} {arXiv:hep-ph/9903289} \BibitemShut
  {NoStop}%
\bibitem [{\citenamefont {Ferreira}\ and\ \citenamefont
  {Jones}(2009)}]{Ferreira:2009jb}%
  \BibitemOpen
  \bibfield  {author} {\bibinfo {author} {\bibfnamefont {P.~M.}\ \bibnamefont
  {Ferreira}}\ and\ \bibinfo {author} {\bibfnamefont {D.~R.~T.}\ \bibnamefont
  {Jones}},\ }\href {\doibase 10.1088/1126-6708/2009/08/069} {\bibfield
  {journal} {\bibinfo  {journal} {JHEP}\ }\textbf {\bibinfo {volume} {08}},\
  \bibinfo {pages} {069} (\bibinfo {year} {2009})},\ \Eprint
  {http://arxiv.org/abs/0903.2856} {arXiv:0903.2856 [hep-ph]} \BibitemShut
  {NoStop}%
\bibitem [{\citenamefont {Herren}\ and\ \citenamefont
  {Steinhauser}(2018)}]{Herren:2017osy}%
  \BibitemOpen
  \bibfield  {author} {\bibinfo {author} {\bibfnamefont {F.}~\bibnamefont
  {Herren}}\ and\ \bibinfo {author} {\bibfnamefont {M.}~\bibnamefont
  {Steinhauser}},\ }\href {\doibase 10.1016/j.cpc.2017.11.014} {\bibfield
  {journal} {\bibinfo  {journal} {Comput. Phys. Commun.}\ }\textbf {\bibinfo
  {volume} {224}},\ \bibinfo {pages} {333} (\bibinfo {year} {2018})},\ \Eprint
  {http://arxiv.org/abs/1703.03751} {arXiv:1703.03751 [hep-ph]} \BibitemShut
  {NoStop}%
\bibitem [{\citenamefont {Wainwright}(2012)}]{Wainwright:2011kj}%
  \BibitemOpen
  \bibfield  {author} {\bibinfo {author} {\bibfnamefont {C.~L.}\ \bibnamefont
  {Wainwright}},\ }\href {\doibase 10.1016/j.cpc.2012.04.004} {\bibfield
  {journal} {\bibinfo  {journal} {Comput. Phys. Commun.}\ }\textbf {\bibinfo
  {volume} {183}},\ \bibinfo {pages} {2006} (\bibinfo {year} {2012})},\ \Eprint
  {http://arxiv.org/abs/1109.4189} {arXiv:1109.4189 [hep-ph]} \BibitemShut
  {NoStop}%
\bibitem [{\citenamefont {Kainulainen}(2021)}]{Kainulainen:2021oqs}%
  \BibitemOpen
  \bibfield  {author} {\bibinfo {author} {\bibfnamefont {K.}~\bibnamefont
  {Kainulainen}},\ }\href {\doibase 10.1088/1475-7516/2021/11/042} {\bibfield
  {journal} {\bibinfo  {journal} {JCAP}\ }\textbf {\bibinfo {volume} {11}},\
  \bibinfo {pages} {042} (\bibinfo {year} {2021})},\ \Eprint
  {http://arxiv.org/abs/2108.08336} {arXiv:2108.08336 [hep-ph]} \BibitemShut
  {NoStop}%
\bibitem [{\citenamefont {Postma}\ \emph {et~al.}(2022)\citenamefont {Postma},
  \citenamefont {van~de Vis},\ and\ \citenamefont {White}}]{Postma:2022dbr}%
  \BibitemOpen
  \bibfield  {author} {\bibinfo {author} {\bibfnamefont {M.}~\bibnamefont
  {Postma}}, \bibinfo {author} {\bibfnamefont {J.}~\bibnamefont {van~de Vis}},
  \ and\ \bibinfo {author} {\bibfnamefont {G.}~\bibnamefont {White}},\ }\href
  {\doibase 10.1007/JHEP12(2022)121} {\bibfield  {journal} {\bibinfo  {journal}
  {JHEP}\ }\textbf {\bibinfo {volume} {12}},\ \bibinfo {pages} {121} (\bibinfo
  {year} {2022})},\ \Eprint {http://arxiv.org/abs/2206.01120} {arXiv:2206.01120
  [hep-ph]} \BibitemShut {NoStop}%
\bibitem [{\citenamefont {Grossman}\ and\ \citenamefont
  {Nir}(1997)}]{Grossman:1997sk}%
  \BibitemOpen
  \bibfield  {author} {\bibinfo {author} {\bibfnamefont {Y.}~\bibnamefont
  {Grossman}}\ and\ \bibinfo {author} {\bibfnamefont {Y.}~\bibnamefont {Nir}},\
  }\href {\doibase 10.1016/S0370-2693(97)00210-4} {\bibfield  {journal}
  {\bibinfo  {journal} {Phys. Lett. B}\ }\textbf {\bibinfo {volume} {398}},\
  \bibinfo {pages} {163} (\bibinfo {year} {1997})},\ \Eprint
  {http://arxiv.org/abs/hep-ph/9701313} {arXiv:hep-ph/9701313} \BibitemShut
  {NoStop}%
\bibitem [{\citenamefont {Moulson}(2020)}]{Moulson:2019ifj}%
  \BibitemOpen
  \bibfield  {author} {\bibinfo {author} {\bibfnamefont {M.}~\bibnamefont
  {Moulson}} (\bibinfo {collaboration} {KLEVER Project}),\ }\href {\doibase
  10.1088/1742-6596/1526/1/012028} {\bibfield  {journal} {\bibinfo  {journal}
  {J. Phys. Conf. Ser.}\ }\textbf {\bibinfo {volume} {1526}},\ \bibinfo {pages}
  {012028} (\bibinfo {year} {2020})},\ \Eprint
  {http://arxiv.org/abs/1912.10037} {arXiv:1912.10037 [hep-ex]} \BibitemShut
  {NoStop}%
\bibitem [{\citenamefont {Buras}\ \emph {et~al.}(2015)\citenamefont {Buras},
  \citenamefont {Buttazzo}, \citenamefont {Girrbach-Noe},\ and\ \citenamefont
  {Knegjens}}]{Buras:2015qea}%
  \BibitemOpen
  \bibfield  {author} {\bibinfo {author} {\bibfnamefont {A.~J.}\ \bibnamefont
  {Buras}}, \bibinfo {author} {\bibfnamefont {D.}~\bibnamefont {Buttazzo}},
  \bibinfo {author} {\bibfnamefont {J.}~\bibnamefont {Girrbach-Noe}}, \ and\
  \bibinfo {author} {\bibfnamefont {R.}~\bibnamefont {Knegjens}},\ }\href
  {\doibase 10.1007/JHEP11(2015)033} {\bibfield  {journal} {\bibinfo  {journal}
  {JHEP}\ }\textbf {\bibinfo {volume} {11}},\ \bibinfo {pages} {033} (\bibinfo
  {year} {2015})},\ \Eprint {http://arxiv.org/abs/1503.02693} {arXiv:1503.02693
  [hep-ph]} \BibitemShut {NoStop}%
\bibitem [{\citenamefont {Batley}\ \emph {et~al.}(2002)\citenamefont {Batley}
  \emph {et~al.}}]{NA48:2002tmj}%
  \BibitemOpen
  \bibfield  {author} {\bibinfo {author} {\bibfnamefont {J.~R.}\ \bibnamefont
  {Batley}} \emph {et~al.} (\bibinfo {collaboration} {NA48}),\ }\href {\doibase
  10.1016/S0370-2693(02)02476-0} {\bibfield  {journal} {\bibinfo  {journal}
  {Phys. Lett. B}\ }\textbf {\bibinfo {volume} {544}},\ \bibinfo {pages} {97}
  (\bibinfo {year} {2002})},\ \Eprint {http://arxiv.org/abs/hep-ex/0208009}
  {arXiv:hep-ex/0208009} \BibitemShut {NoStop}%
\bibitem [{\citenamefont {Alavi-Harati}\ \emph {et~al.}(2003)\citenamefont
  {Alavi-Harati} \emph {et~al.}}]{KTeV:2002qqy}%
  \BibitemOpen
  \bibfield  {author} {\bibinfo {author} {\bibfnamefont {A.}~\bibnamefont
  {Alavi-Harati}} \emph {et~al.} (\bibinfo {collaboration} {KTeV}),\ }\href
  {\doibase 10.1103/PhysRevD.70.079904} {\bibfield  {journal} {\bibinfo
  {journal} {Phys. Rev. D}\ }\textbf {\bibinfo {volume} {67}},\ \bibinfo
  {pages} {012005} (\bibinfo {year} {2003})},\ \bibinfo {note} {[Erratum:
  Phys.Rev.D 70, 079904 (2004)]},\ \Eprint
  {http://arxiv.org/abs/hep-ex/0208007} {arXiv:hep-ex/0208007} \BibitemShut
  {NoStop}%
\bibitem [{\citenamefont {Abouzaid}\ \emph {et~al.}(2011)\citenamefont
  {Abouzaid} \emph {et~al.}}]{KTeV:2010sng}%
  \BibitemOpen
  \bibfield  {author} {\bibinfo {author} {\bibfnamefont {E.}~\bibnamefont
  {Abouzaid}} \emph {et~al.} (\bibinfo {collaboration} {KTeV}),\ }\href
  {\doibase 10.1103/PhysRevD.83.092001} {\bibfield  {journal} {\bibinfo
  {journal} {Phys. Rev. D}\ }\textbf {\bibinfo {volume} {83}},\ \bibinfo
  {pages} {092001} (\bibinfo {year} {2011})},\ \Eprint
  {http://arxiv.org/abs/1011.0127} {arXiv:1011.0127 [hep-ex]} \BibitemShut
  {NoStop}%
\bibitem [{\citenamefont {Blum}\ \emph {et~al.}(2015)\citenamefont {Blum} \emph
  {et~al.}}]{Blum:2015ywa}%
  \BibitemOpen
  \bibfield  {author} {\bibinfo {author} {\bibfnamefont {T.}~\bibnamefont
  {Blum}} \emph {et~al.},\ }\href {\doibase 10.1103/PhysRevD.91.074502}
  {\bibfield  {journal} {\bibinfo  {journal} {Phys. Rev. D}\ }\textbf {\bibinfo
  {volume} {91}},\ \bibinfo {pages} {074502} (\bibinfo {year} {2015})},\
  \Eprint {http://arxiv.org/abs/1502.00263} {arXiv:1502.00263 [hep-lat]}
  \BibitemShut {NoStop}%
\bibitem [{\citenamefont {Abbott}\ \emph {et~al.}(2020)\citenamefont {Abbott}
  \emph {et~al.}}]{RBC:2020kdj}%
  \BibitemOpen
  \bibfield  {author} {\bibinfo {author} {\bibfnamefont {R.}~\bibnamefont
  {Abbott}} \emph {et~al.} (\bibinfo {collaboration} {RBC, UKQCD}),\ }\href
  {\doibase 10.1103/PhysRevD.102.054509} {\bibfield  {journal} {\bibinfo
  {journal} {Phys. Rev. D}\ }\textbf {\bibinfo {volume} {102}},\ \bibinfo
  {pages} {054509} (\bibinfo {year} {2020})},\ \Eprint
  {http://arxiv.org/abs/2004.09440} {arXiv:2004.09440 [hep-lat]} \BibitemShut
  {NoStop}%
\bibitem [{\citenamefont {Cirigliano}\ \emph {et~al.}(2020)\citenamefont
  {Cirigliano}, \citenamefont {Gisbert}, \citenamefont {Pich},\ and\
  \citenamefont {Rodr\'\i{}guez-S\'anchez}}]{Cirigliano:2019cpi}%
  \BibitemOpen
  \bibfield  {author} {\bibinfo {author} {\bibfnamefont {V.}~\bibnamefont
  {Cirigliano}}, \bibinfo {author} {\bibfnamefont {H.}~\bibnamefont {Gisbert}},
  \bibinfo {author} {\bibfnamefont {A.}~\bibnamefont {Pich}}, \ and\ \bibinfo
  {author} {\bibfnamefont {A.}~\bibnamefont {Rodr\'\i{}guez-S\'anchez}},\
  }\href {\doibase 10.1007/JHEP02(2020)032} {\bibfield  {journal} {\bibinfo
  {journal} {JHEP}\ }\textbf {\bibinfo {volume} {02}},\ \bibinfo {pages} {032}
  (\bibinfo {year} {2020})},\ \Eprint {http://arxiv.org/abs/1911.01359}
  {arXiv:1911.01359 [hep-ph]} \BibitemShut {NoStop}%
\bibitem [{\citenamefont {Cline}\ and\ \citenamefont
  {Laurent}(2021)}]{Cline:2021dkf}%
  \BibitemOpen
  \bibfield  {author} {\bibinfo {author} {\bibfnamefont {J.~M.}\ \bibnamefont
  {Cline}}\ and\ \bibinfo {author} {\bibfnamefont {B.}~\bibnamefont
  {Laurent}},\ }\href {\doibase 10.1103/PhysRevD.104.083507} {\bibfield
  {journal} {\bibinfo  {journal} {Phys. Rev. D}\ }\textbf {\bibinfo {volume}
  {104}},\ \bibinfo {pages} {083507} (\bibinfo {year} {2021})},\ \Eprint
  {http://arxiv.org/abs/2108.04249} {arXiv:2108.04249 [hep-ph]} \BibitemShut
  {NoStop}%
\bibitem [{\citenamefont {Aaij}\ \emph {et~al.}(2013)\citenamefont {Aaij} \emph
  {et~al.}}]{LHCb:2012myk}%
  \BibitemOpen
  \bibfield  {author} {\bibinfo {author} {\bibfnamefont {R.}~\bibnamefont
  {Aaij}} \emph {et~al.} (\bibinfo {collaboration} {LHCb}),\ }\href {\doibase
  10.1140/epjc/s10052-013-2373-2} {\bibfield  {journal} {\bibinfo  {journal}
  {Eur. Phys. J. C}\ }\textbf {\bibinfo {volume} {73}},\ \bibinfo {pages}
  {2373} (\bibinfo {year} {2013})},\ \Eprint {http://arxiv.org/abs/1208.3355}
  {arXiv:1208.3355 [hep-ex]} \BibitemShut {NoStop}%
\bibitem [{\citenamefont {Cerri}\ \emph {et~al.}(2019)\citenamefont {Cerri}
  \emph {et~al.}}]{Cerri:2018ypt}%
  \BibitemOpen
  \bibfield  {author} {\bibinfo {author} {\bibfnamefont {A.}~\bibnamefont
  {Cerri}} \emph {et~al.},\ }\href {\doibase 10.23731/CYRM-2019-007.867}
  {\bibfield  {journal} {\bibinfo  {journal} {CERN Yellow Rep. Monogr.}\
  }\textbf {\bibinfo {volume} {7}},\ \bibinfo {pages} {867} (\bibinfo {year}
  {2019})},\ \Eprint {http://arxiv.org/abs/1812.07638} {arXiv:1812.07638
  [hep-ph]} \BibitemShut {NoStop}%
\bibitem [{\citenamefont {Altmannshofer}\ \emph {et~al.}(2019)\citenamefont
  {Altmannshofer} \emph {et~al.}}]{Belle-II:2018jsg}%
  \BibitemOpen
  \bibfield  {author} {\bibinfo {author} {\bibfnamefont {W.}~\bibnamefont
  {Altmannshofer}} \emph {et~al.} (\bibinfo {collaboration} {Belle-II}),\
  }\href {\doibase 10.1093/ptep/ptz106} {\bibfield  {journal} {\bibinfo
  {journal} {PTEP}\ }\textbf {\bibinfo {volume} {2019}},\ \bibinfo {pages}
  {123C01} (\bibinfo {year} {2019})},\ \bibinfo {note} {[Erratum: PTEP 2020,
  029201 (2020)]},\ \Eprint {http://arxiv.org/abs/1808.10567} {arXiv:1808.10567
  [hep-ex]} \BibitemShut {NoStop}%
\bibitem [{\citenamefont {Hou}\ \emph {et~al.}(2020)\citenamefont {Hou},
  \citenamefont {Hsu},\ and\ \citenamefont {Modak}}]{Hou:2020ciy}%
  \BibitemOpen
  \bibfield  {author} {\bibinfo {author} {\bibfnamefont {W.-S.}\ \bibnamefont
  {Hou}}, \bibinfo {author} {\bibfnamefont {T.-H.}\ \bibnamefont {Hsu}}, \ and\
  \bibinfo {author} {\bibfnamefont {T.}~\bibnamefont {Modak}},\ }\href
  {\doibase 10.1103/PhysRevD.102.055006} {\bibfield  {journal} {\bibinfo
  {journal} {Phys. Rev. D}\ }\textbf {\bibinfo {volume} {102}},\ \bibinfo
  {pages} {055006} (\bibinfo {year} {2020})},\ \Eprint
  {http://arxiv.org/abs/2008.02573} {arXiv:2008.02573 [hep-ph]} \BibitemShut
  {NoStop}%
\bibitem [{\citenamefont {Iguro}(2022)}]{Iguro:2022uzz}%
  \BibitemOpen
  \bibfield  {author} {\bibinfo {author} {\bibfnamefont {S.}~\bibnamefont
  {Iguro}},\ }\href {\doibase 10.1103/PhysRevD.105.095011} {\bibfield
  {journal} {\bibinfo  {journal} {Phys. Rev. D}\ }\textbf {\bibinfo {volume}
  {105}},\ \bibinfo {pages} {095011} (\bibinfo {year} {2022})},\ \Eprint
  {http://arxiv.org/abs/2201.06565} {arXiv:2201.06565 [hep-ph]} \BibitemShut
  {NoStop}%
\bibitem [{\citenamefont {Iguro}\ and\ \citenamefont
  {Tobe}(2017)}]{Iguro:2017ysu}%
  \BibitemOpen
  \bibfield  {author} {\bibinfo {author} {\bibfnamefont {S.}~\bibnamefont
  {Iguro}}\ and\ \bibinfo {author} {\bibfnamefont {K.}~\bibnamefont {Tobe}},\
  }\href {\doibase 10.1016/j.nuclphysb.2017.10.014} {\bibfield  {journal}
  {\bibinfo  {journal} {Nucl. Phys. B}\ }\textbf {\bibinfo {volume} {925}},\
  \bibinfo {pages} {560} (\bibinfo {year} {2017})},\ \Eprint
  {http://arxiv.org/abs/1708.06176} {arXiv:1708.06176 [hep-ph]} \BibitemShut
  {NoStop}%
\bibitem [{\citenamefont {Ghosh}\ \emph {et~al.}(2020)\citenamefont {Ghosh},
  \citenamefont {Hou},\ and\ \citenamefont {Modak}}]{Ghosh:2019exx}%
  \BibitemOpen
  \bibfield  {author} {\bibinfo {author} {\bibfnamefont {D.~K.}\ \bibnamefont
  {Ghosh}}, \bibinfo {author} {\bibfnamefont {W.-S.}\ \bibnamefont {Hou}}, \
  and\ \bibinfo {author} {\bibfnamefont {T.}~\bibnamefont {Modak}},\ }\href
  {\doibase 10.1103/PhysRevLett.125.221801} {\bibfield  {journal} {\bibinfo
  {journal} {Phys. Rev. Lett.}\ }\textbf {\bibinfo {volume} {125}},\ \bibinfo
  {pages} {221801} (\bibinfo {year} {2020})},\ \Eprint
  {http://arxiv.org/abs/1912.10613} {arXiv:1912.10613 [hep-ph]} \BibitemShut
  {NoStop}%
\bibitem [{\citenamefont {Ellis}\ \emph {et~al.}(1976)\citenamefont {Ellis},
  \citenamefont {Gaillard},\ and\ \citenamefont {Nanopoulos}}]{Ellis:1975ap}%
  \BibitemOpen
  \bibfield  {author} {\bibinfo {author} {\bibfnamefont {J.~R.}\ \bibnamefont
  {Ellis}}, \bibinfo {author} {\bibfnamefont {M.~K.}\ \bibnamefont {Gaillard}},
  \ and\ \bibinfo {author} {\bibfnamefont {D.~V.}\ \bibnamefont {Nanopoulos}},\
  }\href {\doibase 10.1016/0550-3213(76)90382-5} {\bibfield  {journal}
  {\bibinfo  {journal} {Nucl. Phys. B}\ }\textbf {\bibinfo {volume} {106}},\
  \bibinfo {pages} {292} (\bibinfo {year} {1976})}\BibitemShut {NoStop}%
\bibitem [{\citenamefont {Shifman}\ \emph {et~al.}(1979)\citenamefont
  {Shifman}, \citenamefont {Vainshtein}, \citenamefont {Voloshin},\ and\
  \citenamefont {Zakharov}}]{Shifman:1979eb}%
  \BibitemOpen
  \bibfield  {author} {\bibinfo {author} {\bibfnamefont {M.~A.}\ \bibnamefont
  {Shifman}}, \bibinfo {author} {\bibfnamefont {A.~I.}\ \bibnamefont
  {Vainshtein}}, \bibinfo {author} {\bibfnamefont {M.~B.}\ \bibnamefont
  {Voloshin}}, \ and\ \bibinfo {author} {\bibfnamefont {V.~I.}\ \bibnamefont
  {Zakharov}},\ }\href@noop {} {\bibfield  {journal} {\bibinfo  {journal} {Sov.
  J. Nucl. Phys.}\ }\textbf {\bibinfo {volume} {30}},\ \bibinfo {pages} {711}
  (\bibinfo {year} {1979})}\BibitemShut {NoStop}%
\bibitem [{\citenamefont {Corbin}\ and\ \citenamefont
  {Cornish}(2006)}]{Corbin:2005ny}%
  \BibitemOpen
  \bibfield  {author} {\bibinfo {author} {\bibfnamefont {V.}~\bibnamefont
  {Corbin}}\ and\ \bibinfo {author} {\bibfnamefont {N.~J.}\ \bibnamefont
  {Cornish}},\ }\href {\doibase 10.1088/0264-9381/23/7/014} {\bibfield
  {journal} {\bibinfo  {journal} {Class. Quant. Grav.}\ }\textbf {\bibinfo
  {volume} {23}},\ \bibinfo {pages} {2435} (\bibinfo {year} {2006})},\ \Eprint
  {http://arxiv.org/abs/gr-qc/0512039} {arXiv:gr-qc/0512039} \BibitemShut
  {NoStop}%
\bibitem [{\citenamefont {Cline}\ \emph {et~al.}(2021)\citenamefont {Cline},
  \citenamefont {Friedlander}, \citenamefont {He}, \citenamefont {Kainulainen},
  \citenamefont {Laurent},\ and\ \citenamefont {Tucker-Smith}}]{Cline:2021iff}%
  \BibitemOpen
  \bibfield  {author} {\bibinfo {author} {\bibfnamefont {J.~M.}\ \bibnamefont
  {Cline}}, \bibinfo {author} {\bibfnamefont {A.}~\bibnamefont {Friedlander}},
  \bibinfo {author} {\bibfnamefont {D.-M.}\ \bibnamefont {He}}, \bibinfo
  {author} {\bibfnamefont {K.}~\bibnamefont {Kainulainen}}, \bibinfo {author}
  {\bibfnamefont {B.}~\bibnamefont {Laurent}}, \ and\ \bibinfo {author}
  {\bibfnamefont {D.}~\bibnamefont {Tucker-Smith}},\ }\href {\doibase
  10.1103/PhysRevD.103.123529} {\bibfield  {journal} {\bibinfo  {journal}
  {Phys. Rev. D}\ }\textbf {\bibinfo {volume} {103}},\ \bibinfo {pages}
  {123529} (\bibinfo {year} {2021})},\ \Eprint
  {http://arxiv.org/abs/2102.12490} {arXiv:2102.12490 [hep-ph]} \BibitemShut
  {NoStop}%
\bibitem [{\citenamefont {Sala}(2014)}]{Sala:2013osa}%
  \BibitemOpen
  \bibfield  {author} {\bibinfo {author} {\bibfnamefont {F.}~\bibnamefont
  {Sala}},\ }\href {\doibase 10.1007/JHEP03(2014)061} {\bibfield  {journal}
  {\bibinfo  {journal} {JHEP}\ }\textbf {\bibinfo {volume} {03}},\ \bibinfo
  {pages} {061} (\bibinfo {year} {2014})},\ \Eprint
  {http://arxiv.org/abs/1312.2589} {arXiv:1312.2589 [hep-ph]} \BibitemShut
  {NoStop}%
\bibitem [{\citenamefont {Matsumiya}\ \emph {et~al.}(2022)\citenamefont
  {Matsumiya} \emph {et~al.}}]{TUCAN:2022koi}%
  \BibitemOpen
  \bibfield  {author} {\bibinfo {author} {\bibfnamefont {R.}~\bibnamefont
  {Matsumiya}} \emph {et~al.} (\bibinfo {collaboration} {TUCAN}),\ }in\
  \href@noop {} {\emph {\bibinfo {booktitle} {{24th International Symposium on
  Spin Physics}}}}\ (\bibinfo {year} {2022})\ \Eprint
  {http://arxiv.org/abs/2207.09880} {arXiv:2207.09880 [physics.ins-det]}
  \BibitemShut {NoStop}%
\end{thebibliography}%

\end{document}